%% file: AR_xsec_review.tex
\documentclass[aps,prd,preprint,superscriptaddress,nofootinbib]{revtex4-1}
\pdfoutput=1
\usepackage{url}
\usepackage{footnote}
\usepackage{xspace}
\usepackage{graphicx}
\usepackage{amssymb}
\usepackage[svgnames]{xcolor}
\usepackage[bookmarks=false,colorlinks=true,linkcolor=DarkBlue,citecolor=DarkBlue]{hyperref}
\usepackage{subfigure}
\usepackage{rotating}
\usepackage{units}
\usepackage{amsmath}
\usepackage{bm}
\usepackage{lineno}
\usepackage{listings} 
\usepackage[normalem]{ulem}
\usepackage[section]{placeins}
\usepackage{hepunits}
\usepackage{hepparticles}
\usepackage{cancel}
\usepackage{hepnames}
\usepackage{epstopdf}
\usepackage{mathtools}
\usepackage[capitalise]{cleveref}
\usepackage{braket}
\usepackage{slashed}
\usepackage{xcolor}
\usepackage{multirow}
\usepackage{tablefootnote}

\newcommand{\mb}{MiniBooNE\xspace}
\newcommand{\minerva}{MINER$\nu$A\xspace}

\newcommand{\enuqe}{\ensuremath{E_{\nu}^{\mathrm{QE}}}\xspace}
\newcommand{\qqqe}{\ensuremath{Q^{2}_{\mathrm{QE}}}\xspace}

\newcommand{\ma}{\ensuremath{M_{\mathrm{A}}}\xspace}

\def\bracketbar{\hbox{\kern-7pt\raise1pt%
         \hbox{{\tiny(}{\lower1.5pt\hbox{\bf--}}{\tiny)}}}}

\definecolor{NALblue}{rgb}{0,0.2,0.6}
\definecolor{NALgreen}{rgb}{0.317647,0.603922,0.141176}
\definecolor{NALorange}{rgb}{0.858824,0.447059,0.0470588}

\begin{document}

\title{Progress in measurements of 0.1--10 GeV neutrino-nucleus scattering and anticipated results from future experiments}
\date{\today}
\author{Kendall Mahn}
\affiliation{Department of Physics and Astronomy, Michigan State University, East Lansing, USA, 48823}
\email{mahn@pa.msu.edu}
\author{Chris Marshall}
\affiliation{Lawrence Berkeley National Laboratory, Berkeley, USA, 94720}
\author{Callum Wilkinson}
\affiliation{University of Bern, Albert Einstein Center for Fundamental Physics, Laboratory for High Energy Physics (LHEP), Bern 3012, Switzerland}

\begin{abstract}
Neutrino interactions with nuclei have been the subject of intense interest over the last 15 years. Current and future measurements of neutrino oscillation and exotic physics use order 0.1--10 GeV neutrinos on a range of nuclear targets ($^{12}$C, $^{16}$O, $^{40}$Ar). As the precision of these experiments has increased, information from their detectors and dedicated experiments indicate deficiencies in the modeling of neutrino interactions on nuclear targets. Here, we present the current state of knowledge about neutrino-nucleus interactions, the challenge of extracting the cross section of these processes, and current experimental puzzles in the field. We also look forward to new and novel measurements and efforts in the future which seek to resolve these questions.
\end{abstract}

\maketitle

\tableofcontents

\input{intro}

\input{methodology}
\input{qe}

\input{spp}
\input{rare}

\input{inclusive}

\input{future}
\input{summary}

\section*{Disclosure statement}
While the authors experimental affiliations do not affect the objectivity of this review, we wish to include them for transparency. KM and CW are current members of the T2K and DUNE collaborations. KM is a former member of the K2K, SciBooNE and MiniBooNE collaborations. CM is a former member of MINERvA, and a current member of the DUNE experiment. 
The authors are not aware of any funding, or financial holdings that might be perceived as affecting the objectivity of this review. 

\begin{acknowledgements}
We acknowledge the support of the Office of Science, Office of High Energy Physics, of the U.S. Department of Energy under Contract Number DOE OHEP DE-AC02-05CH11231 and award DE-SC0015903, the Swiss National Science Foundation and SERI, and the Alfred P. Sloan Foundation.

This document was prepared as an account of work sponsored by the United States Government. While this document is believed to contain correct information, neither the United States Government nor any agency thereof, nor the Regents of the University of California, nor any of their employees, makes any warranty, express or implied, or assumes any legal responsibility for the accuracy, completeness, or usefulness of any information, apparatus, product, or process disclosed, or represents that its use would not infringe privately owned rights. Reference herein to any specific commercial product, process, or service by its trade name, trademark, manufacturer, or otherwise, does not necessarily constitute or imply its endorsement, recommendation, or favoring by the United States Government or any agency thereof, or the Regents of the University of California. The views and opinions of authors expressed herein do not necessarily state or reflect those of the United States Government or any agency thereof or the Regents of the University of California.
\end{acknowledgements}

\bibliographystyle{apsrev4-1}
\bibliography{AR_xsec_review}

\appendix
\section{Reference model}\label{app:appendix}
In this review, we used one of the out-of-the-box configuration of GENIE v2.12.8 as a reference model. The configuration is labeled ``ValenciaQEBergerSehgalCOHRES'' in GENIE, and is described below. We note that this configuration is not the default option for this GENIE version, but is provided by the GENIE authors. We chose this configuration option based on internal model consistency and because it includes many recent theoretically motivated improvements. No additional tuning is applied.

{\bf QE:} The QE model is a Local Fermi Gas (LFG), with momentum distributions based on Nieves et. al~\cite{Nieves:2004wx} and uses an axial mass of $M_\mathrm{A}^{\mathrm{QE}}=1.05$ GeV. This model includes the Relativistic Phase Approximation (RPA) and Coulomb effects for the outgoing muon.

{\bf Nuclear model:} For all processes other than QE, the Bodek-Ritchie model~\cite{BodekRitchie} is used to to describe the initial state nucleon momentum distribution.

{\bf 2p2h:} A consistent model with QE is used for multi-nucleon processes~\cite{nieves_2011,Gran:2013kda}, the GENIE implementation for which is described in Ref.~\cite{Schwehr:2016pvn}. 

{\bf $1\pi$:} The resonant model is based on the Rein-Sehgal model~\cite{Rein-Sehgal, Rein:2006di}, which has been updated to include lepton mass corrections to the phase space limits (although not the cross section calculation itself). Details of the GENIE resonant model can be found in Refs.~\cite{Andreopoulos:2009rq, Rodrigues:2016xjj}. The coherent model is based on Berger-Seghal~\cite{Berger:2008xs} in the configuration used in this work.

{\bf Transition region and DIS:} The Bodek-Yang~\cite{bodek-yang} model is used to model the DIS cross section, with hadronization described by the AKGY model~\cite{Yang:2009zx}. These models also provide the non-resonant background under the resonant region for $W \geq 1.7$ GeV.

{\bf FSI model:} Final state interactions are described by a custom GENIE model, tuned to hadron-scattering data, where a single interaction is used to approximate cascade model behavior~\cite{Andreopoulos:2009rq} is applied to all processes.

\end{document}

%% file: intro.tex
\section{Introduction}
\label{sec:intro}

Neutrino interactions in the medium energy 0.1--10 GeV range are of great interest to a variety of physics programs\footnote{An exciting program of physics is also possible with lower energy neutrinos produced from supernova, the Sun, reactors and astrophysical sources, but in general, the interaction rates are better known and outside the scope of this review (see, for example, Ref.~\cite{zeller12} for a recent review.)}. In particular, accelerator neutrino oscillation experiments use intense beams of predominantly muon neutrinos or anti-neutrinos, sent over short (km) or long (hundreds of km) distances in order to measure non-standard or three-flavor oscillations~\cite{Patrignani:2016xqp}. In these experiments, neutrino oscillations are inferred from flavor transitions between the source and detector, either by measuring the disappearance of muon (anti-)neutrinos, or the appearance of electron (anti-)neutrinos. In a disappearance measurement, a deficit of muon (anti-)neutrinos is observed, due to oscillations to electron or tau flavor. In an electron (anti-)neutrino appearance search, an excess of electron-like candidates is observed over background. The determination of the neutrino mass ordering and search for CP violation is possible by comparing neutrino and antineutrino oscillation rates in both modes.

Experiments cannot measure the oscillation probability directly, instead, it is inferred from the charged-current (CC, $\nu^{\bracketbar}_{l} + A \rightarrow l^{-(+)} + X$) and neutral-current (NC, $\nu^{\bracketbar}_{l} + A \rightarrow \nu^{\bracketbar}_{l} + X$) event rates, which can generically be expressed as
\begin{equation}
  R(\vec{\mathbf{x}}) = \sum_{i}^{\mathrm{process}} \sum_{j}^{\mathrm{target}} \int_{E_{\mathrm{min}}}^{E_{\mathrm{max}}}\Phi(E_\nu) \times \sigma_{i}(E_{\nu}, \vec{\mathbf{x}}) \times \epsilon(\vec{\mathbf{x}}) \times N_{j} \times P(\nu_{A} \rightarrow \nu_{B})
  \label{eq:event_rate}
\end{equation}
\noindent where $R(\vec{\mathbf{x}})$ is the total event rate for all processes as a function of the reconstructed kinematic variables $\vec{\mathbf{x}}$, $\Phi_{\nu} (E_{\nu})$ is the neutrino flux as a function of the neutrino energy $E_{\nu}$, $\sigma_{i}$ is the neutrino cross section for a particular mode, $\epsilon$ is the detector efficiency and $N_{j}$ is the number of target nuclei in the detector fiducial volume for target type $j$. To infer the oscillation probability from the event rate, it is necessary to construct an estimator for the neutrino energy from the measurable quantities in the detector, the reconstructed kinematic variables, $\vec{\mathbf{x}}$, and particle composition of the final state. This implicitly relies on the cross section $\sigma_{i}(E_{\nu}, \vec{\mathbf{x}})$, which is why the cross section model has such a central importance for neutrino oscillation experiments.

\begin{figure}[htbp]
  \includegraphics[width=0.6\textwidth]{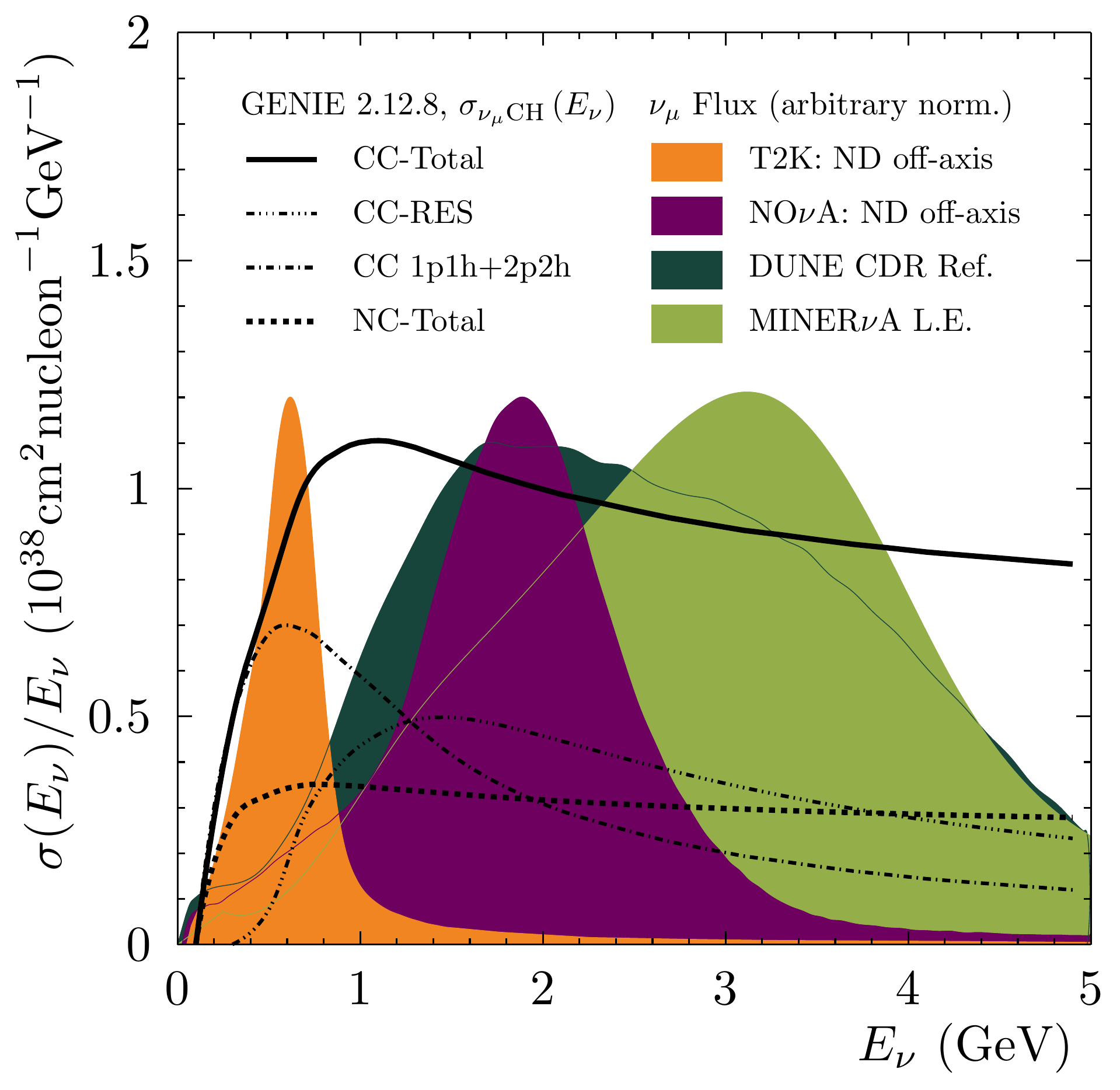}
  \caption[fragile]{Neutrino fluxes, as a function of energy, of current and future accelerator-based neutrino experiments. The fluxes relevant to neutrino interaction measurements are shown, from the T2K experiment off-axis near detector (ND)~\cite{Abe:2012av}, NOvA ND~\cite{novacommunication}, and MINERvA (Low Energy configuration only~\cite{Aliaga:2016oaz}). One future program, DUNE~\cite{Acciarri:2015uup}, is shown; the HK flux is very similar to the T2K ND off-axis flux. Also overlaid are the total CC cross section divided by energy, according to the GENIE 2.12.8 reference model (described in Appendix~{\hyperlink{sec:appendix}{A}}).}
  \label{fig:flux_and_xsec}    
\end{figure}
Figure~\ref{fig:flux_and_xsec} shows the total CC and NC cross sections for $\nu_{\mu}$--CH interactions as a function of neutrino energy, $E_{\nu}$. In CC interactions, the flavor of the neutrino $\nu^{\bracketbar}_{l}$ is by definition the flavor of the charged lepton produced, the charge of which also distinguishes neutrinos from anti-neutrinos. The main contributions to the CC cross section for the GENIE 2.12.8 reference model used throughout this review are also shown. The reference model is described in Appendix~{\hyperlink{sec:appendix}{A}}, and is an up to date Monte Carlo simulation used widely in the field. At low $E_{\nu} \lesssim 1$ GeV, the cross section is dominated by the lowest energy transfer process, charged-current quasi-elastic (CCQE) scattering ($\;\nu^{\bracketbar}_{l} + A \rightarrow l^{-(+)} + A' + p(n)$), which is also called single nucleon knock-out, or 1p1h (one particle, one hole). There are also significant contributions from multiple-nucleon knock-out (2p2h), where the interaction is with more than one nucleon in the nucleus ($\nu_{l} + A(n,p) \rightarrow l^{-} + A'+p+p$ and $\overline{\nu}_{l} + A(n,p) \rightarrow l^{+} + A'+n+n$\footnote{Note this is an approximation for illustration. There are significant questions about the proton, neutron composition of the final state for 2p2h.}). At higher energy transfers, nucleon resonances (RES) can be excited, which decay to produce pions and nucleons in the final state ($\;\nu^{\bracketbar}_{l} + N \rightarrow \nu^{\bracketbar}_{l} + N' + \pi^{+/-/0}$). At large energy transfers, the interaction is with a constituent quark within the nucleon, the so-called deep inelastic scattering (DIS) region. DIS is well understood, and dominates for $E_\nu \gtrsim 10$ GeV. The transition between these different interaction modes is a significant challenge to model because of the change in the underlying physical assumptions between them. Figure~\ref{fig:flux_and_xsec} also shows the neutrino flux distributions for current and planned long baseline oscillation experiments T2K (also Hyper-K), NOvA, and DUNE. It is clear that the fluxes for these experiments include all of these processes, including the difficult transition region.

In addition to the broad neutrino flux, the choice of target material is important for understanding neutrino interactions relevant for oscillation programs. Neutrino experiments use medium to heavy nuclei, including carbon, water, and argon, as interaction targets. As these have a higher cross section than hydrogen, this choice boosts event rates for, typically statistics limited, oscillation experiments.  However, interactions on any target material that is not hydrogen will also include nuclear effects. Measuring neutrino interactions on nuclear targets convolves single nucleon uncertainties with significant uncertainties related to the nucleus. These nuclear effects include initial-state interactions between nucleons, as well as final-state interactions (FSI) between outgoing hadrons and the residual nucleus.

Neutrino interaction uncertainties are currently one of the largest sources of systematic uncertainty for operating experiments. As an example, the T2K experiment's largest source of systematic uncertainty on the $\nu_e$-appearance rate in an initially $\nu_\mu$ dominated beam come from  cross section related uncertainties. The total fractional uncertainty is 5.4\%, with 3.9\% from the interaction model, with comparable uncertainties on $\overline{\nu}_e$-appearance rates~\cite{Abe:2017vif}. 
Bias in the determination oscillation parameters is possible due to mis-modeling of neutrino interactions, as has been demonstrated by multiple groups~\cite{FernandezMartinez:2010dm,Lalakulich:2012hs,Meloni:2012fq,Coloma:2013rqa,Martini:2012uc,Coloma:2013tba,Jen:2014aja,Abe:2015awa,Ankowski:2016bji,Abe:2017vif,Mosel:2013fxa,Ankowski:2015kya}.

Exotic physics searches are also accessible with accelerator-produced neutrino beams. Searches for ``sterile'' neutrinos, neutrinos which do not couple to the weak force, use accelerator neutrino beams, and detectors close to the neutrino production point to search for oscillation inconsistent with three flavors~\cite{Patrignani:2016xqp}. Sterile neutrino appearance and disappearance searches are similar to the three-flavor case, but may have larger backgrounds due to a smaller oscillation probability. NC interactions are also used as the signal channel for exotic physics. In NC disappearance searches, the presence of sterile neutrinos would reduce the observed number of active neutrinos determined from NC interactions~\cite{Adamson:2011ku}. Other exotic physics which use NC events as signal are searches for sub-GeV dark matter particles production, visible as an excess of neutrino NC-like interactions~\cite{Aguilar-Arevalo:2017mqx}. These kinds of analyses rely on a prediction for what NC interactions look like in the detector to characterize the signal.

Searches for proton decay, which is predicted by Grand Unified Theories (GUTs)~\cite{Patrignani:2016xqp}, also rely on a good understanding of neutrino cross sections, as atmospheric neutrino interactions can mimic proton decay signatures, and are the dominant background. The simplest GUTs predict $p \rightarrow e^+ \pi^0$ to be the dominant proton decay channel. For free protons at rest, the signal is a back-to-back positron and $\pi^0$, but the energy and angle of the decay products are smeared by initial-state nucleon momentum in a bound nucleus, as well as by FSI. Neutrinos can also produce an electron and pion in a detector $\nu_{e} n \rightarrow e^- p \pi^0$, where the electron and pion can conspire to give an invariant mass near that of a proton. In Super-Kamiokande (SK), which has the most stringent experimental limits on proton decay for this channel and in general~\cite{SK_epi0,SK_Knu}, this is indistinguishable from signal as SK has a high threshold to detect protons and cannot distinguish electrons from positrons. Another important decay channel is $p \rightarrow K^+ \bar{\nu}$, which is favored by supersymmetric GUTs. Since the outgoing neutrino is not observed, the signal is a single kaon, possibly with a de-excitation photon from the residual nucleus. Neutral-current neutrino production of $K^+$, $\nu^{\bracketbar}_{l} p \rightarrow \nu^{\bracketbar}_{l} n K^{+}$, also has the same signature in SK. Due to the unique nature of the proton decay signal, typical background rates from atmospheric neutrinos are on the order of a few events per Megaton year. Neutrino interactions processes contributing to proton decay backgrounds are very rare, or have very limited phase space relevant for proton decay. Thus, it is important to consider not only the uncertainty on the rate of the underlying interaction process that contributes to the proton decay background, but also uncertainties on the shape, for example in kinematics that affect the invariant mass reconstruction.

Over the last decade, an numerous neutrino scattering measurements have been made with neutrino and antineutrino beams on a range of nuclear targets. Table~\ref{tab:expt} summarizes recent and planned experiments, including the choice of target material, beam properties and detection methods.

\begin{table}[h]
\tabcolsep7.5pt
\caption{Recent and planned near-term neutrino scattering experiments. The peak energy is listed, and in general the mean of the flux and the peak are not the same. Future measurements are denoted with an asterisk.}
\label{tab:expt}
\begin{center}
\begin{tabular}{@{}l|c|c|c|c@{}}
\hline
\hline
Experiment & Flavor & Flux Peak (GeV) & Target & Detection \\
\hline
NOMAD & $\nu_\mu$,$\overline{\nu}_\mu$ & 10, 5  & C & Tracking \\
K2K & $\nu_\mu$ & 1.3  & CH, H$_2$O & Tracking \\
MiniBooNE\tablefootnote{MiniBooNE is also sensitive to a mono-energetic beam of neutrinos from kaon decay at rest.}$^{\rm b}$ & $\nu_\mu$,$\overline{\nu}_\mu$ & 0.6, 0.4  & CH$_2$ & Cherenkov \\
MINOS & $\nu_\mu$ & 3.0 & Fe  & Tracking \\
ArgoNeuT & $\nu_\mu$,$\overline{\nu}_\mu$ &  6.5,5.5 & Ar  & Tracking+Calorimetry \\
SciBooNE & $\nu_\mu$,$\overline{\nu}_\mu$ & 0.6,0.4  & CH & Tracking \\
T2K off-axis & $\nu_\mu$,$\overline{\nu}_\mu$,$\nu_e$,$\overline{\nu}_e$$^{\rm *}$ & 0.6,0.6,1,1 & CH,H$_2$O & Tracking \\
T2K on-axis & $\nu_\mu$,$\overline{\nu}_\mu$$^{\rm *}$ & 1,1 & CH, H$_2$O$^{\rm *}$, Fe & Tracking \\
MINERvA\tablefootnote{MINERvA has the capability to run in different beam energies, and will have data at higher energies.}$^{\rm a}$ & $\nu_\mu$,$\overline{\nu}_\mu$,$\nu_e$ & 3.5 & He$^{\rm *}$, C, CH, & Tracking+Calorimetry \\
   &  &  & H$_2$O$^{\rm *}$, Fe, Pb&  \\
NOvA$^{\rm *}$ & $\nu_\mu$,$\overline{\nu}_\mu$,$\nu_e$, $\overline{\nu}_e$ & 2,$^{\rm *}$,1,$^{\rm *}$ & CH$_2$ & Tracking+Calorimetry \\
MicroBooNE$^{\rm *}$ & $\nu_\mu$ & 0.6  & Ar & Tracking+Calorimetry \\
ANNIE$^{\rm *}$ & $\nu_\mu$ & 0.6  & H$_2$O & Cherenkov \\
WAGASCI$^{\rm *}$ & $\nu_\mu$,$\overline{\nu}_\mu$ & 1,1  & H$_2$O & Tracking \\
NINJA$^{\rm *}$ & $\nu_\mu$,$\overline{\nu}_\mu$ & 1,1  & H$_2$O, Fe & Emulsion \\
\hline
\hline
\end{tabular}
\end{center}
\end{table}

Progress in the field has resulted in improved understanding of experimental techniques and new measurements. In Section~\ref{sec:method}, we give an overview of how neutrino cross-section measurements are generally made, and discuss how these methods can lead to a hidden reliance on the Monte Carlo model used to simulate the experiment during the analysis. Then we go on to give an overview of the available neutrino-nucleus scattering data, and discuss the compatibility of different datasets with each other and with theoretical models. We note there is an important, vibrant program in model development from theory and experimental groups~\cite{Alvarez-Ruso:2014bla,Garvey:2014exa,Mosel:2016cwa,Katori:2016yel,Alvarez-Ruso:2017oui}, not discussed here, but which facilitates the comparisons and discussion in this work. In Sections~\ref{sec:qe} and~\ref{sec:spp} we describe QE-like and single pion production measurements respectively. Rare processes are described in Section~\ref{sec:rare} and inclusive measurements are discussed in Section~\ref{sec:incl}. We make extensive use of a reference theoretical model to facilitate comparisons to data, which is described in Appendix~{\hyperlink{sec:appendix}{A}}. Finally, we discuss expected future results and experiments in Section~\ref{sec:future}, and present a summary in Section~\ref{sec:summary}.

%% file: methodology.tex
\section{Making neutrino interaction measurements}
\label{sec:method}

\subsection{What can we measure?}
Cross section measurements are extracted from the event rate measured in a detector, and rearranging Equation~\ref{eq:event_rate}. Generally they are made close to the neutrino production point, where $P(\nu_{A} \rightarrow \nu_{B}) = 0$, and oscillations can be neglected. Unfortunately, we cannot observe interaction level processes on nucleons. After a neutrino interacts inside a nucleus, particles produced at the vertex have to propagate through the dense nuclear medium, where many outgoing hadrons will re-interact (FSI). As we do not want to assume a model for these processes, we can only measure cross sections for a given post-FSI observable particle content, or topology, $\widetilde{\sigma}_k$:
\begin{equation}
  \widetilde{\sigma}_{k}(\vec{\mathbf{y}}) = \sum_i \int_{E_{\mathrm{min}}}^{E_{\mathrm{max}}} \sigma_{i}(E_{\nu}, \vec{\mathbf{x}}) \times \mathrm{FSI}(\vec{\mathbf{x}}, \vec{\mathbf{y}}) dE_{\nu},
  \label{eq:topology_xsec}
\end{equation}
\noindent where $\sigma_{i}$ is the contribution of true process category (mode) $i$ to the final state topology $k$. For example, we cannot in general measure CCQE, all we can observe are events with a topology of a single charged lepton and no pions (CC0$\pi$). The CC0$\pi$ topology could be further sub-divided, for example by the number of protons in the final states, CC0$\pi$+0p,  CC0$\pi$+1p, and so on.

Clearly there is some disconnect between what we can measure, and what we want to know for oscillation experiments. Firstly, we cannot directly measure the cross section, $\sigma_{i}(E_{\nu}, \vec{\mathbf{x}})$, for the $i$th mode as is required for oscillation analyses, instead, we measure some topological cross section $\widetilde{\sigma}_k(\vec{\mathbf{y}})$, which is averaged over the broad neutrino energy distribution of an experiment\footnote{We note that there is confusion in the literature over whether to call this a ``flux-averaged'' or ``flux-integrated'' cross section and that experiments may calculate this differently.} Secondly, because the incoming neutrino energy is not known on an event by event basis, in general it is not possible to measure cross sections as a function of ``interaction level'' variables, like energy transfer or neutrino energy.
Instead, measurements must be made as a function of outgoing particle kinematics $\vec{\mathbf{y}}$, such as $p_{\mu}$, $\cos\theta_{\mu}$, or some variable derived from a combination of them. 
Theorists and other users of neutrino scattering data (such as oscillation analyzers) want to compare and constrain their predictions for $\sigma_{i}(E_{\nu}, \vec{\mathbf{x}})$ using measurements of $\widetilde{\sigma}_k(\vec{\mathbf{y}})$. Because of the complexity of this comparison, a lot of data is required, ideally from different experiments with different neutrino fluxes.

\subsection{Cross section extraction method}

Cross section measurements typically use the same formula to extract the differential cross section in the $i$th bin of a true kinematic variable $x$, $d\sigma/dx_{i}$:
\begin{equation}
  \frac{d\sigma}{dx_{i}} = \frac{\sum_{j} \widetilde{U}^{-1}_{ij} \left(N_{j} - B_{j}\right)}{\Phi_{\nu}T\Delta x_{i} \epsilon_{i}},
  \label{eq:trad_xsec}
\end{equation}
\noindent where $N_{j}$ ($B_{j}$) is the number of selected (predicted background) events in reconstructed bin $j$, $\Phi_{\nu}$ is the total integrated flux, $T$ is the number of target nuclei, $\Delta x_{i}$ is the width of the true bin, $\epsilon_{i}$ is the selection efficiency, and $\widetilde{U}^{-1}_{ij}$ is the unfolding matrix (the pseudoinverse of the matrix, $U_{ij}$ which describes detector smearing). Equation~\ref{eq:trad_xsec} is readily extended to an $N$-dimensional differential cross section measurement by allowing indices $i$, $j$ to identify a single bin in $N$-d space, and replacing $\Delta x_{i}$ with the product of the $N$ bin widths.  

The experiment's Monte Carlo simulation is used to calculate the efficiency of the detector response to the particle(s) in the interaction.  In principle, efficiency corrections are only dependent on properties of the detector modeling, which is relatively well understood.

\subsubsection{Selection}
Somewhat obviously, the aim of selection cuts is to increase the proportion of signal events. Two important metrics are the selection efficiency, and purity:
\begin{align*}
  \text{Purity} = \frac{\text{\# selected signal}}{\text{\# total selected}}; && \text{Efficiency} = \frac{\text{\# selected signal}}{\text{\# total signal}}.
\end{align*}

A low efficiency selection is more susceptible to variations in the signal or background model, although the efficiency does not have to be high to make a measurement as long as it is well understood. A low purity selection is more susceptible to how background processes are modeled. In a real analysis, there is a trade-off between the efficiency and purity of a selection, and cuts are generally optimized by maximizing a figure of merit which combines the two, such as selection efficiency $\times$ purity. 

For analyses when the purity is low, the model dependence of the background is critical, and additional background-enhanced ``control'' samples are used. Often, using the same language as collider physics, these are referred to as ``sidebands'', which comes from analyses where regions to the side of a peak in some variable could be used to effectively constrain the background under the peak. For neutrino cross section measurements, the same principle applies, a region distinct from the signal selection is selected, and used to constrain the background. However, it is not possible to cut on the variable of interest, so instead some selection cut is inverted. The background samples can then be used to constrain the backgrounds and reduce the modeling uncertainty. 

\subsubsection{Unfolding}
Unfolding, or deconvolution, is the general term for removing the effect of a measuring device from a measurement~\cite{Prosper:2011zz, kuusela_master, kuusela_phd}. In the context of Equation~\ref{eq:trad_xsec}, this means taking a measurement in bins of reconstructed variable, $j$, and producing a result in bins of true variable, $i$, and is performed by $\widetilde{U}^{-1}_{ij}$, the pseudoinverse of the smearing matrix $U_{ij}$, which describes the mapping between true and reconstructed bins. In the simplest case, one can simply invert the smearing matrix, but this often leads to large fluctuations in the data, as statistical uncertainty in the reconstructed space is blown up by the inversion~\cite{Prosper:2011zz, kuusela_master, kuusela_phd}. This is because neighboring bins in true space have very similar responses in reconstructed space, and so a small fluctuation in reconstructed space can be fit with a large fluctuation of one true bin, with compensating changes in neighboring true bins to preserve the total normalization, giving the characteristic bin-to-bin anti-correlations. Various methods exist for regularizing, or smoothing, unfolded results, where the allowed solutions are restricted to those in which the result fits some prior expectation~\cite{Prosper:2011zz, kuusela_master, kuusela_phd}. For example, we do not expect large fluctuations in the cross section on small scales, so smoothing out the large oscillations in the result may seem reasonable. The regularized unfolding matrix is therefore the pseudoinverse of the smearing matrix.

The most popular method used in the field is D'Agostini unfolding~\cite{D'Agostini:1994zf}, which has been characterized as an algorithm for maximum-likelihood estimation with early stopping~\cite{kuusela_master, kuusela_phd}\footnote{We note that although this method is widely known as ``Bayesian unfolding'' in the field, this is a misnomer as it is a fully frequentist method~\cite{kuusela_master}.}. When the algorithm converges, the result is equivalent to a matrix inversion in the Gaussian limit~\cite{kuusela_master, kuusela_phd}. 
Another popular approach to the unfolding problem is template likelihood fitting, where each true bin $i$ has a template in the reconstructed bins $j$ which describes the true to reconstructed smearing, essentially a single column of the matrix $U_{ij}$. By fitting the normalizations of these templates, and minimizing some goodness-of-fit statistic between the prediction in reconstructed space and the data in reconstructed space, the unsmeared data in true space is found~\cite{kuusela_master, kuusela_phd}. In the Gaussian limit, and with no regularization, this method is also functionally equivalent to matrix inversion, but naturally supports non-Gaussian statistics by using a Poisson-Likelihood as the goodness-of-fit statistic~\cite{kuusela_master, kuusela_phd}. It is also easy to add additional regularization terms to the fit. This iterative forward folding technique is very robust for large problems because there is no need to invert a large matrix (which could even be singular). The prediction at each step of the fit is formed by folding the Monte Carlo prediction with the smearing matrix $U_{ij}$, which can be easily calculated.

A third approach is to not unfold; this can be expressed (in comparison with Equation~\ref{eq:trad_xsec}) as
\begin{align}
  \frac{d\sigma}{dx_{j}} = \frac{\left(N_{j} - B_{j}\right)}{\Phi_{\nu}T\Delta x_{j}}, && \frac{d\sigma^{th}}{dx_{j}} = \sum_{i}U_{ij}\;\sigma^{th}_{i},
  \label{eq:smeared_xsec}
\end{align}
\noindent where the cross section result is in the reconstructed space, and a theoretical prediction can be smeared from the true to the reconstructed space with the smearing matrix $U_{ij}$. Examples of results which have been presented in reconstructed space can be found in Refs.~\cite{Tice:2014pgu, AguilarArevalo:2010cx}.

\subsubsection{Systematic uncertainties}
\label{sec:syst}

Flux, detector and cross section systematic uncertainties affect Equation~\ref{eq:trad_xsec} in various ways. 
Uncertainty on the detector response is typically determined by  a combination of calibration samples (e.g, cosmic muons, electrons from pair production) in conjunction with standalone tests of the detector (e.g. measurement of response of the light detection system). In addition, charged particle beam data using a similar or identical detector,  can be used to validate the detector model or reduce detector uncertainties. 

Often the limiting systematic uncertainty for cross section measurements are flux uncertainties. For a conventional neutrino beam, which focuses mesons produced in collisions between accelerated protons off of a target, the uncertainties on the resulting flux prediction are approximately ~10\%, due to dedicated hadron production experiments~\cite{barton_1983, Ambrosini:1998et, Ambrosini:1998th, Raja:2005kow, Catanesi:2007ig, Abgrall:2014xwa}. In-situ constraints from measurements of processes with known cross sections can further reduce uncertainties on the flux prediction for neutrino cross section. Two primary methods have been used: neutrino-electron elastic scattering $\nu e \rightarrow \nu e$, and the low-$\nu$ technique.

{\bf Neutrino-electron elastic scattering} Neutrino-electron elastic scattering is a pure electroweak process, with a known cross section up to radiative corrections. For $\nu_{e}$, a charged-current contribution interferes with the neutral-current contribution that is equal for all flavors, and enhances the cross section by about a factor of 6. The signal is a single, forward electron and no other particles, subject to the the kinematic constraint for the two-body process

\begin{align}
\label{eq:nue_kinematics}
E_{\nu} &= \frac{E_{e}}{1-\frac{E_{e}(1-\cos\theta)}{m}} \approx \frac{E_{e}}{1-\frac{E_{e}\theta^{2}}{2m}}, && E_{e}\theta_{e}^{2} = 2m_{e}(1-y) < 2m_{e}.
\end{align}

\noindent
Backgrounds typically come from charged-current processes initiated by $\nu_{e}$ at very low $Q^{2}$, and from neutral-current $\pi^{0}$ production where a photon from $\pi^{0} \rightarrow \gamma \gamma$ is mistaken for an electron.

MINERvA demonstrated a technique for constraining the NuMI flux with a sample of 127 $\nu e \rightarrow \nu e$ candidate events, on an expected background of 30 and an estimated signal efficiency of 73\%~\cite{minerva_nue}. 
The uncertainty on the total flux was reduced to 6\%. In principle, this technique is somewhat sensitive to the flux shape. However, the $E_{e}$ spectrum does not uniquely determine the neutrino flux energy distribution, and in practice very little improvement to the shape uncertainties is obtained, compared to the \textit{a priori} prediction constrained by hadron production. Better shape resolution could be obtained by using the electron angular information in addition to $E_{e}$ and measuring the neutrino energy directly as in Equation~\ref{eq:nue_kinematics}, but the feasibility of this has not been demonstrated for a realistic detector.

{\bf The Low-$\nu$ method} The differential cross section for charged-current cross neutrino scattering as a function of the hadronic energy, $\nu$, can be written as

\begin{equation}
\label{eq:lownu}
\frac{d\sigma}{d\nu} = A + B\frac{\nu}{E} - C\frac{\nu^{2}}{2E^{2}}
\end{equation}

\noindent
where $E$ is the neutrino energy, and the coefficients $A$, $B$, and $C$ are integrals of structure functions. The cross section is independent of the neutrino energy when $\nu/E$ is small. This can be used to measure the relative shape of the neutrino flux \textit{in situ} by selecting a sample of events $\nu$ below some fixed value $\nu_{0}$. 
This technique was first used by high energy experiments CCFR~\cite{ccfr_lownu} and NuTeV~\cite{nutev_lownu}.

Extending the low-$\nu$ technique to lower neutrino energies is challenging~\cite{BodekSarica_lownu}. If $\nu_{0}$ is too large, the $B$ and $C$ terms in Equation~\ref{eq:lownu} become significant and introduce modeling uncertainties. However, a very small $\nu_{0}$ limits the statistical power of the sample. To use a low-$\nu$ flux constraint in a cross section measurement, it is also important that the $\nu < \nu_{0}$ sample does not overlap significantly with the signal. MINOS~\cite{minos_lownu} and MINERvA~\cite{minerva_lownu} have dealt with these challenges by using different $\nu_{0}$ cuts for different regions of neutrino energy.

\subsubsection{Neutrino interaction model uncertainties and biases in measurements} 
\label{sec:biases}
In an ideal scenario, a neutrino cross section measurement should only depend on data and the detector response model, not the neutrino interaction model used by a particular experiment. We note some of the conditions where the dependence of the experiment's choice of signal or background model enters into cross section measurements, which can result in larger uncertainties on the measurement, or bias the result.  We also suggest possible future approaches to mitigate or minimize these issues.

{\bf Treatment of backgrounds:} The metric of efficiency $\times$ purity is often used, which is equivalent to $S/\sqrt{(S+B)}$ where $S$ ($B$) is the number of signal (background) events. This is only optimal when the systematic uncertainty on the background can be neglected. This is appropriate for many applications in particle physics, such as peak finding with a background which has a well understood shape, but is not the case for medium-energy neutrino cross section measurements, where the background processes are generally not well understood.

The use of background enhanced control regions helps mitigate some of the dependence on the background model, although the extrapolation of the background constraint into the signal region still relies on the model. This model dependence can be mitigated by defining background control regions with similar underlying kinematics to the background in the signal region.

{\bf Efficiency and phase space:} There are a number of ways that the efficiency can couple to the signal interaction model. The most obvious is when there are regions of phase space which have zero efficiency. For example, if you measure a cross section in terms of the outgoing muon kinematics, $p_{\mu}$, $\cos\theta_{\mu}$, but have no sensitivity to backward ($\cos\theta_{\mu} < 0$) events, reporting a single-differential $d\sigma/dp_{\mu}$ cross section without restricting the $\cos\theta_{\mu}$ range would add model dependence, as the efficiency correction would be correcting for missing events without any data at all to inform the backward region. This example serves to highlight a more subtle issue of regions of rapidly changing efficiency. Reporting a $d\sigma/dp_{\mu}$ cross section using Equation~\ref{eq:trad_xsec} where the efficiency correction is integrated over any region in $\cos\theta_{\mu}$ where the efficiency is not flat implicitly adds Monte Carlo bias, as it assumes that the relationship between different regions of $\cos\theta_{\mu}$ for that $p_{\mu}$ bin is correctly modeled.  This problem can be mitigated by restricting the phase space to regions of phase space with flat efficiencies, and can be avoided by extracting a two-dimensional $d^{2}\sigma/dp_{\mu}d\cos\theta_{\mu}$ cross section, and collapsing the $\cos\theta_{\mu}$ axis. Unfortunately, restricting the measured region of phase space has the consequence of reducing the impact of the measurement.

Of course, the problem described above becomes more complicated when measuring cross sections with multiple particles in the final state as the efficiency will depend on all of the particles. It is, of course, not practical to calculate an $N$-dimensional cross section and collapse it on the desired axes, but the general problem needs to be treated with care, and can be mitigated by adding additional dimensions to the cross section extraction method, restricting the measurement to regions of particle kinematics where the efficiency is flat, or by ensuring that the cross section measurement would not be significantly changed by making large distortions to the signal model.

{\bf Unfolding:} All regularization methods add some bias, as unfolding algorithms move away from solutions which are contrary to our expectation. Various methods exist to tune the strength of the regularization (e.g., to what extent the unfolding algorithm is allowed to bias the result), which try to balance the bias in the result with the variance in each bin (known as the bias--variance trade-off). It should be noted that methods for optimizing the regularization strength are generally designed to give the best estimate of the central values, and are not guaranteed to give good coverage, which may lead to underestimated errors for regularized, unfolded, results~\cite{Prosper:2011zz, kuusela_master, kuusela_phd}. We note also that unfolding results necessarily makes all bins strongly correlated, so bias tests can be extremely misleading if only the bin-by-bin bias is investigated. It is necessary to check for biases in the $\chi^{2}$ produced with the full output covariance matrix, or some other goodness-of-fit metric.

In the widely used D'Agostini method~\cite{D'Agostini:1994zf}, each iteration of the algorithm reduces the bias towards the input Monte Carlo simulation, and increase the size of the variance in each bin. The main problem for the D'Agostini method is that the stopping criterion is generally set by Monte Carlo studies of the potential bias. If the simulations used for those studies are substantially wrong, it is likely that the result is strongly affected by the bias towards the input Monte Carlo, in a way which cannot be adequately assessed after the fact. This problem is likely to be particularly bad for a small number ($<10$) of iterations of the algorithm, which are typical in the field. We note that version of the D'Agostini method often does not include the updates described in Ref.~\cite{dagostini2nd}, which includes a simultaneous unfolding of signal and background, and may partially mitigate the pathologies of the original method.

The advantage to unfolding is simply to allow different results to be presented in the same way, and directly compared, although in general, this is not possible for our purposes because different experiments publish results averaged over different fluxes ($\Phi_{\nu}$ in Equation~\ref{eq:trad_xsec}). Regularization is simply helpful for guiding the eye, as it is produces smoother results with reduced correlations. This is useful to understand if the data agrees with a particular model at a coarse level. However, as many of the data sets produced are highly correlated, additional goodness-of-fit tests are desirable where the smoother results do not add any benefit. Problems have been found attempting to fit to existing unfolded (and regularized) data~\cite{Wilkinson:2016wmz}. Indeed, it has been argued that it is always better to smear theory to match data, as in Equation~\ref{eq:smeared_xsec}, rather than trying to unsmear data to match theory~\cite{Cousins:2016ksu}. 

Looking to the future, where underlying cross section parameters will need to be constrained with a large number of datasets, it would be ideal for to have both types of results available, where unregularized results are available if a regularization method has been employed for the main result, and folded results are made available if unfolding is used.

%% file: qe.tex
\section{Quasi-elastic and Quasi-elastic-like scattering (0$\pi$)}
\label{sec:qe}

Charged-current scattering with no final-state pions (CC0$\pi$) is the dominant interaction channel for neutrino energies below 1 GeV (see Figure~\ref{fig:flux_and_xsec}), and is an important contribution to the signal for oscillation experiments in the few GeV regime. In neutrino-nucleus scattering, CC0$\pi$ includes processes with any number of final-state nucleons, but no mesons or photons. In the absence of nuclear effects, it is a clean reaction known as charged-current quasi-elastic (CCQE) scattering:

\begin{equation}
\begin{split}
\label{eq:ccqe}
\nu_{l} n & \rightarrow l^{-} p \\
\bar{\nu}_{l} p & \rightarrow l^{+} n
\end{split}
\end{equation}
\noindent where lepton flavor $l=e,\mu,\tau$. 

The free nucleon CCQE cross section can be calculated from electroweak theory in terms of nucleon form factors, in the Llewellyn-Smith formalism~\cite{LlewellynSmithCCQE}. Where the nucleon form factors are constrained by data from electron-nucleon scattering, pion electroproduction, and neutrino-deuterium scattering~\cite{BBBAaxial}.
In measurements of $\nu_{\mu} n \rightarrow \mu^{-} p$ by deuterium bubble chambers, the muon and proton are both observed, as well as the spectator proton (not involved in the initial interaction) in a subset of events~\cite{ANL_QE,BNL_QE}. The data are in reasonable agreement with a dipole axial form factor and axial mass $\ma\sim 1$~GeV, however, new efforts rooted in lattice QCD are revisiting whether or not a dipole approximation is sufficient for the uncertainties~\cite{Meyer:2016oeg}. 

Historically, the most common neutrino interaction simulation packages (``event generators") GENIE~\cite{Andreopoulos:2009rq}, NEUT~\cite{Hayato:2009zz}, and NUANCE~\cite{nuanceMC}, extended the Llewellyn-Smith formula to larger nuclei with the relativistic Fermi Gas (RFG) model, which treats nucleons as quasi-free with Fermi momentum, in the mean field of the nucleus~\cite{Smith-Moniz}, where the nuclear parameters are tuned to electron-scattering data~\cite{moniz_eA_RFG_1971}. However, recent data do not agree with this prediction; fits to \ma assuming the RFG model with a dipole axial form factor on heavy targets suggest a larger cross section than what is expected based on deuterium data~\cite{K2K_QE,MiniBoone_MA,MINOS_MA, t2kCCQE}.

For a target nucleon at rest, the neutrino energy, $E_{\nu}$, and four-momentum transfer, $Q^{2}$, can be calculated from the kinematics of the outgoing lepton, assuming a nuclear binding energy $E_{\mathrm{b}}$:

\begin{equation}
\begin{split}
\label{eq:enuqe}
\enuqe & = \frac{M_{f}^{2} - (M_{i}-E_{\mathrm{b}})^{2} - m_{l}^2 +2 (M_{i}-E_{\mathrm{b}}) E_{l}}{2(M_{i}-E_{\mathrm{b}}-E_{l}+p_{l}\cos\theta_{l})}, \\
\qqqe & = 2\enuqe(E_{l}-p_{l}\cos\theta_{l})-m_{l},
\end{split}
\end{equation}

\noindent
where $M_{i}$ ($M_{f}$) is the mass of the struck (final-state) nucleon, and $E_{l}$, $p_{l}$, $\theta_{l}$, and $m_{l}$ are the energy, three-momentum, angle with respect to the neutrino, and mass of the final-state lepton, respectively. 

On heavy nuclei, CC0$\pi$ is complicated by the presence of many nucleons within the nucleus. In addition to CCQE interactions with a single nucleon (1p1h), the scattering target can be a bound state of two (2p2h) or more (npnh) correlated nucleons. FSI between scattered nucleons and the residual nucleus also modify the final state. Pions can be absorbed or produced in FSI, moving events into or out of a CC0$\pi$ sample. The relationship between detector observables \enuqe or \qqqe, and the underlying $E_{\nu}$ or $Q^{2}$, depends on the nuclear model.

As modern measurements are given in terms of flux-averaged differential cross sections (see Equation~\ref{eq:topology_xsec}), it is difficult to compare results from different experiments on a single figure. Differences between experiments in target nucleus, phase space restrictions, and signal definitions further complicate comparisons. Instead, results of various measurements are shown as fractional deviations between data and a reference model, in this case the widely-used GENIE event generator~\cite{Andreopoulos:2009rq} is used with models enabled in version 2.12.8 (details in Appendix~{\hyperlink{sec:appendix}{A}}). Experimental differences are taken out by incorporating into the reference model the flux, target, phase space restrictions, and signal definition of each measurement using NUISANCE~\cite{Stowell:2016jfr}. The main disadvantage of this is the necessity of choosing a particular reference model for comparison; we chose a consistent 1p1h+2p2h model, and updated resonance model for these comparisons.

\begin{figure}[htbp]
  \includegraphics[width=0.8\textwidth]{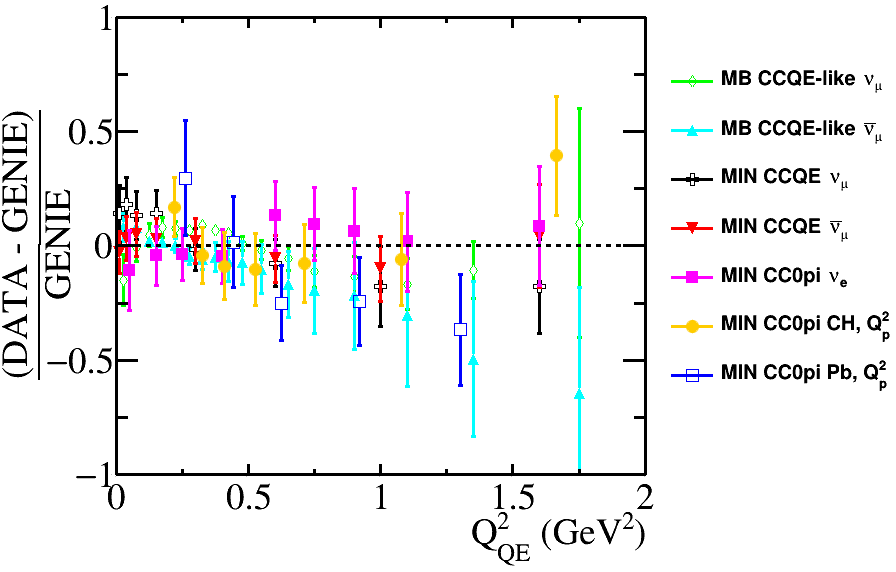}
  \caption{Differential cross sections for QE and 0$\pi$, measured by MiniBooNE (MB) and MINERvA (MIN), are shown as a function of \qqqe as fractional residuals to the reference model, using NUISANCE~\cite{Stowell:2016jfr}. The datasets are taken from Refs.~\cite{mbCCQE, mbAntiCCQE, minerva-antinu-ccqe, Fiorentini:2013ezn, Walton:2014esl, Wolcott:2015hda, Betancourt:2017uso}.}
  \label{fig:cc0pi_q2}    
\end{figure}
Figure~\ref{fig:cc0pi_q2} shows shape-only ratios of flux-integrated differential cross section measurements to the reference GENIE model, as a function of \qqqe. MiniBooNE has a peak energy of 0.6 (0.4) GeV for $\nu_{\mu}$ ($\bar{\nu}_{\mu}$), and has $4\pi$ angular acceptance down to muon kinetic energy of 200 MeV. The signal definition includes all processes which have no final-state pions~\cite{mbCCQE, mbAntiCCQE}. The detector has no sensitivity to final-state protons due to the high Cherenkov threshold. MINERvA is sensitive to final-state protons, and has published cross sections with respect to $d\sigma/d\qqqe$ as measured from both the muon and proton side\footnote{\qqqe is reconstructed from the proton kinematics with $\qqqe = (M_{i}-E_{\mathrm{b}})^{2} - M_{f}^{2} + 2(M_{i}-E_{\mathrm{b}}(T_p + M_f - M_i + E_{\mathrm{b}})$, where $T_p$ is the proton kinetic energy~\cite{Walton:2014esl}.} with a peak neutrino energy of 3.5 GeV~\cite{minerva-antinu-ccqe, Fiorentini:2013ezn, Walton:2014esl, Betancourt:2017uso}. The muon-side measurement for both neutrinos and antineutrinos is corrected to CCQE, using the GENIE version 2.6.2 model, tuned to control samples, to subtract non-QE backgrounds. The version of GENIE used did not include any npnh processes. To mitigate this, the region around the interaction vertex, where additional proton energy would be expected in these events, is deliberately excluded from the analysis, however, it is not clear to what extent the measurement includes npnh effects. Acceptance is limited to muons above about 1 GeV and at forward angles below about 20$^{\circ}$ due to the requirement that tracks enter the magnetized MINOS near detector. The proton-side measurement is of CC0$\pi$ on $^{12}$C, CH, $^{56}$Fe, and $^{208}$Pb targets (CH and $^{208}$Pb are shown). It includes CCQE-like processes, and requires at least one proton with momentum greater than 450 MeV to exclude the region below the proton tracking threshold. Also shown in Fig.~\ref{fig:cc0pi_q2} is MINERvA's measurement of CC0$\pi$ for $\nu_{e}$, using the roughly 1\% beam ``contamination'' in the mostly-$\nu_{\mu}$ configuration~\cite{Wolcott:2015hda}. The MINERvA result is in agreement with reference model and lepton universality, but lacks the precision to test potential unsimulated cross section differences, such as those proposed in~\cite{Day:2012gb}. The data are in reasonable agreement with the reference GENIE model. MiniBooNE observes a larger suppression at very low \qqqe than is predicted by the reference GENIE model, and an enhancement at moderate \qqqe. There is some tension in the shape between MINERvA and MiniBooNE.

\begin{figure}[htbp]
  \includegraphics[width=0.8\textwidth]{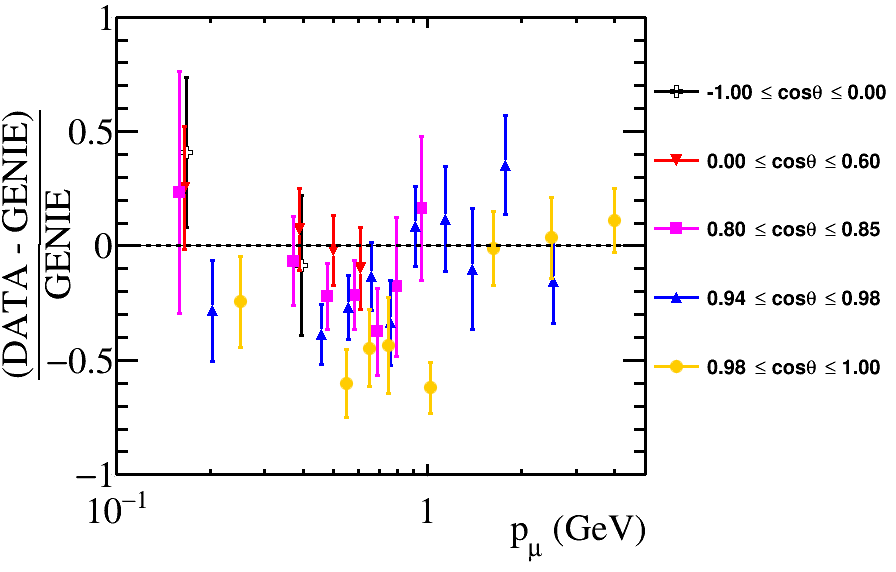}
  \caption{The T2K CC0$\pi$ double-differential cross section $d\sigma/dp_{\mu}d\cos\theta_{\mu}$ is shown as a function of $p_{\mu}$ in angular bins, as a fractional difference between the published data and the reference GENIE model, produced using NUISANCE~\cite{Stowell:2016jfr}. The data used are taken from Ref.~\cite{Abe:2016tmq}.}
  \label{fig:cc0pi_dd}    
\end{figure}
Double differential cross sections for CC0$\pi$ with respect to muon energy and angle, $d^{2}\sigma/dT_{\mu}d\cos\theta_{\mu}$, have been published by T2K~\cite{Abe:2016tmq, Abe:2017rfw}, MiniBooNE~\cite{mbCCQE, mbAntiCCQE};  MINERvA produced their result in the transverse and longitudinal components of the muon momentum relative to the incident beam~\cite{minerva_0pi_2d}. 
This is experimentally less model dependent than measuring $d\sigma/d\qqqe$. Detector acceptance typically depends strongly on muon kinematics, with experiments having minimum muon energies, maximum muon angles, or both. Since differential cross section measurements integrate over neutrino energy, a given bin of \qqqe contains events at a wide range of muon energy and angle. Ratios to the reference model of $d^{2}\sigma/dT_{\mu}d\cos\theta_{\mu}$ for CC0$\pi$ $\nu_{\mu}$-C$_8$H$_8$ interactions, measured by T2K are shown in Fig.~\ref{fig:cc0pi_dd} as a function of $p_{\mu}$ in selected bins of $\cos\theta_{\mu}$. The T2K data are from the ND280 off-axis near detector, with peak neutrino energy of 0.6 GeV. A deficit is observed with respect to the reference GENIE model in the most forward angular bins, especially at low muon momentum. The data agree with the reference model for backward and high-angle muons.

\begin{figure}[htbp]
  \includegraphics[width=0.9\textwidth]{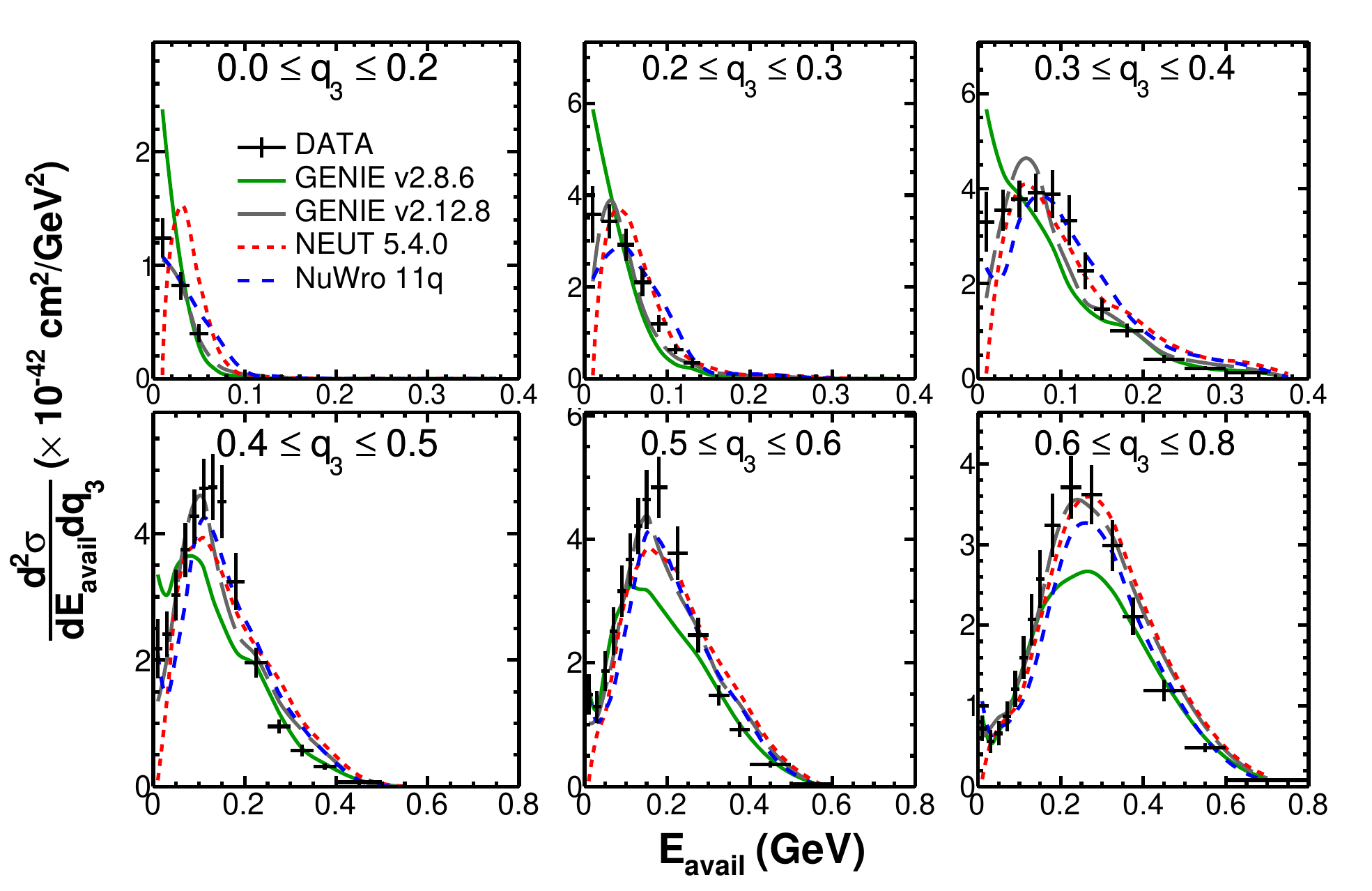}
  \caption{MINERvA charged-current inclusive data as a function of available energy, $E_{\mathrm{avail}}$ (described in the text), in slices of three-momentum transfer, $q_{3}$, compared to four predictions from different simulations~\cite{Hayato:2009zz, Juszczak:2009qa, Andreopoulos:2009rq}, produced using NUISANCE~\cite{Stowell:2016jfr}. The data is taken from Ref.~\cite{Rodrigues:2015hik}. 
  }
  \label{fig:cc0pi_q0q3}    
\end{figure}
MINERvA has also made a double-differential measurement with respect to energy and three-momentum transfer, $d\sigma/dE_{\mathrm{avail}}dq_{3}$, for charged-current (inclusive) reactions~\cite{Rodrigues:2015hik}. While this is not explicitly CC0$\pi$, it covers a region of three-momentum transfer ($q_{3} \leq 0.8$~GeV) dominated by CC0$\pi$ and CC1$\pi$. The quantity $E_{\mathrm{avail}}$ is defined as the sum over kinetic energy of all protons and $\pi^\pm$, and total energy of all photons and $\pi^0$. It is essentially the observable hadronic energy in the detector, and serves as a proxy for true energy transfer $q_{0} = E_\nu - E_\mu$ while avoiding model dependence that would be introduced by correcting for unobserved hadronic energy, for example from neutrons. The reconstruction of $q_{3}$ introduces some model dependence by correcting to true neutrino energy, including unobserved hadrons. The data are compared with four different models in Figure~\ref{fig:cc0pi_q0q3}: GENIE version 2.8.6 and the reference GENIE version 2.12.8 model, NEUT v5.4.0~\cite{Hayato:2009zz}, and NuWro v11q~\cite{Juszczak:2009qa}. The different panels in Figure~\ref{fig:cc0pi_q0q3} show different slices of $q_{3}$. At fixed $q_{3}$, quasi-elastic (1p1h) events appear at low $E_{\mathrm{avail}}$, with resonance pion production at high $E_{\mathrm{avail}}$. Multi-nucleon processes fill in the ``dip'' region between the QE and $\Delta$ peaks. The data disfavor GENIE 2.8.6, which lacks any npnh simulation. The data are well described by the reference GENIE model, although an excess is observed at very low $E_{\mathrm{avail}}$.

Additional measurements are also important for a complete picture of CCQE-like scattering. A new analysis on MiniBooNE uses a kaon decay at rest beam to provide information from a mono-energetic beam~\cite{Aguilar-Arevalo:2018ylq}. This provides a unique case of known energy transfer in neutrino interactions, which will provide new constraints on models. Neutral current elastic (NCE) scattering, $\nu (n,p) \rightarrow \nu (n,p)$, is sensitive to
the strange quark contribution to the nucleus
and may be used as a signal sample in searches for oscillations to sterile neutrinos. The only observable particles in the final state are protons and neutrons. The signal is typically a single proton track. In addition to backgrounds from charged-current processes, experiments face an additional challenge from neutrons produced outside the detector, which knock out protons inside the detector. The NCE differential cross section $d\sigma/dQ^{2}$ was measured at BNL~\cite{BNL_NCE}, and more recently by MiniBooNE~\cite{miniboone_NCE, mbAntiNCEL}, who also published ratios of NCE to CCQE.

%% file: spp.tex
\section{Single pion production}
\label{sec:spp}

With sufficiently large energy transfer to the struck nucleon, a nucleon resonance is produced, which typically decays to produce a proton or neutron, and either a charged or neutral pion. Resonances can also decay to produce multi-pion final states, photons, or other mesons. The latter cases are discussed in Section~\ref{sec:rare}. Resonant production is the dominant single pion production process in the 0.1--10 GeV energy range of interest for neutrino scattering experiments. However, non-resonant processes also produce single pion final states, which can either come from non-resonant interactions with the whole nucleon, or at higher energy transfers, soft-DIS processes where the interaction is with a constituent quark, and the subsequent hadronization yields a single pion. Modeling the transition between non-resonant nucleon-level and quark-level interactions is a significant challenge for the community.

As is familiar from the discussion of CCQE/CC0$\pi$ in Section~\ref{sec:qe}, the relationship between the experimental and theoretical definitions of single pion production is complicated by FSI. Multi-pion or soft-DIS interactions can contribute events where only a single pion escapes the nucleus. Additionally, FSI may modify the outgoing pion charge, momentum and direction for interactions for which a single pion was produced at the vertex. Indeed, some such events will enter the CC0$\pi$ signal if the pion is absorbed, as has been discussed previously. Due to the complications of FSI, and the number of interaction processes which can contribute to single {\it final-state} pion production in an experiment, it is very challenging to use single {\it final-state} pion production data to constrain theoretical model parameters without use of a Monte Carlo event generator. It is also challenging to attribute disagreement with data to any specific model deficiency.

Other datasets can provide constraints on the single pion production model outlined above. Much use is made of the rather limited set of $\nu_{\mu}$-deuterium bubble chamber data from the 1970s and 1980s~\cite{ANL_Barish_1979,ANL_Radecky_1982,BNL_Kitagaki_1986, BNL_Kitagaki_1990} to test and tune models of single pion production at the vertex. Although a long-standing discrepancy between some of these datasets appears to have recently been resolved~\cite{anl_bnl_reanalysis}, the low-statistics involved mean that these datasets can only constrain model parameters at the $\sim$5--10\% level~\cite{Rodrigues:2016xjj}. Additionally, it is not clear that FSI effects can be safely neglected for deuterium~\cite{sato_2014}. A large body of pion-nucleus and nucleon-nucleus scattering measurements are used to constrain the absorption, inelastic, elastic and charge exchange interaction probabilities which underpin models of FSI, although a model is required to relate the scattering of external particles on ground-state nuclei to intra-nuclear scattering within an excited nuclear remnant~\cite{Salcedo-Oset}. Again, the datasets used to constrain the FSI models typically date from the 1960s-1980s (for a review, see, for example, Ref.~\cite{Abe:2015awa}), and are often incomplete, e.g., correlations between data points and for different measurements produced by the same experiment are not available. A notable recent exception are new datasets from the DUET experiment~\cite{Ieki:2015okz, PinzonGuerra:2016uae}, which recently published results of $\pi^{+}$-$^{12}$C charge exchange and absorption for pion momentum $200 \leq p_{\pi} \leq 300$ MeV, including correlations between all data points~\cite{PinzonGuerra:2016uae}.

In order to satisfy the stringent error budgets of future neutrino oscillation experiments~\cite{lbne,t2k_sensitivity_2014}, it is necessary both to have a theoretical model which adequately describes single pion production in neutrino-nucleus scattering, and which has smaller uncertainties than can be obtained by tuning the aspects of the model to the independent datasets outlined above. Therefore, it is important that we have a model which consistently describes all of the currently available neutrino--nucleus scattering data, and that we have sufficient data to constrain the model. This discussion will occupy much of the remainder of this section.

\subsection{Resonance pion production}

In recent years, many new high-statistics neutrino--nucleus single pion production datasets on medium mass targets (hydrocarbon and water) differential in a number of variables have become available from the \mb~\cite{AguilarArevalo:2009ww, AguilarArevalo:2010xt, AguilarArevalo:2010bm}, \minerva~\cite{Eberly:2014mra, Aliaga:2015wva, McGivern:2016bwh} and T2K~\cite{Abe:2016aoo} experiments. Additionally, the first measurements of neutrino--argon single pion production have been published from the ArgoNeuT experiment~\cite{Acciarri:2015ncl}, although without sufficient statistics to produce differential results.

\begin{figure}[htbp]
  \includegraphics[width=0.8\textwidth]{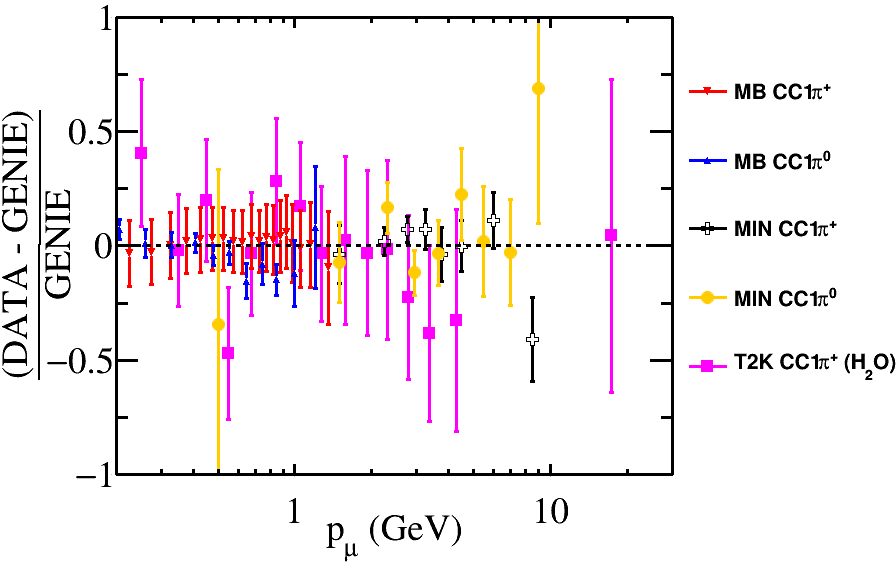}
  \caption{Flux-averaged differential cross section predictions are made using the reference GENIE model and NUISANCE~\cite{Stowell:2016jfr} for various measurements. The fractional difference between the published data and the reference model is shown to allow results from multiple experiments to be plotted on the same axis as a function of $p_{\mu}$. The datasets used are taken from Refs.~\cite{AguilarArevalo:2010xt, AguilarArevalo:2010bm, McGivern:2016bwh, Abe:2016aoo}.}
  \label{fig:spp_GENIE_pmu_ratio}    
\end{figure}
\begin{figure}[htbp]
  \includegraphics[width=0.8\textwidth]{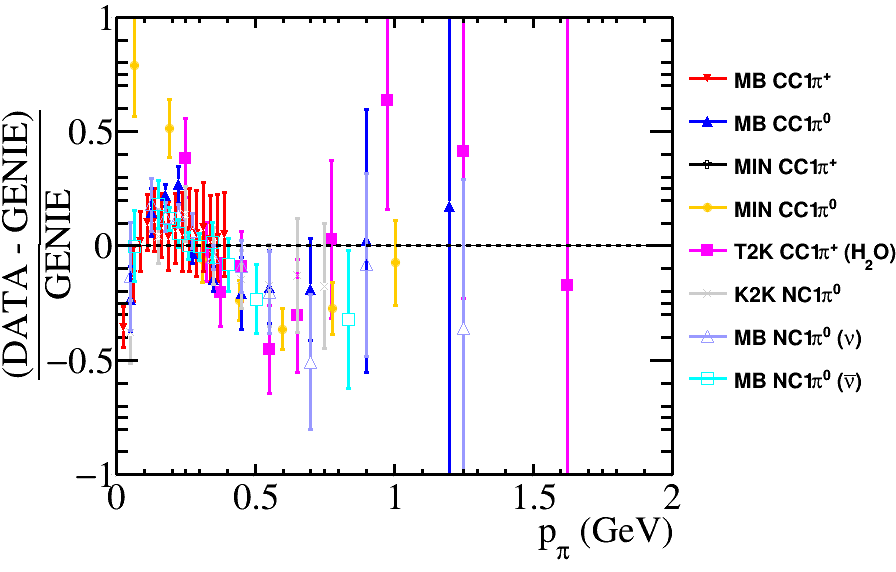}
  \caption{Flux-averaged differential cross section predictions are made using the reference GENIE model and NUISANCE~\cite{Stowell:2016jfr} for various measurements. The fractional difference between the published data and the reference model is shown to allow results from multiple experiments to be plotted on the same axis as a function of $p_{\pi}$. The datasets used are taken from Refs.~\cite{Nakayama:2004dp, AguilarArevalo:2009ww, AguilarArevalo:2010xt, AguilarArevalo:2010bm, McGivern:2016bwh, Abe:2016aoo}.}
  \label{fig:spp_GENIE_ppi_ratio}    
\end{figure}

Figures~\ref{fig:spp_GENIE_pmu_ratio} and~\ref{fig:spp_GENIE_ppi_ratio} show fractional deviations between data and the reference GENIE model (see Appendix~{\hyperlink{sec:appendix}{A}} for details).  As with CC0$\pi$ in Section~\ref{sec:qe}, the reference model prediction incorporates the flux, signal definition, and phase space of each measurement, and the comparisons shown in Figures~\ref{fig:spp_GENIE_pmu_ratio} and~\ref{fig:spp_GENIE_ppi_ratio} are shape-only, as each experiment has a large normalization uncertainty due to uncertainty in the neutrino flux. We can see from Figure~\ref{fig:spp_GENIE_pmu_ratio} that there is broad agreement between the reference model and various datasets in terms of the momentum of the outgoing muon, $p_{\mu}$; and we can see from Figure~\ref{fig:spp_GENIE_ppi_ratio} that the same agreement is not seen in terms of the momentum of the outgoing pion, $p_{\pi}$. However, it is interesting that the deviations in terms of pion kinematics appears to be relatively consistent across experiments, and between CC1$\pi^{+}$, CC1$\pi^{0}$ and NC1$\pi^{0}$ measurements, with a deficit of strength in data deficit relative to GENIE at very low momentum, an excess of strength in data for the intermediate $0.2\leq p_{\pi} \leq 0.4$ GeV region, where much of the data lies, and with similarly consistent structure at higher $p_{\pi}$. While NC measurements are limited to NC$\pi^0$ only, CC$\pi^+$ and NC$\pi^0$ are similar in shape-only comparisons. When normalizations are included, CC and NC are less consistent.

\subsection{Coherent pion production}
At very low energy transfer, neutrinos can interact inelastically with the entire nucleus in a coherent manner, producing either a charged or neutral pion, but leaving the nucleus intact and in its ground state. Coherent pion production is of particular interest to neutrino oscillation experiments because NC coherent interactions produce a $\pi^{0}$, and are a background to $\nu^{\bracketbar}_{e}$-appearance measurements with a large uncertainty. CC production, with minimal recoil of the nucleus, provides an interesting handle on neutrino energy for oscillation experiments, but such analyses suffer from significant resonant background contributions. Various historical measurements of coherent pion production exist for both CC and NC processes on a number of targets at neutrino energies $E_{\nu} > 2$ GeV, which are reviewed in Refs.~\cite{Vilain:1993sf,zeller12, pdg_2014}, but at low neutrino energies ($E_{\nu} < 2$ GeV), results presented a confusing picture. Measurements of NC coherent $\pi^{0}$ with average neutrino energies of $\sim$1 GeV were reported~\cite{AguilarArevalo:2008xs, Kurimoto:2009wq}, but measurements of CC coherent pion production found no evidence~\cite{Hasegawa:2005td, Hiraide:2008eu, Tanaka:2009ag}. Furthermore, the apparent difference between NC and CC coherent pion production conflicted with the available theoretical models~\cite{Kurimoto:2010rc}.

However, recent MINERvA~\cite{minerva_coh_2014, Mislivec:2017qfz} and T2K~\cite{Abe:2016fic} measurements of CC coherent pion production for $\nu^{\bracketbar}_{\mu}$--$^{12}$C at low energies observe coherent production, which disagrees with the previous null results. The high-statistics MINERvA dataset is of particular interest as the results are presented as single differential cross sections as a function of the outgoing pion momentum, angle with respect to the incoming neutrino beam, and the neutrino energy~\cite{minerva_coh_2014}. Additionally, a recent measurement of NC $\nu_{\mu}$--$^{56}$Fe coherent pion production has been produced by the MINOS collaboration~\cite{Adamson:2016hyz}, and the first measurements of CC $\nu^{\bracketbar}_{\mu}$--$^{40}$Ar coherent pion production have been presented by ArgoNeuT~\cite{Acciarri:2014eit}. These results are important to test how the coherent pion production cross section scales between nuclei, but both the MINOS and ArgoNeuT results have average neutrino energies of $\sim$5 GeV or greater.

Diffractive pion production is a similar process to coherent pion production where the interaction is with a free proton, and the proton remains in an unexcited state. The first experimental evidence for NC diffractive $\pi^{0}$ production has recently been presented by the MINERvA collaboration~\cite{Wolcott:2016hws}. As with coherent pion production, NC diffractive $\pi^{0}$ production constitutes a background for $\nu^{\bracketbar}_{e}$-appearance measurements in experiments with a hydrogen component. The potential problem is highlighted by the fact that the MINERvA diffractive sample was isolated after anomalies were found during the analysis of $\nu_{e}$ events~\cite{Wolcott:2015hda}. Currently, diffractive pion production is only modeled theoretically for invariant masses $W \geq 2$ GeV~\cite{Rein:1986cd}, so the potential size of the effect in $0.1 \leq E_{\nu} \leq 10$ GeV oscillation experiments is not under control.

%% file: rare.tex
\section{Rare processes}
\label{sec:rare}
As described in Section~\ref{sec:intro}, charged kaon production by neutrinos is a background in searches for the proton decay channel $p \rightarrow K^{+} \nu$. The most stringent limits on $p \rightarrow K^{+} \nu$ come from SK, where the largest background is from neutral-current $K^{+}$ by atmospheric neutrinos where no final-state particles are above Cherenkov threshold~\cite{SK_Knu}. Recently, differential cross sections for CC and NC $K^{+}$ production have been measured by MINERvA~\cite{minerva_cckplus,minerva_nckplus}. The MINERvA data for NC $K^{+}$ is the first high-statistics test of the cross section models employed by experiments. 

In charged-current reactions, there are also Cabibbo-suppressed processes with a single exotic (anti)quark in the final state, for example $\nu_{\mu} p \rightarrow \mu^{-} K^{+} n$ or $\bar{\nu}_{\mu} p \rightarrow \mu^{+} \Lambda$. These events have been seen in bubble chambers but exclusive cross sections have not been measured with high statistics. Cabibbo-suppressed CCQE $\bar{\nu}_{\mu} p \rightarrow \mu^{+} \Lambda$ could be a few percent of the total CCQE cross section for antineutrino scattering, with no neutrino analogue. MINERvA observed the first evidence for the Cabibbo-suppressed charged-current coherent $K^{+}$ production at $3\sigma$ significance~\cite{minerva_cohkplus}.

%% file: inclusive.tex
\section{Inclusive scattering, other notable processes}
\label{sec:incl}
Charged-current inclusive scattering, to which all processes which produce a charged lepton at the vertex contribute, has the simplest experimental definition. It is used as the signal for neutrino oscillation experiments which have peak beam energies over $\sim$1 GeV, which are not dominated by CCQE. Because of this, the energy cannot be reconstructed using only the outgoing lepton information, as in Equation~\ref{eq:enuqe}, but instead has to use both leptonic and hadronic information
\begin{equation}
\label{eq:enuhad}
E_{\nu}^{\mathrm{rec}} = E_{l} + E_{\mathrm{had}},
\end{equation}
\noindent where $E_{l}$ is the energy of the outgoing lepton, and $E_{\mathrm{had}}$ is the reconstructed hadronic energy. In a perfect detector, the reconstructed hadronic energy would be the sum of the kinetic energy of nucleons, and the total energy of all other outgoing particles. This would result in accurate neutrino energy reconstruction, assuming that nucleons are not produced through hadronization, and neglecting initial state binding and Fermi motion effects. In a realistic detector, neutron reconstruction is typically so poor that they can be neglected, and many particles will not be sufficiently energetic to identify individually. Model dependent corrections are therefore used to correct for the missing energy due to neutrons, and to estimate average multiplicities for particles which could not be reconstructed, but which still deposit energy in the detector.

Theoretically, CC-inclusive processes can seem easier to model than exclusive processes. Theorists are used to inclusive electron scattering measurements, where only the outgoing lepton kinematics are detected. Inclusive neutrino scattering cross section measurements differential in the outgoing charged lepton kinematics are very similar, and allow the hadronic state to be integrated over, thus avoiding the issues of FSI which make it difficult to predict exclusive final states such as CC0$\pi$ or CC1$\pi$. However, for neutrino oscillation experiments, the relative energy fraction lost to neutrons, and the average pion and proton multiplicities are very important for the reconstruction method outlined in Equation~\ref{eq:enuhad}. Unfortunately, the task of producing a CC-inclusive model suitable for neutrino oscillation experiments is significantly more challenging than for much of the electron-scattering data, and does not sidestep the issue of FSI.

As well as the CC0$\pi$ and CC single pion production channels, multi-pion and heavy meson production processes contribute to the CC-inclusive cross section (for example, the rare processes described in Section~\ref{sec:rare}). These additional processes are at predominantly DIS at larger invariant masses, where the interaction is with a constituent quark within the nucleon, rather than with the nucleon itself. Although DIS is well defined theoretically, and is the dominant interaction for $E_{\nu} \geq 3$ GeV, it is not a well-defined topology, so cannot easily be selected experimentally. For example, DIS can produce a single pion and no other mesons in the final state, and thus is indistinguishable from resonant pion production, and conversely, it is possible for a nucleon resonance to decay to produce multiple pions. Hadronization models for DIS are tuned to high energy neutrino scattering data, and are not valid for low invariant masses~\cite{Tzanov:2005kr}. This is readily understandable, as the hadronization models are only valid for neutrino--quark interactions. Unfortunately, the transition region between neutrino--nucleon and neutrino--quark interactions is not well understood, and neutrino interaction generators typically extrapolate both models outside the regions where they are expected to be valid, and blend the two in an unphysical way to predict the cross section. A significant number of events for current and future neutrino oscillation experiments will lie in this transition region, so inclusive data, which can be used to improve and validate models for this region is very important.

It is, however, very challenging to use neutrino CC-inclusive data to validate single aspects of a model. Broad band neutrino fluxes, and the inability to reconstruct neutrino energy on an event-by-event basis mean that the model has to be integrated over all interaction kinematics. Figure~\ref{fig:cc0pi_q0q3} demonstrated how difficult it is to identify the cause of data--MC disagreement as a function of interaction kinematics for a broad band neutrino flux. CC-inclusive results as a function of outgoing particle kinematics are even more challenging to disentangle. Taking similar measurements with different neutrino fluxes and different target configurations provides a way to break degeneracies, and so inclusive measurements from multiple experiments is desirable.

\begin{figure}[htbp]
  \includegraphics[width=0.6\textwidth]{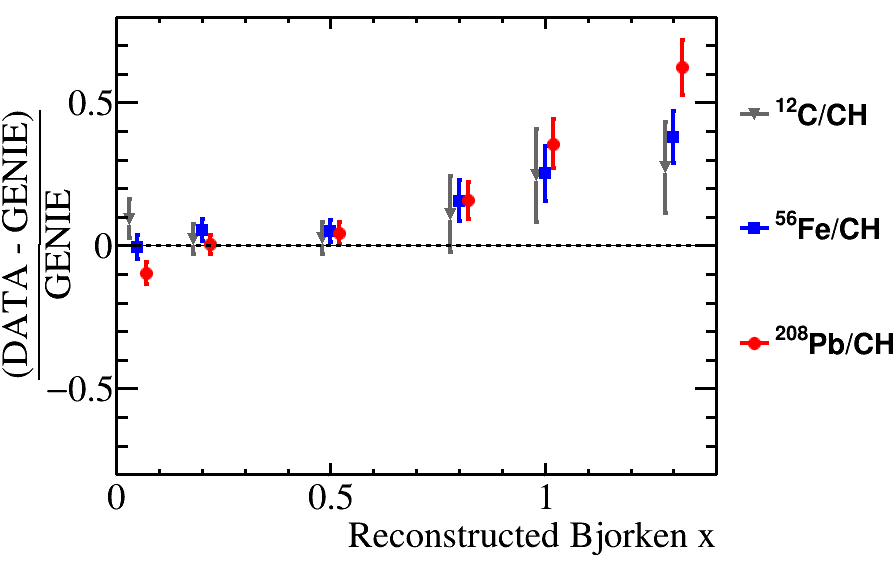}
  \caption{Predictions for the flux-averaged CC-inclusive heavy target ratio measurements from MINERvA as a function of reconstructed Bjorken $x$ are produced using the reference GENIE model and NUISANCE~\cite{Stowell:2016jfr}. The fractional difference between the published data and the reference GENIE model is shown for the published rations. The data used are taken from Ref.~\cite{Tice:2014pgu}. Smearing matrices which relate true $x$ and $x$ as reconstructed by the MINERvA detector are also provided in the data used, and are used in NUISANCE to fold detector effects into the GENIE prediction and make an appropriate comparison with data.}
  \label{fig:GENIE_xratios}
\end{figure}

There are a number of issues which make it more challenging to measure an inclusive cross section than exclusive channels like CC0$\pi$ or single pion production. Typically, events with more final state particles are more difficult to reconstruct and enter the selection with a lower efficiency, which may also depend on the kinematics of those final state particles. This can lead to the implicit input Monto Carlo bias outlined in Section~\ref{sec:biases}. It also makes it challenging to measure a cross section differential in the number of final state particles with modern experiments. The limited data that exists on track multiplicity all comes from old bubble chamber experiments. Older measurements typically give results as a function of neutrino energy~\cite{Tzanov:2005kr, Wu:2007ab, Adamson:2009ju}, but recent measurements have mostly been made differential in muon kinematics~\cite{Anderson:2011ce, Abe:2013jth, Acciarri:2014isz, Nakajima:2010fp}, or as single bin measurements~\cite{Abe:2015biq, Abe:2017ufe}. Recent exceptions have been from the MINERvA experiment which have attempted to reconstruct interaction-level kinematics in various ways~\cite{Tice:2014pgu, Rodrigues:2015hik, DeVan:2016rkm, Ren:2017xov}. One such example is a series of heavy target ratios, as a function of reconstructed Bjorken $x$~\cite{Tice:2014pgu}, which is reconstructed using reconstructed $E_{\mu}$, $\theta_\mu$ and $\nu$: $E_{\nu} = E_{\mu} + \nu$, $Q^2 = 4E_{\nu}E_{\mu}\sin^{2}\left(\frac{\theta_{\mu}}{2}\right)$, $x = \frac{Q^{2}}{2M_{\mathrm{N}}\nu}$, where $M_{\mathrm{N}}$ is the average of the proton and neutron masses. Comparisons with the reference GENIE model are shown in Figure~\ref{fig:GENIE_xratios}, from which it is clear that there are discrepancies at both high and low $x$, which increase with the size of nuclear target used in the numerator. However, in making the comparison to data using the reference GENIE model, the smearing matrix published by MINERvA to relate reconstructed $x$ with true $x$ was used. That matrix was created using an earlier version of GENIE (v2.6.2), which may add model dependence to the smearing matrix through the relationship between reconstructed and true energy transfer.

%% file: future.tex
\section{Future experiment possibilities} 
\label{sec:future}
In the next five years, the current generation of neutrino scattering experiments (T2K, MINERvA, NOvA, MicroBooNE~\cite{Chen:2007ae}) programs will produce final, high statistics, measurements for many topologies with neutrino and antineutrino configurations. Future detectors in the T2K beam (T2K ND Upgrade Project, WAGASCI~\cite{Kin:2017way}, NINJA~\cite{Fukuda:2017ezx}) will expand the phase space available to measure or provide new detection methods. There is a burgeoning program of measurements of not only the leptonic side but the hadronic side. These experiments are also developing creative new methods or measurements to expand our understanding. Thoughtful projections of observables may yield new information about nuclear effects~\cite{Lu:2015tcr}, and such measurements are underway on T2K and MINERvA experiments. Over this time period, all experiments will include significant improvements to sources of systematic uncertainty. The most clear case is that of the neutrino flux following methods discussed in Section~\ref{sec:syst}; neutrino-electron scattering, for example, benefits from as long a run period (increased statistics) as possible. There may also be improvements to other sources of systematic uncertainty. For example, 
data from LArIAT~\cite{Cavanna:2014iqa} and ProtoDUNE~\cite{Abi:2017aow}  will benefit the short baseline program at Fermilab~\cite{Antonello:2015lea} and DUNE. 

Another critical new category of measurement not readily possible before will come from measurements of neutrons out of neutrino interactions.  Neutron multiplicity and energy varies enormously from generator to generator, these measurements, if feasible, could help refine where and how model implementation breaks down, especially in conjunction with other hadronic state measurements. MINERvA and NOvA calorimeter experiments can use neutron scatters in the detector as an estimate of neutron energy and multiplicity. There are also dedicated experiments, CAPTAIN ($^{40}$Ar) and ANNIE (H$_2$O). CAPTAIN~\cite{Berns:2013usa} is a small LAr detector placed near a neutron source used to determine if neutron scatters can be tagged. ANNIE~\cite{Anghel:2015xxt} will use neutron capture on Gadolinium in a neutrino beam to infer neutron multiplicity. 

 One can also use the neutrino beam itself to yield new information about the cross section model. T2K recently produced a measurement using a comparison of neutrino interactions across the beam~\cite{Abe:2015biq}, as the beam changes. The flux differences provide information about the energy dependence of the cross section for nearly identical detectors. This technique (``NuPRISM'')~\cite{Bhadra:2014oma} can be expanded and is under consideration for the future near detectors of the Hyper-Kamiokande and DUNE experiments.

%% file: summary.tex
\section{Summary} 
\label{sec:summary}
Neutrino interactions play a critical role in neutrino oscillation experiments, which has motivated an extensive experimental program to measure neutrino scattering cross sections. Recent theoretical improvements over a number of years have produced CC0$\pi$ models which show broad qualitative agreement with data, but more effort is needed to understand regions of disagreements between models and experiments, especially in the forward scattering region. There has been less theoretical work on single pion production, and there is considerable tension between neutrino-nucleus scattering data and theoretical predictions, particularly for the outgoing pion kinematics. That said, the discrepancies with data appear to be reasonably consistent between charged and neutral current and between charged and neutral pions. Less data exists for inclusive measurements, particularly at the energies most relevant for NOvA and DUNE, who use CC-inclusive events as the signal for oscillation measurements. Modeling the transition region between resonant pion production, and DIS interactions with a quark inside the nucleus is a significant theoretical challenge for the future, and there are fewer recent datasets which are appropriate for tuning models of this higher invariant mass region.

Other critical neutrino scattering issues for oscillation experiments are the lack of $\nu^{\bracketbar}_{e}$ cross section measurements, so models tuned with $\nu_{\mu}^{\bracketbar}$ data must rely on theoretical arguments to relate the two (vital for appearance measurements). Similarly, most of the current measurements are taken on hydrocarbon or water targets, a critical problem for the short baseline and DUNE oscillation programs, which use argon targets. There is an exciting future program of neutrino scattering to address many of these issues, with a host of new technologies and techniques being used. In particular, many new results are anticipated which will probe QE-like interactions in detail, measurements on $^{40}$Ar targets, neutrons, or which use novel projections of existing data sets.

%% file: AR_xsec_review.bbl
\begin{thebibliography}{150}%
\makeatletter
\providecommand \@ifxundefined [1]{%
 \@ifx{#1\undefined}
}%
\providecommand \@ifnum [1]{%
 \ifnum #1\expandafter \@firstoftwo
 \else \expandafter \@secondoftwo
 \fi
}%
\providecommand \@ifx [1]{%
 \ifx #1\expandafter \@firstoftwo
 \else \expandafter \@secondoftwo
 \fi
}%
\providecommand \natexlab [1]{#1}%
\providecommand \enquote  [1]{``#1''}%
\providecommand \bibnamefont  [1]{#1}%
\providecommand \bibfnamefont [1]{#1}%
\providecommand \citenamefont [1]{#1}%
\providecommand \href@noop [0]{\@secondoftwo}%
\providecommand \href [0]{\begingroup \@sanitize@url \@href}%
\providecommand \@href[1]{\@@startlink{#1}\@@href}%
\providecommand \@@href[1]{\endgroup#1\@@endlink}%
\providecommand \@sanitize@url [0]{\catcode `\\12\catcode `\$12\catcode
  `\&12\catcode `\#12\catcode `\^12\catcode `\_12\catcode `\%12\relax}%
\providecommand \@@startlink[1]{}%
\providecommand \@@endlink[0]{}%
\providecommand \url  [0]{\begingroup\@sanitize@url \@url }%
\providecommand \@url [1]{\endgroup\@href {#1}{\urlprefix }}%
\providecommand \urlprefix  [0]{URL }%
\providecommand \Eprint [0]{\href }%
\providecommand \doibase [0]{http://dx.doi.org/}%
\providecommand \selectlanguage [0]{\@gobble}%
\providecommand \bibinfo  [0]{\@secondoftwo}%
\providecommand \bibfield  [0]{\@secondoftwo}%
\providecommand \translation [1]{[#1]}%
\providecommand \BibitemOpen [0]{}%
\providecommand \bibitemStop [0]{}%
\providecommand \bibitemNoStop [0]{.\EOS\space}%
\providecommand \EOS [0]{\spacefactor3000\relax}%
\providecommand \BibitemShut  [1]{\csname bibitem#1\endcsname}%
\let\auto@bib@innerbib\@empty
\bibitem [{\citenamefont {Formaggio}\ and\ \citenamefont
  {Zeller}(2012)}]{zeller12}%
  \BibitemOpen
  \bibfield  {author} {\bibinfo {author} {\bibfnamefont {J.~A.}\ \bibnamefont
  {Formaggio}}\ and\ \bibinfo {author} {\bibfnamefont {G.~P.}\ \bibnamefont
  {Zeller}},\ }\href {\doibase 10.1103/RevModPhys.84.1307} {\bibfield
  {journal} {\bibinfo  {journal} {Rev. Mod. Phys.}\ }\textbf {\bibinfo {volume}
  {84}},\ \bibinfo {pages} {1307} (\bibinfo {year} {2012})}\BibitemShut
  {NoStop}%
\bibitem [{\citenamefont {Patrignani}\ \emph {et~al.}(2016)\citenamefont
  {Patrignani} \emph {et~al.}}]{Patrignani:2016xqp}%
  \BibitemOpen
  \bibfield  {author} {\bibinfo {author} {\bibfnamefont {C.}~\bibnamefont
  {Patrignani}} \emph {et~al.} (\bibinfo {collaboration} {Particle Data
  Group}),\ }\href {\doibase 10.1088/1674-1137/40/10/100001} {\bibfield
  {journal} {\bibinfo  {journal} {Chin. Phys.}\ }\textbf {\bibinfo {volume}
  {C40}},\ \bibinfo {pages} {100001} (\bibinfo {year} {2016})}\BibitemShut
  {NoStop}%
\bibitem [{\citenamefont {Abe}\ \emph {et~al.}(2013{\natexlab{a}})\citenamefont
  {Abe} \emph {et~al.}}]{Abe:2012av}%
  \BibitemOpen
  \bibfield  {author} {\bibinfo {author} {\bibfnamefont {K.}~\bibnamefont
  {Abe}} \emph {et~al.} (\bibinfo {collaboration} {T2K}),\ }\href {\doibase
  10.1103/PhysRevD.87.012001, 10.1103/PhysRevD.87.019902} {\bibfield  {journal}
  {\bibinfo  {journal} {Phys. Rev.}\ }\textbf {\bibinfo {volume} {D87}},\
  \bibinfo {pages} {012001} (\bibinfo {year} {2013}{\natexlab{a}})},\ \bibinfo
  {note} {[Addendum: Phys. Rev.D87,no.1,019902(2013)]},\ \Eprint
  {http://arxiv.org/abs/1211.0469} {arXiv:1211.0469 [hep-ex]} \BibitemShut
  {NoStop}%
\bibitem [{nov()}]{novacommunication}%
  \BibitemOpen
  \href@noop {} {}\bibinfo {note} {{NOvA collaboration, publicly available flux
  prediction}}\BibitemShut {NoStop}%
\bibitem [{\citenamefont {Aliaga}\ \emph {et~al.}(2016)\citenamefont {Aliaga}
  \emph {et~al.}}]{Aliaga:2016oaz}%
  \BibitemOpen
  \bibfield  {author} {\bibinfo {author} {\bibfnamefont {L.}~\bibnamefont
  {Aliaga}} \emph {et~al.} (\bibinfo {collaboration} {MINERvA}),\ }\href
  {\doibase 10.1103/PhysRevD.94.092005, 10.1103/PhysRevD.95.039903} {\bibfield
  {journal} {\bibinfo  {journal} {Phys. Rev.}\ }\textbf {\bibinfo {volume}
  {D94}},\ \bibinfo {pages} {092005} (\bibinfo {year} {2016})},\ \bibinfo
  {note} {[Addendum: Phys. Rev.D95,no.3,039903(2017)]},\ \Eprint
  {http://arxiv.org/abs/1607.00704} {arXiv:1607.00704 [hep-ex]} \BibitemShut
  {NoStop}%
\bibitem [{\citenamefont {Acciarri}\ \emph {et~al.}(2015)\citenamefont
  {Acciarri} \emph {et~al.}}]{Acciarri:2015uup}%
  \BibitemOpen
  \bibfield  {author} {\bibinfo {author} {\bibfnamefont {R.}~\bibnamefont
  {Acciarri}} \emph {et~al.} (\bibinfo {collaboration} {DUNE}),\ }\href@noop {}
  {\  (\bibinfo {year} {2015})},\ \Eprint {http://arxiv.org/abs/1512.06148}
  {arXiv:1512.06148 [physics.ins-det]} \BibitemShut {NoStop}%
\bibitem [{\citenamefont {Abe}\ \emph {et~al.}(2017{\natexlab{a}})\citenamefont
  {Abe} \emph {et~al.}}]{Abe:2017vif}%
  \BibitemOpen
  \bibfield  {author} {\bibinfo {author} {\bibfnamefont {K.}~\bibnamefont
  {Abe}} \emph {et~al.} (\bibinfo {collaboration} {T2K}),\ }\href {\doibase
  10.1103/PhysRevD.96.092006} {\bibfield  {journal} {\bibinfo  {journal} {Phys.
  Rev.}\ }\textbf {\bibinfo {volume} {D96}},\ \bibinfo {pages} {092006}
  (\bibinfo {year} {2017}{\natexlab{a}})},\ \Eprint
  {http://arxiv.org/abs/1707.01048} {arXiv:1707.01048 [hep-ex]} \BibitemShut
  {NoStop}%
\bibitem [{\citenamefont {Fernandez-Martinez}\ and\ \citenamefont
  {Meloni}(2011)}]{FernandezMartinez:2010dm}%
  \BibitemOpen
  \bibfield  {author} {\bibinfo {author} {\bibfnamefont {E.}~\bibnamefont
  {Fernandez-Martinez}}\ and\ \bibinfo {author} {\bibfnamefont
  {D.}~\bibnamefont {Meloni}},\ }\href {\doibase
  10.1016/j.physletb.2011.02.043} {\bibfield  {journal} {\bibinfo  {journal}
  {Phys. Lett.}\ }\textbf {\bibinfo {volume} {B697}},\ \bibinfo {pages} {477}
  (\bibinfo {year} {2011})},\ \Eprint {http://arxiv.org/abs/1010.2329}
  {arXiv:1010.2329 [hep-ph]} \BibitemShut {NoStop}%
\bibitem [{\citenamefont {Lalakulich}\ \emph {et~al.}(2012)\citenamefont
  {Lalakulich}, \citenamefont {Mosel},\ and\ \citenamefont
  {Gallmeister}}]{Lalakulich:2012hs}%
  \BibitemOpen
  \bibfield  {author} {\bibinfo {author} {\bibfnamefont {O.}~\bibnamefont
  {Lalakulich}}, \bibinfo {author} {\bibfnamefont {U.}~\bibnamefont {Mosel}}, \
  and\ \bibinfo {author} {\bibfnamefont {K.}~\bibnamefont {Gallmeister}},\
  }\href {\doibase 10.1103/PhysRevC.86.054606} {\bibfield  {journal} {\bibinfo
  {journal} {Phys. Rev.}\ }\textbf {\bibinfo {volume} {C86}},\ \bibinfo {pages}
  {054606} (\bibinfo {year} {2012})},\ \Eprint {http://arxiv.org/abs/1208.3678}
  {arXiv:1208.3678 [nucl-th]} \BibitemShut {NoStop}%
\bibitem [{\citenamefont {Meloni}\ and\ \citenamefont
  {Martini}(2012)}]{Meloni:2012fq}%
  \BibitemOpen
  \bibfield  {author} {\bibinfo {author} {\bibfnamefont {D.}~\bibnamefont
  {Meloni}}\ and\ \bibinfo {author} {\bibfnamefont {M.}~\bibnamefont
  {Martini}},\ }\href {\doibase 10.1016/j.physletb.2012.08.007} {\bibfield
  {journal} {\bibinfo  {journal} {Phys. Lett.}\ }\textbf {\bibinfo {volume}
  {B716}},\ \bibinfo {pages} {186} (\bibinfo {year} {2012})},\ \Eprint
  {http://arxiv.org/abs/1203.3335} {arXiv:1203.3335 [hep-ph]} \BibitemShut
  {NoStop}%
\bibitem [{\citenamefont {Coloma}\ and\ \citenamefont
  {Huber}(2013)}]{Coloma:2013rqa}%
  \BibitemOpen
  \bibfield  {author} {\bibinfo {author} {\bibfnamefont {P.}~\bibnamefont
  {Coloma}}\ and\ \bibinfo {author} {\bibfnamefont {P.}~\bibnamefont {Huber}},\
  }\href {\doibase 10.1103/PhysRevLett.111.221802} {\bibfield  {journal}
  {\bibinfo  {journal} {Phys. Rev. Lett.}\ }\textbf {\bibinfo {volume} {111}},\
  \bibinfo {pages} {221802} (\bibinfo {year} {2013})},\ \Eprint
  {http://arxiv.org/abs/1307.1243} {arXiv:1307.1243 [hep-ph]} \BibitemShut
  {NoStop}%
\bibitem [{\citenamefont {Martini}\ \emph {et~al.}(2013)\citenamefont
  {Martini}, \citenamefont {Ericson},\ and\ \citenamefont
  {Chanfray}}]{Martini:2012uc}%
  \BibitemOpen
  \bibfield  {author} {\bibinfo {author} {\bibfnamefont {M.}~\bibnamefont
  {Martini}}, \bibinfo {author} {\bibfnamefont {M.}~\bibnamefont {Ericson}}, \
  and\ \bibinfo {author} {\bibfnamefont {G.}~\bibnamefont {Chanfray}},\ }\href
  {\doibase 10.1103/PhysRevD.87.013009} {\bibfield  {journal} {\bibinfo
  {journal} {Phys. Rev.}\ }\textbf {\bibinfo {volume} {D87}},\ \bibinfo {pages}
  {013009} (\bibinfo {year} {2013})},\ \Eprint {http://arxiv.org/abs/1211.1523}
  {arXiv:1211.1523 [hep-ph]} \BibitemShut {NoStop}%
\bibitem [{\citenamefont {Coloma}\ \emph {et~al.}(2014)\citenamefont {Coloma},
  \citenamefont {Huber}, \citenamefont {Jen},\ and\ \citenamefont
  {Mariani}}]{Coloma:2013tba}%
  \BibitemOpen
  \bibfield  {author} {\bibinfo {author} {\bibfnamefont {P.}~\bibnamefont
  {Coloma}}, \bibinfo {author} {\bibfnamefont {P.}~\bibnamefont {Huber}},
  \bibinfo {author} {\bibfnamefont {C.-M.}\ \bibnamefont {Jen}}, \ and\
  \bibinfo {author} {\bibfnamefont {C.}~\bibnamefont {Mariani}},\ }\href
  {\doibase 10.1103/PhysRevD.89.073015} {\bibfield  {journal} {\bibinfo
  {journal} {Phys. Rev.}\ }\textbf {\bibinfo {volume} {D89}},\ \bibinfo {pages}
  {073015} (\bibinfo {year} {2014})},\ \Eprint {http://arxiv.org/abs/1311.4506}
  {arXiv:1311.4506 [hep-ph]} \BibitemShut {NoStop}%
\bibitem [{\citenamefont {Jen}\ \emph {et~al.}(2014)\citenamefont {Jen},
  \citenamefont {Ankowski}, \citenamefont {Benhar}, \citenamefont {Furmanski},
  \citenamefont {Kalousis},\ and\ \citenamefont {Mariani}}]{Jen:2014aja}%
  \BibitemOpen
  \bibfield  {author} {\bibinfo {author} {\bibfnamefont {C.~M.}\ \bibnamefont
  {Jen}}, \bibinfo {author} {\bibfnamefont {A.}~\bibnamefont {Ankowski}},
  \bibinfo {author} {\bibfnamefont {O.}~\bibnamefont {Benhar}}, \bibinfo
  {author} {\bibfnamefont {A.~P.}\ \bibnamefont {Furmanski}}, \bibinfo {author}
  {\bibfnamefont {L.~N.}\ \bibnamefont {Kalousis}}, \ and\ \bibinfo {author}
  {\bibfnamefont {C.}~\bibnamefont {Mariani}},\ }\href {\doibase
  10.1103/PhysRevD.90.093004} {\bibfield  {journal} {\bibinfo  {journal} {Phys.
  Rev.}\ }\textbf {\bibinfo {volume} {D90}},\ \bibinfo {pages} {093004}
  (\bibinfo {year} {2014})},\ \Eprint {http://arxiv.org/abs/1402.6651}
  {arXiv:1402.6651 [hep-ex]} \BibitemShut {NoStop}%
\bibitem [{\citenamefont {Abe}\ \emph {et~al.}(2015{\natexlab{a}})\citenamefont
  {Abe} \emph {et~al.}}]{Abe:2015awa}%
  \BibitemOpen
  \bibfield  {author} {\bibinfo {author} {\bibfnamefont {K.}~\bibnamefont
  {Abe}} \emph {et~al.} (\bibinfo {collaboration} {T2K}),\ }\href {\doibase
  10.1103/PhysRevD.91.072010} {\bibfield  {journal} {\bibinfo  {journal} {Phys.
  Rev.}\ }\textbf {\bibinfo {volume} {D91}},\ \bibinfo {pages} {072010}
  (\bibinfo {year} {2015}{\natexlab{a}})},\ \Eprint
  {http://arxiv.org/abs/1502.01550} {arXiv:1502.01550 [hep-ex]} \BibitemShut
  {NoStop}%
\bibitem [{\citenamefont {Ankowski}\ \emph {et~al.}(2016)\citenamefont
  {Ankowski}, \citenamefont {Benhar}, \citenamefont {Mariani},\ and\
  \citenamefont {Vagnoni}}]{Ankowski:2016bji}%
  \BibitemOpen
  \bibfield  {author} {\bibinfo {author} {\bibfnamefont {A.~M.}\ \bibnamefont
  {Ankowski}}, \bibinfo {author} {\bibfnamefont {O.}~\bibnamefont {Benhar}},
  \bibinfo {author} {\bibfnamefont {C.}~\bibnamefont {Mariani}}, \ and\
  \bibinfo {author} {\bibfnamefont {E.}~\bibnamefont {Vagnoni}},\ }\href
  {\doibase 10.1103/PhysRevD.93.113004} {\bibfield  {journal} {\bibinfo
  {journal} {Phys. Rev.}\ }\textbf {\bibinfo {volume} {D93}},\ \bibinfo {pages}
  {113004} (\bibinfo {year} {2016})},\ \Eprint
  {http://arxiv.org/abs/1603.01072} {arXiv:1603.01072 [hep-ph]} \BibitemShut
  {NoStop}%
\bibitem [{\citenamefont {Mosel}\ \emph {et~al.}(2014)\citenamefont {Mosel},
  \citenamefont {Lalakulich},\ and\ \citenamefont
  {Gallmeister}}]{Mosel:2013fxa}%
  \BibitemOpen
  \bibfield  {author} {\bibinfo {author} {\bibfnamefont {U.}~\bibnamefont
  {Mosel}}, \bibinfo {author} {\bibfnamefont {O.}~\bibnamefont {Lalakulich}}, \
  and\ \bibinfo {author} {\bibfnamefont {K.}~\bibnamefont {Gallmeister}},\
  }\href {\doibase 10.1103/PhysRevLett.112.151802} {\bibfield  {journal}
  {\bibinfo  {journal} {Phys. Rev. Lett.}\ }\textbf {\bibinfo {volume} {112}},\
  \bibinfo {pages} {151802} (\bibinfo {year} {2014})},\ \Eprint
  {http://arxiv.org/abs/1311.7288} {arXiv:1311.7288 [nucl-th]} \BibitemShut
  {NoStop}%
\bibitem [{\citenamefont {Ankowski}\ \emph {et~al.}(2015)\citenamefont
  {Ankowski}, \citenamefont {Coloma}, \citenamefont {Huber}, \citenamefont
  {Mariani},\ and\ \citenamefont {Vagnoni}}]{Ankowski:2015kya}%
  \BibitemOpen
  \bibfield  {author} {\bibinfo {author} {\bibfnamefont {A.~M.}\ \bibnamefont
  {Ankowski}}, \bibinfo {author} {\bibfnamefont {P.}~\bibnamefont {Coloma}},
  \bibinfo {author} {\bibfnamefont {P.}~\bibnamefont {Huber}}, \bibinfo
  {author} {\bibfnamefont {C.}~\bibnamefont {Mariani}}, \ and\ \bibinfo
  {author} {\bibfnamefont {E.}~\bibnamefont {Vagnoni}},\ }\href {\doibase
  10.1103/PhysRevD.92.091301} {\bibfield  {journal} {\bibinfo  {journal} {Phys.
  Rev.}\ }\textbf {\bibinfo {volume} {D92}},\ \bibinfo {pages} {091301}
  (\bibinfo {year} {2015})},\ \Eprint {http://arxiv.org/abs/1507.08561}
  {arXiv:1507.08561 [hep-ph]} \BibitemShut {NoStop}%
\bibitem [{\citenamefont {Adamson}\ \emph {et~al.}(2011)\citenamefont {Adamson}
  \emph {et~al.}}]{Adamson:2011ku}%
  \BibitemOpen
  \bibfield  {author} {\bibinfo {author} {\bibfnamefont {P.}~\bibnamefont
  {Adamson}} \emph {et~al.} (\bibinfo {collaboration} {MINOS}),\ }\href
  {\doibase 10.1103/PhysRevLett.107.011802} {\bibfield  {journal} {\bibinfo
  {journal} {Phys. Rev. Lett.}\ }\textbf {\bibinfo {volume} {107}},\ \bibinfo
  {pages} {011802} (\bibinfo {year} {2011})},\ \Eprint
  {http://arxiv.org/abs/1104.3922} {arXiv:1104.3922 [hep-ex]} \BibitemShut
  {NoStop}%
\bibitem [{\citenamefont {Aguilar-Arevalo}\ \emph {et~al.}(2017)\citenamefont
  {Aguilar-Arevalo} \emph {et~al.}}]{Aguilar-Arevalo:2017mqx}%
  \BibitemOpen
  \bibfield  {author} {\bibinfo {author} {\bibfnamefont {A.~A.}\ \bibnamefont
  {Aguilar-Arevalo}} \emph {et~al.} (\bibinfo {collaboration} {MiniBooNE}),\
  }\href {\doibase 10.1103/PhysRevLett.118.221803} {\bibfield  {journal}
  {\bibinfo  {journal} {Phys. Rev. Lett.}\ }\textbf {\bibinfo {volume} {118}},\
  \bibinfo {pages} {221803} (\bibinfo {year} {2017})},\ \Eprint
  {http://arxiv.org/abs/1702.02688} {arXiv:1702.02688 [hep-ex]} \BibitemShut
  {NoStop}%
\bibitem [{\citenamefont {Abe}\ \emph {et~al.}(2017{\natexlab{b}})\citenamefont
  {Abe}, \citenamefont {Haga}, \citenamefont {Hayato}, \citenamefont {Ikeda},
  \citenamefont {Iyogi}, \citenamefont {Kameda}, \citenamefont {Kishimoto},
  \citenamefont {Miura}, \citenamefont {Moriyama}, \citenamefont {Nakahata},
  \citenamefont {Nakajima}, \citenamefont {Nakano}, \citenamefont {Nakayama},
  \citenamefont {Orii}, \citenamefont {Sekiya}, \citenamefont {Shiozawa},
  \citenamefont {Takeda}, \citenamefont {Tanaka}, \citenamefont {Tomura},
  \citenamefont {Wendell}, \citenamefont {Akutsu}, \citenamefont {Irvine},
  \citenamefont {Kajita}, \citenamefont {Kaneyuki}, \citenamefont {Nishimura},
  \citenamefont {Richard}, \citenamefont {Okumura}, \citenamefont {Labarga},
  \citenamefont {Fernandez}, \citenamefont {Gustafson}, \citenamefont
  {Kachulis}, \citenamefont {Kearns}, \citenamefont {Raaf}, \citenamefont
  {Stone}, \citenamefont {Sulak}, \citenamefont {Berkman}, \citenamefont
  {Nantais}, \citenamefont {Tanaka}, \citenamefont {Tobayama}, \citenamefont
  {Goldhaber}, \citenamefont {Kropp}, \citenamefont {Mine}, \citenamefont
  {Weatherly}, \citenamefont {Smy}, \citenamefont {Sobel}, \citenamefont
  {Takhistov}, \citenamefont {Ganezer}, \citenamefont {Hartfiel}, \citenamefont
  {Hill}, \citenamefont {Hong}, \citenamefont {Kim}, \citenamefont {Lim},
  \citenamefont {Park}, \citenamefont {Himmel}, \citenamefont {Li},
  \citenamefont {O'Sullivan}, \citenamefont {Scholberg}, \citenamefont
  {Walter}, \citenamefont {Wongjirad}, \citenamefont {Ishizuka}, \citenamefont
  {Tasaka}, \citenamefont {Jang}, \citenamefont {Learned}, \citenamefont
  {Matsuno}, \citenamefont {Smith}, \citenamefont {Friend}, \citenamefont
  {Hasegawa}, \citenamefont {Ishida}, \citenamefont {Ishii}, \citenamefont
  {Kobayashi}, \citenamefont {Nakadaira}, \citenamefont {Nakamura},
  \citenamefont {Oyama}, \citenamefont {Sakashita}, \citenamefont {Sekiguchi},
  \citenamefont {Tsukamoto}, \citenamefont {Suzuki}, \citenamefont {Takeuchi},
  \citenamefont {Yano}, \citenamefont {Cao}, \citenamefont {Hiraki},
  \citenamefont {Hirota}, \citenamefont {Huang}, \citenamefont {Kikawa},
  \citenamefont {Minamino}, \citenamefont {Nakaya}, \citenamefont {Suzuki},
  \citenamefont {Fukuda}, \citenamefont {Choi}, \citenamefont {Itow},
  \citenamefont {Suzuki}, \citenamefont {Mijakowski}, \citenamefont
  {Frankiewicz}, \citenamefont {Hignight}, \citenamefont {Imber}, \citenamefont
  {Jung}, \citenamefont {Li}, \citenamefont {Palomino}, \citenamefont
  {Wilking}, \citenamefont {Yanagisawa}, \citenamefont {Fukuda}, \citenamefont
  {Ishino}, \citenamefont {Kayano}, \citenamefont {Kibayashi}, \citenamefont
  {Koshio}, \citenamefont {Mori}, \citenamefont {Sakuda}, \citenamefont {Xu},
  \citenamefont {Kuno}, \citenamefont {Tacik}, \citenamefont {Kim},
  \citenamefont {Okazawa}, \citenamefont {Choi}, \citenamefont {Nishijima},
  \citenamefont {Koshiba}, \citenamefont {Totsuka}, \citenamefont {Suda},
  \citenamefont {Yokoyama}, \citenamefont {Bronner}, \citenamefont {Hartz},
  \citenamefont {Martens}, \citenamefont {Marti}, \citenamefont {Suzuki},
  \citenamefont {Vagins}, \citenamefont {Martin}, \citenamefont {Konaka},
  \citenamefont {Chen}, \citenamefont {Zhang},\ and\ \citenamefont
  {Wilkes}}]{SK_epi0}%
  \BibitemOpen
  \bibfield  {author} {\bibinfo {author} {\bibfnamefont {K.}~\bibnamefont
  {Abe}}, \bibinfo {author} {\bibfnamefont {Y.}~\bibnamefont {Haga}}, \bibinfo
  {author} {\bibfnamefont {Y.}~\bibnamefont {Hayato}}, \bibinfo {author}
  {\bibfnamefont {M.}~\bibnamefont {Ikeda}},  \emph {et~al.} (\bibinfo
  {collaboration} {Super-Kamiokande Collaboration}),\ }\href {\doibase
  10.1103/PhysRevD.95.012004} {\bibfield  {journal} {\bibinfo  {journal} {Phys.
  Rev. D}\ }\textbf {\bibinfo {volume} {95}},\ \bibinfo {pages} {012004}
  (\bibinfo {year} {2017}{\natexlab{b}})}\BibitemShut {NoStop}%
\bibitem [{\citenamefont {Abe}\ \emph {et~al.}(2014{\natexlab{a}})\citenamefont
  {Abe}, \citenamefont {Hayato}, \citenamefont {Iyogi}, \citenamefont {Kameda},
  \citenamefont {Miura}, \citenamefont {Moriyama}, \citenamefont {Nakahata},
  \citenamefont {Nakayama}, \citenamefont {Wendell}, \citenamefont {Sekiya},
  \citenamefont {Shiozawa}, \citenamefont {Suzuki}, \citenamefont {Takeda},
  \citenamefont {Takenaga}, \citenamefont {Ueno}, \citenamefont {Yokozawa},
  \citenamefont {Kaji}, \citenamefont {Kajita}, \citenamefont {Kaneyuki},
  \citenamefont {Lee}, \citenamefont {Okumura}, \citenamefont {McLachlan},
  \citenamefont {Labarga}, \citenamefont {Kearns}, \citenamefont {Raaf},
  \citenamefont {Stone}, \citenamefont {Sulak}, \citenamefont {Goldhaber},
  \citenamefont {Bays}, \citenamefont {Carminati}, \citenamefont {Kropp},
  \citenamefont {Mine}, \citenamefont {Renshaw}, \citenamefont {Smy},
  \citenamefont {Sobel}, \citenamefont {Ganezer}, \citenamefont {Hill},
  \citenamefont {Keig}, \citenamefont {Jang}, \citenamefont {Kim},
  \citenamefont {Lim}, \citenamefont {Albert}, \citenamefont {Scholberg},
  \citenamefont {Walter}, \citenamefont {Wongjirad}, \citenamefont {Ishizuka},
  \citenamefont {Tasaka}, \citenamefont {Learned}, \citenamefont {Matsuno},
  \citenamefont {Smith}, \citenamefont {Hasegawa}, \citenamefont {Ishida},
  \citenamefont {Ishii}, \citenamefont {Kobayashi}, \citenamefont {Nakadaira},
  \citenamefont {Nakamura}, \citenamefont {Nishikawa}, \citenamefont {Oyama},
  \citenamefont {Sakashita}, \citenamefont {Sekiguchi}, \citenamefont
  {Tsukamoto}, \citenamefont {Suzuki}, \citenamefont {Takeuchi}, \citenamefont
  {Ieki}, \citenamefont {Ikeda}, \citenamefont {Kubo}, \citenamefont
  {Minamino}, \citenamefont {Murakami}, \citenamefont {Nakaya}, \citenamefont
  {Fukuda}, \citenamefont {Choi}, \citenamefont {Itow}, \citenamefont
  {Mitsuka}, \citenamefont {Miyake}, \citenamefont {Mijakowski}, \citenamefont
  {Hignight}, \citenamefont {Imber}, \citenamefont {Jung}, \citenamefont
  {Taylor}, \citenamefont {Yanagisawa}, \citenamefont {Ishino}, \citenamefont
  {Kibayashi}, \citenamefont {Koshio}, \citenamefont {Mori}, \citenamefont
  {Sakuda}, \citenamefont {Takeuchi}, \citenamefont {Kuno}, \citenamefont
  {Kim}, \citenamefont {Okazawa}, \citenamefont {Choi}, \citenamefont
  {Nishijima}, \citenamefont {Koshiba}, \citenamefont {Totsuka}, \citenamefont
  {Yokoyama}, \citenamefont {Martens}, \citenamefont {Marti}, \citenamefont
  {Obayashi}, \citenamefont {Vagins}, \citenamefont {Chen}, \citenamefont
  {Sui}, \citenamefont {Yang}, \citenamefont {Zhang}, \citenamefont {Connolly},
  \citenamefont {Dziomba},\ and\ \citenamefont {Wilkes}}]{SK_Knu}%
  \BibitemOpen
  \bibfield  {author} {\bibinfo {author} {\bibfnamefont {K.}~\bibnamefont
  {Abe}}, \bibinfo {author} {\bibfnamefont {Y.}~\bibnamefont {Hayato}},
  \bibinfo {author} {\bibfnamefont {K.}~\bibnamefont {Iyogi}}, \bibinfo
  {author} {\bibfnamefont {J.}~\bibnamefont {Kameda}},  \emph {et~al.}
  (\bibinfo {collaboration} {Super-Kamiokande Collaboration}),\ }\href
  {\doibase 10.1103/PhysRevD.90.072005} {\bibfield  {journal} {\bibinfo
  {journal} {Phys. Rev. D}\ }\textbf {\bibinfo {volume} {90}},\ \bibinfo
  {pages} {072005} (\bibinfo {year} {2014}{\natexlab{a}})}\BibitemShut
  {NoStop}%
\bibitem [{\citenamefont {Alvarez-Ruso}\ \emph {et~al.}(2014)\citenamefont
  {Alvarez-Ruso}, \citenamefont {Hayato},\ and\ \citenamefont
  {Nieves}}]{Alvarez-Ruso:2014bla}%
  \BibitemOpen
  \bibfield  {author} {\bibinfo {author} {\bibfnamefont {L.}~\bibnamefont
  {Alvarez-Ruso}}, \bibinfo {author} {\bibfnamefont {Y.}~\bibnamefont
  {Hayato}}, \ and\ \bibinfo {author} {\bibfnamefont {J.}~\bibnamefont
  {Nieves}},\ }\href {\doibase 10.1088/1367-2630/16/7/075015} {\bibfield
  {journal} {\bibinfo  {journal} {New J. Phys.}\ }\textbf {\bibinfo {volume}
  {16}},\ \bibinfo {pages} {075015} (\bibinfo {year} {2014})},\ \Eprint
  {http://arxiv.org/abs/1403.2673} {arXiv:1403.2673 [hep-ph]} \BibitemShut
  {NoStop}%
\bibitem [{\citenamefont {Garvey}\ \emph {et~al.}(2015)\citenamefont {Garvey},
  \citenamefont {Harris}, \citenamefont {Tanaka}, \citenamefont {Tayloe},\ and\
  \citenamefont {Zeller}}]{Garvey:2014exa}%
  \BibitemOpen
  \bibfield  {author} {\bibinfo {author} {\bibfnamefont {G.~T.}\ \bibnamefont
  {Garvey}}, \bibinfo {author} {\bibfnamefont {D.~A.}\ \bibnamefont {Harris}},
  \bibinfo {author} {\bibfnamefont {H.~A.}\ \bibnamefont {Tanaka}}, \bibinfo
  {author} {\bibfnamefont {R.}~\bibnamefont {Tayloe}}, \ and\ \bibinfo {author}
  {\bibfnamefont {G.~P.}\ \bibnamefont {Zeller}},\ }\href {\doibase
  10.1016/j.physrep.2015.04.001} {\bibfield  {journal} {\bibinfo  {journal}
  {Phys. Rept.}\ }\textbf {\bibinfo {volume} {580}},\ \bibinfo {pages} {1}
  (\bibinfo {year} {2015})},\ \Eprint {http://arxiv.org/abs/1412.4294}
  {arXiv:1412.4294 [hep-ex]} \BibitemShut {NoStop}%
\bibitem [{\citenamefont {Mosel}(2016)}]{Mosel:2016cwa}%
  \BibitemOpen
  \bibfield  {author} {\bibinfo {author} {\bibfnamefont {U.}~\bibnamefont
  {Mosel}},\ }\href {\doibase 10.1146/annurev-nucl-102115-044720} {\bibfield
  {journal} {\bibinfo  {journal} {Ann. Rev. Nucl. Part. Sci.}\ }\textbf
  {\bibinfo {volume} {66}},\ \bibinfo {pages} {171} (\bibinfo {year} {2016})},\
  \Eprint {http://arxiv.org/abs/1602.00696} {arXiv:1602.00696 [nucl-th]}
  \BibitemShut {NoStop}%
\bibitem [{\citenamefont {Katori}\ and\ \citenamefont
  {Martini}(2018)}]{Katori:2016yel}%
  \BibitemOpen
  \bibfield  {author} {\bibinfo {author} {\bibfnamefont {T.}~\bibnamefont
  {Katori}}\ and\ \bibinfo {author} {\bibfnamefont {M.}~\bibnamefont
  {Martini}},\ }\href {\doibase 10.1088/1361-6471/aa8bf7} {\bibfield  {journal}
  {\bibinfo  {journal} {J. Phys.}\ }\textbf {\bibinfo {volume} {G45}},\
  \bibinfo {pages} {013001} (\bibinfo {year} {2018})},\ \Eprint
  {http://arxiv.org/abs/1611.07770} {arXiv:1611.07770 [hep-ph]} \BibitemShut
  {NoStop}%
\bibitem [{\citenamefont {Alvarez-Ruso}\ \emph {et~al.}(2017)\citenamefont
  {Alvarez-Ruso} \emph {et~al.}}]{Alvarez-Ruso:2017oui}%
  \BibitemOpen
  \bibfield  {author} {\bibinfo {author} {\bibfnamefont {L.}~\bibnamefont
  {Alvarez-Ruso}} \emph {et~al.},\ }\href@noop {} {\  (\bibinfo {year}
  {2017})},\ \Eprint {http://arxiv.org/abs/1706.03621} {arXiv:1706.03621
  [hep-ph]} \BibitemShut {NoStop}%
\bibitem [{\citenamefont {Prosper}\ and\ \citenamefont
  {Lyons}(2011)}]{Prosper:2011zz}%
  \BibitemOpen
  \bibinfo {editor} {\bibfnamefont {H.~B.}\ \bibnamefont {Prosper}}\ and\
  \bibinfo {editor} {\bibfnamefont {L.}~\bibnamefont {Lyons}},\ eds.,\ \href
  {\doibase 10.5170/CERN-2011-006} {\emph {\bibinfo {title} {{Proceedings,
  PHYSTAT 2011 Workshop on Statistical Issues Related to Discovery Claims in
  Search Experiments and Unfolding, CERN,\,Geneva, Switzerland 17-20 January
  2011}}}},\ \bibinfo {organization} {CERN}\ (\bibinfo  {publisher} {CERN},\
  \bibinfo {address} {Geneva},\ \bibinfo {year} {2011})\BibitemShut {NoStop}%
\bibitem [{\citenamefont {Kuusela}(2012)}]{kuusela_master}%
  \BibitemOpen
  \bibfield  {author} {\bibinfo {author} {\bibfnamefont {M.}~\bibnamefont
  {Kuusela}},\ }\emph {\bibinfo {title} {{ Statistical Issues in Unfolding
  Methods for High Energy Physics}}},\ \href@noop {} {Master's thesis},\
  \bibinfo  {school} {Aalto University} (\bibinfo {year} {2012})\BibitemShut
  {NoStop}%
\bibitem [{\citenamefont {Kuusela}(2016)}]{kuusela_phd}%
  \BibitemOpen
  \bibfield  {author} {\bibinfo {author} {\bibfnamefont {M.~J.}\ \bibnamefont
  {Kuusela}},\ }\emph {\bibinfo {title} {Uncertainty quantification in
  unfolding elementary particle spectra at the {L}arge {H}adron {C}ollider}},\
  \href {\doibase 10.5075/epfl-thesis-7118} {Ph.D. thesis},\ \bibinfo  {school}
  {SB}, \bibinfo {address} {Lausanne} (\bibinfo {year} {2016})\BibitemShut
  {NoStop}%
\bibitem [{\citenamefont {D'Agostini}(1995)}]{D'Agostini:1994zf}%
  \BibitemOpen
  \bibfield  {author} {\bibinfo {author} {\bibfnamefont {G.}~\bibnamefont
  {D'Agostini}},\ }\href {\doibase 10.1016/0168-9002(95)00274-X} {\bibfield
  {journal} {\bibinfo  {journal} {Nucl. Instrum. Meth.}\ }\textbf {\bibinfo
  {volume} {A362}},\ \bibinfo {pages} {487} (\bibinfo {year}
  {1995})}\BibitemShut {NoStop}%
\bibitem [{\citenamefont {Tice}\ \emph {et~al.}(2014)\citenamefont {Tice} \emph
  {et~al.}}]{Tice:2014pgu}%
  \BibitemOpen
  \bibfield  {author} {\bibinfo {author} {\bibfnamefont {B.~G.}\ \bibnamefont
  {Tice}} \emph {et~al.} (\bibinfo {collaboration} {MINERvA}),\ }\href@noop {}
  {\bibfield  {journal} {\bibinfo  {journal} {Phys. Rev. Lett.}\ } (\bibinfo
  {year} {2014})},\ \Eprint {http://arxiv.org/abs/1403.2103} {arXiv:1403.2103
  [hep-ex]} \BibitemShut {NoStop}%
\bibitem [{\citenamefont {Aguilar-Arevalo}\ \emph
  {et~al.}(2010{\natexlab{a}})\citenamefont {Aguilar-Arevalo} \emph
  {et~al.}}]{AguilarArevalo:2010cx}%
  \BibitemOpen
  \bibfield  {author} {\bibinfo {author} {\bibfnamefont {A.~A.}\ \bibnamefont
  {Aguilar-Arevalo}} \emph {et~al.} (\bibinfo {collaboration} {MiniBooNE}),\
  }\href@noop {} {\bibfield  {journal} {\bibinfo  {journal} {Phys. Rev.}\
  }\textbf {\bibinfo {volume} {D82}},\ \bibinfo {pages} {092005} (\bibinfo
  {year} {2010}{\natexlab{a}})},\ \Eprint {http://arxiv.org/abs/1007.4730}
  {arXiv:1007.4730 [hep-ex]} \BibitemShut {NoStop}%
\bibitem [{\citenamefont {Barton}\ \emph {et~al.}(1983)\citenamefont {Barton},
  \citenamefont {Brandenburg}, \citenamefont {Busza}, \citenamefont
  {Dobrowolski}, \citenamefont {Friedman}, \citenamefont {Halliwell},
  \citenamefont {Kendall}, \citenamefont {Lyons}, \citenamefont {Nelson},
  \citenamefont {Rosenson}, \citenamefont {Verdier}, \citenamefont {Chiaradia},
  \citenamefont {DeMarzo}, \citenamefont {Favuzzi}, \citenamefont {Germinario},
  \citenamefont {Guerriero}, \citenamefont {LaVopa}, \citenamefont {Maggi},
  \citenamefont {Posa}, \citenamefont {Selvaggi}, \citenamefont {Spinelli},
  \citenamefont {Waldner}, \citenamefont {Cutts}, \citenamefont {Dulude},
  \citenamefont {Hughlock}, \citenamefont {Lanou}, \citenamefont {Massimo},
  \citenamefont {Brenner}, \citenamefont {Carey}, \citenamefont {Elias},
  \citenamefont {Garbincius}, \citenamefont {Polychronakos}, \citenamefont
  {Nassalski},\ and\ \citenamefont {Siemiarczuk}}]{barton_1983}%
  \BibitemOpen
  \bibfield  {author} {\bibinfo {author} {\bibfnamefont {D.~S.}\ \bibnamefont
  {Barton}}, \bibinfo {author} {\bibfnamefont {G.~W.}\ \bibnamefont
  {Brandenburg}}, \bibinfo {author} {\bibfnamefont {W.}~\bibnamefont {Busza}},
  \bibinfo {author} {\bibfnamefont {T.}~\bibnamefont {Dobrowolski}},  \emph
  {et~al.},\ }\href {\doibase 10.1103/PhysRevD.27.2580} {\bibfield  {journal}
  {\bibinfo  {journal} {Phys. Rev. D}\ }\textbf {\bibinfo {volume} {27}},\
  \bibinfo {pages} {2580} (\bibinfo {year} {1983})}\BibitemShut {NoStop}%
\bibitem [{\citenamefont {Ambrosini}\ \emph
  {et~al.}(1998{\natexlab{a}})\citenamefont {Ambrosini} \emph
  {et~al.}}]{Ambrosini:1998et}%
  \BibitemOpen
  \bibfield  {author} {\bibinfo {author} {\bibfnamefont {G.}~\bibnamefont
  {Ambrosini}} \emph {et~al.} (\bibinfo {collaboration} {SPY}),\ }\href
  {\doibase 10.1016/S0370-2693(97)01572-4} {\bibfield  {journal} {\bibinfo
  {journal} {Phys. Lett.}\ }\textbf {\bibinfo {volume} {B420}},\ \bibinfo
  {pages} {225} (\bibinfo {year} {1998}{\natexlab{a}})}\BibitemShut {NoStop}%
\bibitem [{\citenamefont {Ambrosini}\ \emph
  {et~al.}(1998{\natexlab{b}})\citenamefont {Ambrosini} \emph
  {et~al.}}]{Ambrosini:1998th}%
  \BibitemOpen
  \bibfield  {author} {\bibinfo {author} {\bibfnamefont {G.}~\bibnamefont
  {Ambrosini}} \emph {et~al.} (\bibinfo {collaboration} {SPY}),\ }\href
  {\doibase 10.1016/S0370-2693(98)00237-8} {\bibfield  {journal} {\bibinfo
  {journal} {Phys. Lett.}\ }\textbf {\bibinfo {volume} {B425}},\ \bibinfo
  {pages} {208} (\bibinfo {year} {1998}{\natexlab{b}})}\BibitemShut {NoStop}%
\bibitem [{\citenamefont {Raja}(2005)}]{Raja:2005kow}%
  \BibitemOpen
  \bibfield  {author} {\bibinfo {author} {\bibfnamefont {R.}~\bibnamefont
  {Raja}},\ }\bibfield  {booktitle} {\emph {\bibinfo {booktitle} {{Proceedings,
  5th International Workshop on Ring Imaging Cherenkov Detectors (RICH 2004):
  Playa de Carmen, Yucatan Peninsula, Mexico, November 30-December 5, 2004}}},\
  }\href {\doibase 10.1016/j.nima.2005.08.087} {\bibfield  {journal} {\bibinfo
  {journal} {Nucl. Instrum. Meth.}\ }\textbf {\bibinfo {volume} {A553}},\
  \bibinfo {pages} {225} (\bibinfo {year} {2005})}\BibitemShut {NoStop}%
\bibitem [{\citenamefont {Catanesi}\ \emph {et~al.}(2007)\citenamefont
  {Catanesi} \emph {et~al.}}]{Catanesi:2007ig}%
  \BibitemOpen
  \bibfield  {author} {\bibinfo {author} {\bibfnamefont {M.~G.}\ \bibnamefont
  {Catanesi}} \emph {et~al.} (\bibinfo {collaboration} {HARP}),\ }\href
  {\doibase 10.1016/j.nima.2006.08.132} {\bibfield  {journal} {\bibinfo
  {journal} {Nucl. Instrum. Meth.}\ }\textbf {\bibinfo {volume} {A571}},\
  \bibinfo {pages} {527} (\bibinfo {year} {2007})}\BibitemShut {NoStop}%
\bibitem [{\citenamefont {Abgrall}\ \emph {et~al.}(2014)\citenamefont {Abgrall}
  \emph {et~al.}}]{Abgrall:2014xwa}%
  \BibitemOpen
  \bibfield  {author} {\bibinfo {author} {\bibfnamefont {N.}~\bibnamefont
  {Abgrall}} \emph {et~al.} (\bibinfo {collaboration} {NA61}),\ }\href
  {\doibase 10.1088/1748-0221/9/06/P06005} {\bibfield  {journal} {\bibinfo
  {journal} {JINST}\ }\textbf {\bibinfo {volume} {9}},\ \bibinfo {pages}
  {P06005} (\bibinfo {year} {2014})},\ \Eprint {http://arxiv.org/abs/1401.4699}
  {arXiv:1401.4699 [physics.ins-det]} \BibitemShut {NoStop}%
\bibitem [{\citenamefont {Park}\ \emph {et~al.}(2016)\citenamefont {Park},
  \citenamefont {Aliaga}, \citenamefont {Altinok}, \citenamefont {Bellantoni},
  \citenamefont {Bercellie}, \citenamefont {Betancourt}, \citenamefont {Bodek},
  \citenamefont {Bravar}, \citenamefont {Budd}, \citenamefont {Cai},
  \citenamefont {Carneiro}, \citenamefont {Christy}, \citenamefont {Chvojka},
  \citenamefont {da~Motta}, \citenamefont {Dytman}, \citenamefont {D\'{\i}az},
  \citenamefont {Eberly}, \citenamefont {Felix}, \citenamefont {Fields},
  \citenamefont {Fine}, \citenamefont {Gago}, \citenamefont {Galindo},
  \citenamefont {Ghosh}, \citenamefont {Golan}, \citenamefont {Gran},
  \citenamefont {Harris}, \citenamefont {Higuera}, \citenamefont {Kleykamp},
  \citenamefont {Kordosky}, \citenamefont {Le}, \citenamefont {Maher},
  \citenamefont {Manly}, \citenamefont {Mann}, \citenamefont {Marshall},
  \citenamefont {Martinez~Caicedo}, \citenamefont {McFarland}, \citenamefont
  {McGivern}, \citenamefont {McGowan}, \citenamefont {Messerly}, \citenamefont
  {Miller}, \citenamefont {Mislivec}, \citenamefont {Morf\'{\i}n},
  \citenamefont {Mousseau}, \citenamefont {Naples}, \citenamefont {Nelson},
  \citenamefont {Norrick}, \citenamefont {Nuruzzaman}, \citenamefont {Osta},
  \citenamefont {Paolone}, \citenamefont {Patrick}, \citenamefont {Perdue},
  \citenamefont {Rakotondravohitra}, \citenamefont {Ramirez}, \citenamefont
  {Ray}, \citenamefont {Ren}, \citenamefont {Rimal}, \citenamefont {Rodrigues},
  \citenamefont {Ruterbories}, \citenamefont {Schellman}, \citenamefont
  {Solano~Salinas}, \citenamefont {Tagg}, \citenamefont {Tice}, \citenamefont
  {Valencia}, \citenamefont {Walton}, \citenamefont {Wolcott}, \citenamefont
  {Wospakrik}, \citenamefont {Zavala},\ and\ \citenamefont
  {Zhang}}]{minerva_nue}%
  \BibitemOpen
  \bibfield  {author} {\bibinfo {author} {\bibfnamefont {J.}~\bibnamefont
  {Park}}, \bibinfo {author} {\bibfnamefont {L.}~\bibnamefont {Aliaga}},
  \bibinfo {author} {\bibfnamefont {O.}~\bibnamefont {Altinok}}, \bibinfo
  {author} {\bibfnamefont {L.}~\bibnamefont {Bellantoni}},  \emph {et~al.}
  (\bibinfo {collaboration} {MINERvA}),\ }\href {\doibase
  10.1103/PhysRevD.93.112007} {\bibfield  {journal} {\bibinfo  {journal} {Phys.
  Rev. D}\ }\textbf {\bibinfo {volume} {93}},\ \bibinfo {pages} {112007}
  (\bibinfo {year} {2016})}\BibitemShut {NoStop}%
\bibitem [{\citenamefont {Seligman}\ \emph {et~al.}(1997)\citenamefont
  {Seligman}, \citenamefont {Arroyo}, \citenamefont {de~Barbaro}, \citenamefont
  {de~Barbaro}, \citenamefont {Bazarko}, \citenamefont {Bernstein},
  \citenamefont {Bodek}, \citenamefont {Bolton}, \citenamefont {Budd},
  \citenamefont {Conrad}, \citenamefont {Harris}, \citenamefont {Johnson},
  \citenamefont {Kim}, \citenamefont {King}, \citenamefont {Kinnel},
  \citenamefont {Lamm}, \citenamefont {Lefmann}, \citenamefont {Marsh},
  \citenamefont {McFarland}, \citenamefont {McNulty}, \citenamefont {Mishra},
  \citenamefont {Naples}, \citenamefont {Quintas}, \citenamefont {Romosan},
  \citenamefont {Sakumoto}, \citenamefont {Schellman}, \citenamefont {Sciulli},
  \citenamefont {Shaevitz}, \citenamefont {Smith}, \citenamefont {Spentzouris},
  \citenamefont {Stern}, \citenamefont {Vakili}, \citenamefont {Yang},\ and\
  \citenamefont {Yu}}]{ccfr_lownu}%
  \BibitemOpen
  \bibfield  {author} {\bibinfo {author} {\bibfnamefont {W.~G.}\ \bibnamefont
  {Seligman}}, \bibinfo {author} {\bibfnamefont {C.~G.}\ \bibnamefont
  {Arroyo}}, \bibinfo {author} {\bibfnamefont {L.}~\bibnamefont {de~Barbaro}},
  \bibinfo {author} {\bibfnamefont {P.}~\bibnamefont {de~Barbaro}},  \emph
  {et~al.},\ }\href {\doibase 10.1103/PhysRevLett.79.1213} {\bibfield
  {journal} {\bibinfo  {journal} {Phys. Rev. Lett.}\ }\textbf {\bibinfo
  {volume} {79}},\ \bibinfo {pages} {1213} (\bibinfo {year}
  {1997})}\BibitemShut {NoStop}%
\bibitem [{\citenamefont {Tzanov}\ \emph
  {et~al.}(2006{\natexlab{a}})\citenamefont {Tzanov}, \citenamefont {Naples},
  \citenamefont {Boyd}, \citenamefont {McDonald}, \citenamefont {Radescu},
  \citenamefont {Johnson}, \citenamefont {Suwonjandee}, \citenamefont {Vakili},
  \citenamefont {Conrad}, \citenamefont {Fleming}, \citenamefont {Formaggio},
  \citenamefont {Kim}, \citenamefont {Koutsoliotas}, \citenamefont {McNulty},
  \citenamefont {Romosan}, \citenamefont {Shaevitz}, \citenamefont
  {Spentzouris}, \citenamefont {Stern}, \citenamefont {Vaitaitis},
  \citenamefont {Zimmerman}, \citenamefont {Bernstein}, \citenamefont {Bugel},
  \citenamefont {Lamm}, \citenamefont {Marsh}, \citenamefont {Nienaber},
  \citenamefont {Tobien}, \citenamefont {Yu}, \citenamefont {Adams},
  \citenamefont {Alton}, \citenamefont {Bolton}, \citenamefont {Goldman},
  \citenamefont {Goncharov}, \citenamefont {de~Barbaro}, \citenamefont
  {Buchholz}, \citenamefont {Schellman}, \citenamefont {Zeller}, \citenamefont
  {Brau}, \citenamefont {Drucker}, \citenamefont {Frey}, \citenamefont {Mason},
  \citenamefont {Avvakumov}, \citenamefont {de~Barbaro}, \citenamefont {Bodek},
  \citenamefont {Budd}, \citenamefont {Harris}, \citenamefont {McFarland},
  \citenamefont {Sakumoto},\ and\ \citenamefont {Yang}}]{nutev_lownu}%
  \BibitemOpen
  \bibfield  {author} {\bibinfo {author} {\bibfnamefont {M.}~\bibnamefont
  {Tzanov}}, \bibinfo {author} {\bibfnamefont {D.}~\bibnamefont {Naples}},
  \bibinfo {author} {\bibfnamefont {S.}~\bibnamefont {Boyd}}, \bibinfo {author}
  {\bibfnamefont {J.}~\bibnamefont {McDonald}},  \emph {et~al.},\ }\href
  {\doibase 10.1103/PhysRevD.74.012008} {\bibfield  {journal} {\bibinfo
  {journal} {Phys. Rev. D}\ }\textbf {\bibinfo {volume} {74}},\ \bibinfo
  {pages} {012008} (\bibinfo {year} {2006}{\natexlab{a}})}\BibitemShut
  {NoStop}%
\bibitem [{\citenamefont {Bodek}\ \emph {et~al.}(2012)\citenamefont {Bodek},
  \citenamefont {Sarica}, \citenamefont {Naples},\ and\ \citenamefont
  {Ren}}]{BodekSarica_lownu}%
  \BibitemOpen
  \bibfield  {author} {\bibinfo {author} {\bibfnamefont {A.}~\bibnamefont
  {Bodek}}, \bibinfo {author} {\bibfnamefont {U.}~\bibnamefont {Sarica}},
  \bibinfo {author} {\bibfnamefont {D.}~\bibnamefont {Naples}}, \ and\ \bibinfo
  {author} {\bibfnamefont {L.}~\bibnamefont {Ren}},\ }\href {\doibase
  10.1140/epjc/s10052-012-1973-6} {\bibfield  {journal} {\bibinfo  {journal}
  {The European Physical Journal C}\ }\textbf {\bibinfo {volume} {72}},\
  \bibinfo {pages} {1973} (\bibinfo {year} {2012})}\BibitemShut {NoStop}%
\bibitem [{\citenamefont {Adamson}\ \emph
  {et~al.}(2010{\natexlab{a}})\citenamefont {Adamson}, \citenamefont
  {Andreopoulos}, \citenamefont {Arms}, \citenamefont {Armstrong},
  \citenamefont {Auty}, \citenamefont {Ayres}, \citenamefont {Backhouse},
  \citenamefont {Barnes}, \citenamefont {Barr}, \citenamefont {Barrett},
  \citenamefont {Bhattacharya}, \citenamefont {Bishai}, \citenamefont {Blake},
  \citenamefont {Bock}, \citenamefont {Boehnlein}, \citenamefont {Bogert},
  \citenamefont {Bower}, \citenamefont {Cavanaugh}, \citenamefont {Chapman},
  \citenamefont {Cherdack}, \citenamefont {Childress}, \citenamefont
  {Choudhary}, \citenamefont {Coelho}, \citenamefont {Coleman}, \citenamefont
  {Cronin-Hennessy}, \citenamefont {Culling}, \citenamefont {Danko},
  \citenamefont {de~Jong}, \citenamefont {Devenish}, \citenamefont {Diwan},
  \citenamefont {Dorman}, \citenamefont {Erwin}, \citenamefont {Escobar},
  \citenamefont {Evans}, \citenamefont {Falk}, \citenamefont {Feldman},
  \citenamefont {Frohne}, \citenamefont {Gallagher}, \citenamefont {Godley},
  \citenamefont {Goodman}, \citenamefont {Gouffon}, \citenamefont {Gran},
  \citenamefont {Grashorn}, \citenamefont {Grzelak}, \citenamefont {Habig},
  \citenamefont {Harris}, \citenamefont {Harris}, \citenamefont {Hartnell},
  \citenamefont {Hatcher}, \citenamefont {Heller}, \citenamefont {Himmel},
  \citenamefont {Holin}, \citenamefont {Hylen}, \citenamefont {Irwin},
  \citenamefont {Isvan}, \citenamefont {Jaffe}, \citenamefont {James},
  \citenamefont {Jensen}, \citenamefont {Kafka}, \citenamefont {Kasahara},
  \citenamefont {Kim}, \citenamefont {Koizumi}, \citenamefont {Kopp},
  \citenamefont {Kordosky}, \citenamefont {Koskinen}, \citenamefont {Krahn},
  \citenamefont {Kreymer}, \citenamefont {Lang}, \citenamefont {Ling},
  \citenamefont {Litchfield}, \citenamefont {Litchfield}, \citenamefont
  {Loiacono}, \citenamefont {Lucas}, \citenamefont {Ma}, \citenamefont {Mann},
  \citenamefont {Marshak}, \citenamefont {Marshall}, \citenamefont {Mayer},
  \citenamefont {McGowan}, \citenamefont {Mehdiyev}, \citenamefont {Meier},
  \citenamefont {Messier}, \citenamefont {Metelko}, \citenamefont {Michael},
  \citenamefont {Miller}, \citenamefont {Mishra}, \citenamefont {Mitchell},
  \citenamefont {Moore}, \citenamefont {Morf\'{\i}n}, \citenamefont {Mualem},
  \citenamefont {Mufson}, \citenamefont {Musser}, \citenamefont {Naples},
  \citenamefont {Nelson}, \citenamefont {Newman}, \citenamefont {Nichol},
  \citenamefont {Nicholls}, \citenamefont {Ochoa-Ricoux}, \citenamefont
  {Oliver}, \citenamefont {Osiecki}, \citenamefont {Ospanov}, \citenamefont
  {Paley}, \citenamefont {Paolone}, \citenamefont {Patterson}, \citenamefont
  {Pavlovi\ifmmode~\acute{c}\else \'{c}\fi{}}, \citenamefont {Pawloski},
  \citenamefont {Pearce}, \citenamefont {Petyt}, \citenamefont {Pittam},
  \citenamefont {Plunkett}, \citenamefont {Rahaman}, \citenamefont {Rameika},
  \citenamefont {Raufer}, \citenamefont {Rebel}, \citenamefont {Rodrigues},
  \citenamefont {Rosenfeld}, \citenamefont {Rubin}, \citenamefont {Ryabov},
  \citenamefont {Sanchez}, \citenamefont {Saoulidou}, \citenamefont {Schneps},
  \citenamefont {Schreiner}, \citenamefont {Semenov}, \citenamefont {Shanahan},
  \citenamefont {Smart}, \citenamefont {Smith}, \citenamefont {Sousa},
  \citenamefont {Stamoulis}, \citenamefont {Strait}, \citenamefont {Tagg},
  \citenamefont {Talaga}, \citenamefont {Thomas}, \citenamefont {Thomson},
  \citenamefont {Tinti}, \citenamefont {Toner}, \citenamefont {Tsarev},
  \citenamefont {Tzanakos}, \citenamefont {Urheim}, \citenamefont {Vahle},
  \citenamefont {Viren}, \citenamefont {Watabe}, \citenamefont {Weber},
  \citenamefont {Webb}, \citenamefont {West}, \citenamefont {White},
  \citenamefont {Whitehead}, \citenamefont {Wojcicki}, \citenamefont {Wright},
  \citenamefont {Yang}, \citenamefont {Zois}, \citenamefont {Zhang},\ and\
  \citenamefont {Zwaska}}]{minos_lownu}%
  \BibitemOpen
  \bibfield  {author} {\bibinfo {author} {\bibfnamefont {P.}~\bibnamefont
  {Adamson}}, \bibinfo {author} {\bibfnamefont {C.}~\bibnamefont
  {Andreopoulos}}, \bibinfo {author} {\bibfnamefont {K.~E.}\ \bibnamefont
  {Arms}}, \bibinfo {author} {\bibfnamefont {R.}~\bibnamefont {Armstrong}},
  \emph {et~al.} (\bibinfo {collaboration} {MINOS Collaboration}),\ }\href
  {\doibase 10.1103/PhysRevD.81.072002} {\bibfield  {journal} {\bibinfo
  {journal} {Phys. Rev. D}\ }\textbf {\bibinfo {volume} {81}},\ \bibinfo
  {pages} {072002} (\bibinfo {year} {2010}{\natexlab{a}})}\BibitemShut
  {NoStop}%
\bibitem [{\citenamefont {Devan}\ \emph
  {et~al.}(2016{\natexlab{a}})\citenamefont {Devan}, \citenamefont {Ren},
  \citenamefont {Aliaga}, \citenamefont {Altinok}, \citenamefont {Bellantoni},
  \citenamefont {Bercellie}, \citenamefont {Betancourt}, \citenamefont {Bodek},
  \citenamefont {Budd}, \citenamefont {Cai}, \citenamefont {Carneiro},
  \citenamefont {da~Motta}, \citenamefont {Dytman}, \citenamefont {D\'{\i}az},
  \citenamefont {Eberly}, \citenamefont {Endress}, \citenamefont {Felix},
  \citenamefont {Fields}, \citenamefont {Fine}, \citenamefont {Gago},
  \citenamefont {Galindo}, \citenamefont {Gallagher}, \citenamefont {Ghosh},
  \citenamefont {Gran}, \citenamefont {Harris}, \citenamefont {Higuera},
  \citenamefont {Hurtado}, \citenamefont {Kleykamp}, \citenamefont {Kordosky},
  \citenamefont {Le}, \citenamefont {Maher}, \citenamefont {Manly},
  \citenamefont {Mann}, \citenamefont {Marshall}, \citenamefont
  {Martinez~Caicedo}, \citenamefont {McFarland}, \citenamefont {McGivern},
  \citenamefont {McGowan}, \citenamefont {Messerly}, \citenamefont {Miller},
  \citenamefont {Mislivec}, \citenamefont {Morf\'{\i}n}, \citenamefont
  {Mousseau}, \citenamefont {Naples}, \citenamefont {Nelson}, \citenamefont
  {Norrick}, \citenamefont {Nuruzzaman}, \citenamefont {Paolone}, \citenamefont
  {Park}, \citenamefont {Patrick}, \citenamefont {Perdue}, \citenamefont
  {Ramirez}, \citenamefont {Ransome}, \citenamefont {Ray}, \citenamefont
  {Rimal}, \citenamefont {Rodrigues}, \citenamefont {Ruterbories},
  \citenamefont {Schellman}, \citenamefont {Solano~Salinas}, \citenamefont
  {Tice}, \citenamefont {Valencia}, \citenamefont {Wolcott},\ and\
  \citenamefont {Wospakrik}}]{minerva_lownu}%
  \BibitemOpen
  \bibfield  {author} {\bibinfo {author} {\bibfnamefont {J.}~\bibnamefont
  {Devan}}, \bibinfo {author} {\bibfnamefont {L.}~\bibnamefont {Ren}}, \bibinfo
  {author} {\bibfnamefont {L.}~\bibnamefont {Aliaga}}, \bibinfo {author}
  {\bibfnamefont {O.}~\bibnamefont {Altinok}},  \emph {et~al.} (\bibinfo
  {collaboration} {The MINERvA Collaboration}),\ }\href {\doibase
  10.1103/PhysRevD.94.112007} {\bibfield  {journal} {\bibinfo  {journal} {Phys.
  Rev. D}\ }\textbf {\bibinfo {volume} {94}},\ \bibinfo {pages} {112007}
  (\bibinfo {year} {2016}{\natexlab{a}})}\BibitemShut {NoStop}%
\bibitem [{\citenamefont {D'Agostini}(2010)}]{dagostini2nd}%
  \BibitemOpen
  \bibfield  {author} {\bibinfo {author} {\bibfnamefont {G.}~\bibnamefont
  {D'Agostini}},\ }\href@noop {} {\  (\bibinfo {year} {2010})},\ \Eprint
  {http://arxiv.org/abs/1010.0632} {arXiv:1010.0632 [physics]} \BibitemShut
  {NoStop}%
\bibitem [{\citenamefont {Wilkinson}\ \emph {et~al.}(2016)\citenamefont
  {Wilkinson} \emph {et~al.}}]{Wilkinson:2016wmz}%
  \BibitemOpen
  \bibfield  {author} {\bibinfo {author} {\bibfnamefont {C.}~\bibnamefont
  {Wilkinson}} \emph {et~al.},\ }\href {\doibase 10.1103/PhysRevD.93.072010}
  {\bibfield  {journal} {\bibinfo  {journal} {Phys. Rev.}\ }\textbf {\bibinfo
  {volume} {D93}},\ \bibinfo {pages} {072010} (\bibinfo {year} {2016})},\
  \Eprint {http://arxiv.org/abs/1601.05592} {arXiv:1601.05592 [hep-ex]}
  \BibitemShut {NoStop}%
\bibitem [{\citenamefont {Cousins}\ \emph {et~al.}(2016)\citenamefont {Cousins}
  \emph {et~al.}}]{Cousins:2016ksu}%
  \BibitemOpen
  \bibfield  {author} {\bibinfo {author} {\bibfnamefont {R.~D.}\ \bibnamefont
  {Cousins}} \emph {et~al.},\ }\href@noop {} {\  (\bibinfo {year} {2016})},\
  \Eprint {http://arxiv.org/abs/1607.07038} {arXiv:1607.07038
  [physics.data-an]} \BibitemShut {NoStop}%
\bibitem [{\citenamefont {Llewellyn~Smith}(1972)}]{LlewellynSmithCCQE}%
  \BibitemOpen
  \bibfield  {author} {\bibinfo {author} {\bibfnamefont {C.~H.}\ \bibnamefont
  {Llewellyn~Smith}},\ }\bibfield  {booktitle} {\emph {\bibinfo {booktitle}
  {{Gauge Theories and Neutrino Physics, Jacob, 1978:0175}}},\ }\href {\doibase
  10.1016/0370-1573(72)90010-5} {\bibfield  {journal} {\bibinfo  {journal}
  {Phys. Rept.}\ }\textbf {\bibinfo {volume} {3}},\ \bibinfo {pages} {261}
  (\bibinfo {year} {1972})}\BibitemShut {NoStop}%
\bibitem [{\citenamefont {Bodek}\ \emph {et~al.}(2008)\citenamefont {Bodek},
  \citenamefont {Avvakumov}, \citenamefont {Bradford},\ and\ \citenamefont
  {Budd}}]{BBBAaxial}%
  \BibitemOpen
  \bibfield  {author} {\bibinfo {author} {\bibfnamefont {A.}~\bibnamefont
  {Bodek}}, \bibinfo {author} {\bibfnamefont {S.}~\bibnamefont {Avvakumov}},
  \bibinfo {author} {\bibfnamefont {R.}~\bibnamefont {Bradford}}, \ and\
  \bibinfo {author} {\bibfnamefont {H.~S.}\ \bibnamefont {Budd}},\ }\bibfield
  {booktitle} {\emph {\bibinfo {booktitle} {{High energy physics. Proceedings,
  Europhysics Conference, HEP 2007, Manchester, UK, July 19-25, 2007}}},\
  }\href {\doibase 10.1088/1742-6596/110/8/082004} {\bibfield  {journal}
  {\bibinfo  {journal} {J. Phys. Conf. Ser.}\ }\textbf {\bibinfo {volume}
  {110}},\ \bibinfo {pages} {082004} (\bibinfo {year} {2008})},\ \Eprint
  {http://arxiv.org/abs/0709.3538} {arXiv:0709.3538 [hep-ex]} \BibitemShut
  {NoStop}%
\bibitem [{\citenamefont {Barish}\ \emph {et~al.}(1977)\citenamefont {Barish},
  \citenamefont {Campbell}, \citenamefont {Charlton}, \citenamefont {Cho},
  \citenamefont {Derrick}, \citenamefont {Engelmann}, \citenamefont {Hyman},
  \citenamefont {Jankowski}, \citenamefont {Mann}, \citenamefont {Musgrave},
  \citenamefont {Schreiner}, \citenamefont {Schultz}, \citenamefont {Singer},
  \citenamefont {Szczekowski}, \citenamefont {Wangler}, \citenamefont {Yuta},
  \citenamefont {Barnes}, \citenamefont {Carmony}, \citenamefont {Garfinkel},\
  and\ \citenamefont {Radecky}}]{ANL_QE}%
  \BibitemOpen
  \bibfield  {author} {\bibinfo {author} {\bibfnamefont {S.~J.}\ \bibnamefont
  {Barish}}, \bibinfo {author} {\bibfnamefont {J.}~\bibnamefont {Campbell}},
  \bibinfo {author} {\bibfnamefont {G.}~\bibnamefont {Charlton}}, \bibinfo
  {author} {\bibfnamefont {Y.}~\bibnamefont {Cho}},  \emph {et~al.},\ }\href
  {\doibase 10.1103/PhysRevD.16.3103} {\bibfield  {journal} {\bibinfo
  {journal} {Phys. Rev. D}\ }\textbf {\bibinfo {volume} {16}},\ \bibinfo
  {pages} {3103} (\bibinfo {year} {1977})}\BibitemShut {NoStop}%
\bibitem [{\citenamefont {Baker}\ \emph {et~al.}(1981)\citenamefont {Baker},
  \citenamefont {Cnops}, \citenamefont {Connolly}, \citenamefont {Kahn},
  \citenamefont {Kirk}, \citenamefont {Murtagh}, \citenamefont {Palmer},
  \citenamefont {Samios},\ and\ \citenamefont {Tanaka}}]{BNL_QE}%
  \BibitemOpen
  \bibfield  {author} {\bibinfo {author} {\bibfnamefont {N.~J.}\ \bibnamefont
  {Baker}}, \bibinfo {author} {\bibfnamefont {A.~M.}\ \bibnamefont {Cnops}},
  \bibinfo {author} {\bibfnamefont {P.~L.}\ \bibnamefont {Connolly}}, \bibinfo
  {author} {\bibfnamefont {S.~A.}\ \bibnamefont {Kahn}},  \emph {et~al.},\
  }\href {\doibase 10.1103/PhysRevD.23.2499} {\bibfield  {journal} {\bibinfo
  {journal} {Phys. Rev. D}\ }\textbf {\bibinfo {volume} {23}},\ \bibinfo
  {pages} {2499} (\bibinfo {year} {1981})}\BibitemShut {NoStop}%
\bibitem [{\citenamefont {Meyer}\ \emph {et~al.}(2016)\citenamefont {Meyer},
  \citenamefont {Betancourt}, \citenamefont {Gran},\ and\ \citenamefont
  {Hill}}]{Meyer:2016oeg}%
  \BibitemOpen
  \bibfield  {author} {\bibinfo {author} {\bibfnamefont {A.~S.}\ \bibnamefont
  {Meyer}}, \bibinfo {author} {\bibfnamefont {M.}~\bibnamefont {Betancourt}},
  \bibinfo {author} {\bibfnamefont {R.}~\bibnamefont {Gran}}, \ and\ \bibinfo
  {author} {\bibfnamefont {R.~J.}\ \bibnamefont {Hill}},\ }\href {\doibase
  10.1103/PhysRevD.93.113015} {\bibfield  {journal} {\bibinfo  {journal} {Phys.
  Rev.}\ }\textbf {\bibinfo {volume} {D93}},\ \bibinfo {pages} {113015}
  (\bibinfo {year} {2016})},\ \Eprint {http://arxiv.org/abs/1603.03048}
  {arXiv:1603.03048 [hep-ph]} \BibitemShut {NoStop}%
\bibitem [{\citenamefont {Andreopoulos}\ \emph {et~al.}(2010)\citenamefont
  {Andreopoulos} \emph {et~al.}}]{Andreopoulos:2009rq}%
  \BibitemOpen
  \bibfield  {author} {\bibinfo {author} {\bibfnamefont {C.}~\bibnamefont
  {Andreopoulos}} \emph {et~al.},\ }\href {\doibase 10.1016/j.nima.2009.12.009}
  {\bibfield  {journal} {\bibinfo  {journal} {Nucl. Instrum. Meth.}\ }\textbf
  {\bibinfo {volume} {A614}},\ \bibinfo {pages} {87} (\bibinfo {year}
  {2010})},\ \Eprint {http://arxiv.org/abs/0905.2517} {arXiv:0905.2517
  [hep-ph]} \BibitemShut {NoStop}%
\bibitem [{\citenamefont {Hayato}(2009)}]{Hayato:2009zz}%
  \BibitemOpen
  \bibfield  {author} {\bibinfo {author} {\bibfnamefont {Y.}~\bibnamefont
  {Hayato}},\ }\bibfield  {booktitle} {\emph {\bibinfo {booktitle} {{Neutrino
  interactions: From theory to Monte Carlo simulations. Proceedings, 45th
  Karpacz Winter School in Theoretical Physics, Ladek-Zdroj, Poland, February
  2-11, 2009}}},\ }\href@noop {} {\bibfield  {journal} {\bibinfo  {journal}
  {Acta Phys. Polon.}\ }\textbf {\bibinfo {volume} {B40}},\ \bibinfo {pages}
  {2477} (\bibinfo {year} {2009})}\BibitemShut {NoStop}%
\bibitem [{\citenamefont {Casper}(2002)}]{nuanceMC}%
  \BibitemOpen
  \bibfield  {author} {\bibinfo {author} {\bibfnamefont {D.}~\bibnamefont
  {Casper}},\ }\href {\doibase 10.1016/S0920-5632(02)01756-5} {\bibfield
  {journal} {\bibinfo  {journal} {Nucl. Phys. Proc. Suppl.}\ }\textbf {\bibinfo
  {volume} {112}},\ \bibinfo {pages} {161} (\bibinfo {year} {2002})},\ \Eprint
  {http://arxiv.org/abs/hep-ph/0208030} {arXiv:hep-ph/0208030 [hep-ph]}
  \BibitemShut {NoStop}%
\bibitem [{\citenamefont {Smith}\ and\ \citenamefont
  {Moniz}(1972)}]{Smith-Moniz}%
  \BibitemOpen
  \bibfield  {author} {\bibinfo {author} {\bibfnamefont {R.~A.}\ \bibnamefont
  {Smith}}\ and\ \bibinfo {author} {\bibfnamefont {E.~J.}\ \bibnamefont
  {Moniz}},\ }\href {\doibase 10.1016/0550-3213(75)90612-4,
  10.1016/0550-3213(72)90040-5} {\bibfield  {journal} {\bibinfo  {journal}
  {Nucl. Phys.}\ }\textbf {\bibinfo {volume} {B43}},\ \bibinfo {pages} {605}
  (\bibinfo {year} {1972})},\ \bibinfo {note} {[Erratum: Nucl.
  Phys.B101,547(1975)]}\BibitemShut {NoStop}%
\bibitem [{\citenamefont {Moniz}\ \emph {et~al.}(1971)\citenamefont {Moniz},
  \citenamefont {Sick}, \citenamefont {Whitney}, \citenamefont {Ficenec},
  \citenamefont {Kephart},\ and\ \citenamefont {Trower}}]{moniz_eA_RFG_1971}%
  \BibitemOpen
  \bibfield  {author} {\bibinfo {author} {\bibfnamefont {E.~J.}\ \bibnamefont
  {Moniz}}, \bibinfo {author} {\bibfnamefont {I.}~\bibnamefont {Sick}},
  \bibinfo {author} {\bibfnamefont {R.~R.}\ \bibnamefont {Whitney}}, \bibinfo
  {author} {\bibfnamefont {J.~R.}\ \bibnamefont {Ficenec}}, \bibinfo {author}
  {\bibfnamefont {R.~D.}\ \bibnamefont {Kephart}}, \ and\ \bibinfo {author}
  {\bibfnamefont {W.~P.}\ \bibnamefont {Trower}},\ }\href {\doibase
  10.1103/PhysRevLett.26.445} {\bibfield  {journal} {\bibinfo  {journal} {Phys.
  Rev. Lett.}\ }\textbf {\bibinfo {volume} {26}},\ \bibinfo {pages} {445}
  (\bibinfo {year} {1971})}\BibitemShut {NoStop}%
\bibitem [{\citenamefont {Gran}\ \emph {et~al.}(2006)\citenamefont {Gran},
  \citenamefont {Jeon}, \citenamefont {Aliu}, \citenamefont {Andringa},
  \citenamefont {Aoki}, \citenamefont {Argyriades}, \citenamefont {Asakura},
  \citenamefont {Ashie}, \citenamefont {Berghaus}, \citenamefont {Berns},
  \citenamefont {Bhang}, \citenamefont {Blondel}, \citenamefont {Borghi},
  \citenamefont {Bouchez}, \citenamefont {Burguet-Castell}, \citenamefont
  {Casper}, \citenamefont {Catala}, \citenamefont {Cavata}, \citenamefont
  {Cervera}, \citenamefont {Chen}, \citenamefont {Cho}, \citenamefont {Choi},
  \citenamefont {Dore}, \citenamefont {Espinal}, \citenamefont {Fechner},
  \citenamefont {Fernandez}, \citenamefont {Fukuda}, \citenamefont
  {Gomez-Cadenas}, \citenamefont {Hara}, \citenamefont {Hasegawa},
  \citenamefont {Hasegawa}, \citenamefont {Hayashi}, \citenamefont {Hayato},
  \citenamefont {Helmer}, \citenamefont {Hiraide}, \citenamefont {Hosaka},
  \citenamefont {Ichikawa}, \citenamefont {Iinuma}, \citenamefont {Ikeda},
  \citenamefont {Inagaki}, \citenamefont {Ishida}, \citenamefont {Ishihara},
  \citenamefont {Ishii}, \citenamefont {Ishitsuka}, \citenamefont {Itow},
  \citenamefont {Iwashita}, \citenamefont {Jang}, \citenamefont {Jeong},
  \citenamefont {Joo}, \citenamefont {Jover}, \citenamefont {Jung},
  \citenamefont {Kajita}, \citenamefont {Kameda}, \citenamefont {Kaneyuki},
  \citenamefont {Kato}, \citenamefont {Kearns}, \citenamefont {Kerr},
  \citenamefont {Kim}, \citenamefont {Khabibullin}, \citenamefont
  {Khotjantsev}, \citenamefont {Kielczewska}, \citenamefont {Kim},
  \citenamefont {Kim}, \citenamefont {Kitching}, \citenamefont {Kobayashi},
  \citenamefont {Kobayashi}, \citenamefont {Konaka}, \citenamefont {Koshio},
  \citenamefont {Kropp}, \citenamefont {Kubota}, \citenamefont {Kudenko},
  \citenamefont {Kuno}, \citenamefont {Kurimoto}, \citenamefont {Kutter},
  \citenamefont {Learned}, \citenamefont {Likhoded}, \citenamefont {Lim},
  \citenamefont {Loverre}, \citenamefont {Ludovici}, \citenamefont {Maesaka},
  \citenamefont {Mallet}, \citenamefont {Mariani}, \citenamefont {Matsuno},
  \citenamefont {Matveev}, \citenamefont {McConnel~Mahn}, \citenamefont
  {McGrew}, \citenamefont {Mikheyev}, \citenamefont {Minamino}, \citenamefont
  {Mine}, \citenamefont {Mineev}, \citenamefont {Mitsuda}, \citenamefont
  {Miura}, \citenamefont {Moriguchi}, \citenamefont {Morita}, \citenamefont
  {Moriyama}, \citenamefont {Nakadaira}, \citenamefont {Nakahata},
  \citenamefont {Nakamura}, \citenamefont {Nakano}, \citenamefont {Nakaya},
  \citenamefont {Nakayama}, \citenamefont {Namba}, \citenamefont {Nambu},
  \citenamefont {Nawang}, \citenamefont {Nishikawa}, \citenamefont {Nitta},
  \citenamefont {Nova}, \citenamefont {Novella}, \citenamefont {Obayashi},
  \citenamefont {Okada}, \citenamefont {Okumura}, \citenamefont {Oser},
  \citenamefont {Oyama}, \citenamefont {Pac}, \citenamefont {Pierre},
  \citenamefont {Rodriguez}, \citenamefont {Saji}, \citenamefont {Sakuda},
  \citenamefont {Sanchez}, \citenamefont {Sarrat}, \citenamefont {Sasaki},
  \citenamefont {Sato}, \citenamefont {Scholberg}, \citenamefont {Schroeter},
  \citenamefont {Sekiguchi}, \citenamefont {Shiozawa}, \citenamefont
  {Shiraishi}, \citenamefont {Sitjes}, \citenamefont {Smy}, \citenamefont
  {Sobel}, \citenamefont {Sorel}, \citenamefont {Stone}, \citenamefont {Sulak},
  \citenamefont {Suzuki}, \citenamefont {Suzuki}, \citenamefont {Takahashi},
  \citenamefont {Takenaga}, \citenamefont {Takeuchi}, \citenamefont {Taki},
  \citenamefont {Takubo}, \citenamefont {Tamura}, \citenamefont {Tanaka},
  \citenamefont {Terri}, \citenamefont {T'Jampens}, \citenamefont
  {Tornero-Lopez}, \citenamefont {Totsuka}, \citenamefont {Ueda}, \citenamefont
  {Vagins}, \citenamefont {Whitehead}, \citenamefont {Walter}, \citenamefont
  {Wang}, \citenamefont {Wilkes}, \citenamefont {Yamada}, \citenamefont
  {Yamamoto}, \citenamefont {Yanagisawa}, \citenamefont {Yershov},
  \citenamefont {Yokoyama}, \citenamefont {Yokoyama}, \citenamefont {Yoo},
  \citenamefont {Yoshida},\ and\ \citenamefont {Zalipska}}]{K2K_QE}%
  \BibitemOpen
  \bibfield  {author} {\bibinfo {author} {\bibfnamefont {R.}~\bibnamefont
  {Gran}}, \bibinfo {author} {\bibfnamefont {E.~J.}\ \bibnamefont {Jeon}},
  \bibinfo {author} {\bibfnamefont {E.}~\bibnamefont {Aliu}}, \bibinfo {author}
  {\bibfnamefont {S.}~\bibnamefont {Andringa}},  \emph {et~al.} (\bibinfo
  {collaboration} {K2K Collaboration}),\ }\href {\doibase
  10.1103/PhysRevD.74.052002} {\bibfield  {journal} {\bibinfo  {journal} {Phys.
  Rev. D}\ }\textbf {\bibinfo {volume} {74}},\ \bibinfo {pages} {052002}
  (\bibinfo {year} {2006})}\BibitemShut {NoStop}%
\bibitem [{\citenamefont {Aguilar-Arevalo}\ \emph {et~al.}(2008)\citenamefont
  {Aguilar-Arevalo}, \citenamefont {Bazarko}, \citenamefont {Brice},
  \citenamefont {Brown}, \citenamefont {Bugel}, \citenamefont {Cao},
  \citenamefont {Coney}, \citenamefont {Conrad}, \citenamefont {Cox},
  \citenamefont {Curioni}, \citenamefont {Djurcic}, \citenamefont {Finley},
  \citenamefont {Fleming}, \citenamefont {Ford}, \citenamefont {Garcia},
  \citenamefont {Garvey}, \citenamefont {Green}, \citenamefont {Green},
  \citenamefont {Hart}, \citenamefont {Hawker}, \citenamefont {Imlay},
  \citenamefont {Johnson}, \citenamefont {Kasper}, \citenamefont {Katori},
  \citenamefont {Kobilarcik}, \citenamefont {Kourbanis}, \citenamefont
  {Koutsoliotas}, \citenamefont {Laird}, \citenamefont {Link}, \citenamefont
  {Liu}, \citenamefont {Liu}, \citenamefont {Louis}, \citenamefont {Mahn},
  \citenamefont {Marsh}, \citenamefont {Martin}, \citenamefont {McGregor},
  \citenamefont {Metcalf}, \citenamefont {Meyers}, \citenamefont {Mills},
  \citenamefont {Mills}, \citenamefont {Monroe}, \citenamefont {Moore},
  \citenamefont {Nelson}, \citenamefont {Nienaber}, \citenamefont {Ouedraogo},
  \citenamefont {Patterson}, \citenamefont {Perevalov}, \citenamefont {Polly},
  \citenamefont {Prebys}, \citenamefont {Raaf}, \citenamefont {Ray},
  \citenamefont {Roe}, \citenamefont {Russell}, \citenamefont {Sandberg},
  \citenamefont {Schirato}, \citenamefont {Schmitz}, \citenamefont {Shaevitz},
  \citenamefont {Shoemaker}, \citenamefont {Smith}, \citenamefont {Sorel},
  \citenamefont {Spentzouris}, \citenamefont {Stancu}, \citenamefont
  {Stefanski}, \citenamefont {Sung}, \citenamefont {Tanaka}, \citenamefont
  {Tayloe}, \citenamefont {Tzanov}, \citenamefont {Van~de Water}, \citenamefont
  {Wascko}, \citenamefont {White}, \citenamefont {Wilking}, \citenamefont
  {Yang}, \citenamefont {Zeller},\ and\ \citenamefont
  {Zimmerman}}]{MiniBoone_MA}%
  \BibitemOpen
  \bibfield  {author} {\bibinfo {author} {\bibfnamefont {A.~A.}\ \bibnamefont
  {Aguilar-Arevalo}}, \bibinfo {author} {\bibfnamefont {A.~O.}\ \bibnamefont
  {Bazarko}}, \bibinfo {author} {\bibfnamefont {S.~J.}\ \bibnamefont {Brice}},
  \bibinfo {author} {\bibfnamefont {B.~C.}\ \bibnamefont {Brown}},  \emph
  {et~al.} (\bibinfo {collaboration} {MiniBooNE Collaboration}),\ }\href
  {\doibase 10.1103/PhysRevLett.100.032301} {\bibfield  {journal} {\bibinfo
  {journal} {Phys. Rev. Lett.}\ }\textbf {\bibinfo {volume} {100}},\ \bibinfo
  {pages} {032301} (\bibinfo {year} {2008})}\BibitemShut {NoStop}%
\bibitem [{\citenamefont {Adamson}\ \emph {et~al.}(2015)\citenamefont
  {Adamson}, \citenamefont {Anghel}, \citenamefont {Aurisano}, \citenamefont
  {Barr}, \citenamefont {Bishai}, \citenamefont {Blake}, \citenamefont {Bock},
  \citenamefont {Bogert}, \citenamefont {Cao}, \citenamefont {Castromonte},
  \citenamefont {Childress}, \citenamefont {Coelho}, \citenamefont {Corwin},
  \citenamefont {Cronin-Hennessy}, \citenamefont {de~Jong}, \citenamefont
  {Devan}, \citenamefont {Devenish}, \citenamefont {Diwan}, \citenamefont
  {Escobar}, \citenamefont {Evans}, \citenamefont {Falk}, \citenamefont
  {Feldman}, \citenamefont {Frohne}, \citenamefont {Gallagher}, \citenamefont
  {Gomes}, \citenamefont {Goodman}, \citenamefont {Gouffon}, \citenamefont
  {Graf}, \citenamefont {Gran}, \citenamefont {Grzelak}, \citenamefont {Habig},
  \citenamefont {Hahn}, \citenamefont {Hartnell}, \citenamefont {Hatcher},
  \citenamefont {Holin}, \citenamefont {Huang}, \citenamefont {Hylen},
  \citenamefont {Irwin}, \citenamefont {Isvan}, \citenamefont {James},
  \citenamefont {Jensen}, \citenamefont {Kafka}, \citenamefont {Kasahara},
  \citenamefont {Koizumi}, \citenamefont {Kordosky}, \citenamefont {Kreymer},
  \citenamefont {Lang}, \citenamefont {Ling}, \citenamefont {Litchfield},
  \citenamefont {Lucas}, \citenamefont {Mann}, \citenamefont {Marshak},
  \citenamefont {Mayer}, \citenamefont {McGivern}, \citenamefont {Medeiros},
  \citenamefont {Mehdiyev}, \citenamefont {Meier}, \citenamefont {Messier},
  \citenamefont {Miller}, \citenamefont {Mishra}, \citenamefont {Moed~Sher},
  \citenamefont {Moore}, \citenamefont {Mualem}, \citenamefont {Musser},
  \citenamefont {Naples}, \citenamefont {Nelson}, \citenamefont {Newman},
  \citenamefont {Nichol}, \citenamefont {Nowak}, \citenamefont {O'Connor},
  \citenamefont {Orchanian}, \citenamefont {Pahlka}, \citenamefont {Paley},
  \citenamefont {Patterson}, \citenamefont {Pawloski}, \citenamefont {Perch},
  \citenamefont {Pf\"utzner}, \citenamefont {Phan-Budd}, \citenamefont
  {Plunkett}, \citenamefont {Poonthottathil}, \citenamefont {Qiu},
  \citenamefont {Radovic}, \citenamefont {Rebel}, \citenamefont {Rosenfeld},
  \citenamefont {Rubin}, \citenamefont {Sanchez}, \citenamefont {Schneps},
  \citenamefont {Schreckenberger}, \citenamefont {Schreiner}, \citenamefont
  {Sharma}, \citenamefont {Sousa}, \citenamefont {Tagg}, \citenamefont
  {Talaga}, \citenamefont {Thomas}, \citenamefont {Thomson}, \citenamefont
  {Tian}, \citenamefont {Timmons}, \citenamefont {Tognini}, \citenamefont
  {Toner}, \citenamefont {Torretta}, \citenamefont {Urheim}, \citenamefont
  {Vahle}, \citenamefont {Viren}, \citenamefont {Walding}, \citenamefont
  {Weber}, \citenamefont {Webb}, \citenamefont {White}, \citenamefont
  {Whitehead}, \citenamefont {Whitehead}, \citenamefont {Wojcicki},\ and\
  \citenamefont {Zwaska}}]{MINOS_MA}%
  \BibitemOpen
  \bibfield  {author} {\bibinfo {author} {\bibfnamefont {P.}~\bibnamefont
  {Adamson}}, \bibinfo {author} {\bibfnamefont {I.}~\bibnamefont {Anghel}},
  \bibinfo {author} {\bibfnamefont {A.}~\bibnamefont {Aurisano}}, \bibinfo
  {author} {\bibfnamefont {G.}~\bibnamefont {Barr}},  \emph {et~al.} (\bibinfo
  {collaboration} {MINOS Collaboration}),\ }\href {\doibase
  10.1103/PhysRevD.91.012005} {\bibfield  {journal} {\bibinfo  {journal} {Phys.
  Rev. D}\ }\textbf {\bibinfo {volume} {91}},\ \bibinfo {pages} {012005}
  (\bibinfo {year} {2015})}\BibitemShut {NoStop}%
\bibitem [{\citenamefont {Abe}\ \emph {et~al.}(2014{\natexlab{b}})\citenamefont
  {Abe} \emph {et~al.}}]{t2kCCQE}%
  \BibitemOpen
  \bibfield  {author} {\bibinfo {author} {\bibfnamefont {K.}~\bibnamefont
  {Abe}} \emph {et~al.} (\bibinfo {collaboration} {T2K Collaboration}),\
  }\href@noop {} {\  (\bibinfo {year} {2014}{\natexlab{b}})},\ \Eprint
  {http://arxiv.org/abs/1411.6264} {arXiv:1411.6264 [hep-ex]} \BibitemShut
  {NoStop}%
\bibitem [{\citenamefont {Stowell}\ \emph {et~al.}(2017)\citenamefont {Stowell}
  \emph {et~al.}}]{Stowell:2016jfr}%
  \BibitemOpen
  \bibfield  {author} {\bibinfo {author} {\bibfnamefont {P.}~\bibnamefont
  {Stowell}} \emph {et~al.},\ }\href {\doibase 10.1088/1748-0221/12/01/P01016}
  {\bibfield  {journal} {\bibinfo  {journal} {JINST}\ }\textbf {\bibinfo
  {volume} {12}},\ \bibinfo {pages} {P01016} (\bibinfo {year} {2017})},\
  \Eprint {http://arxiv.org/abs/1612.07393} {arXiv:1612.07393 [hep-ex]}
  \BibitemShut {NoStop}%
\bibitem [{\citenamefont {Aguilar-Arevalo}\ \emph
  {et~al.}(2010{\natexlab{b}})\citenamefont {Aguilar-Arevalo} \emph
  {et~al.}}]{mbCCQE}%
  \BibitemOpen
  \bibfield  {author} {\bibinfo {author} {\bibfnamefont {A.}~\bibnamefont
  {Aguilar-Arevalo}} \emph {et~al.} (\bibinfo {collaboration} {MiniBooNE
  Collaboration}),\ }\href {\doibase 10.1103/PhysRevD.81.092005} {\bibfield
  {journal} {\bibinfo  {journal} {Phys. Rev.}\ }\textbf {\bibinfo {volume}
  {D81}},\ \bibinfo {pages} {092005} (\bibinfo {year} {2010}{\natexlab{b}})},\
  \Eprint {http://arxiv.org/abs/1002.2680} {arXiv:1002.2680 [hep-ex]}
  \BibitemShut {NoStop}%
\bibitem [{\citenamefont {Aguilar-Arevalo}\ \emph {et~al.}(2013)\citenamefont
  {Aguilar-Arevalo} \emph {et~al.}}]{mbAntiCCQE}%
  \BibitemOpen
  \bibfield  {author} {\bibinfo {author} {\bibfnamefont {A.}~\bibnamefont
  {Aguilar-Arevalo}} \emph {et~al.} (\bibinfo {collaboration} {MiniBooNE
  Collaboration}),\ }\href {\doibase 10.1103/PhysRevD.88.032001} {\bibfield
  {journal} {\bibinfo  {journal} {Phys. Rev.}\ }\textbf {\bibinfo {volume}
  {D88}},\ \bibinfo {pages} {032001} (\bibinfo {year} {2013})},\ \Eprint
  {http://arxiv.org/abs/1301.7067} {arXiv:1301.7067 [hep-ex]} \BibitemShut
  {NoStop}%
\bibitem [{\citenamefont {Fields}\ \emph {et~al.}(2013)\citenamefont {Fields}
  \emph {et~al.}}]{minerva-antinu-ccqe}%
  \BibitemOpen
  \bibfield  {author} {\bibinfo {author} {\bibfnamefont {L.}~\bibnamefont
  {Fields}} \emph {et~al.} (\bibinfo {collaboration} {{MINERvA}}),\ }\href
  {\doibase 10.1103/PhysRevLett.111.022501} {\bibfield  {journal} {\bibinfo
  {journal} {Phys. Rev. Lett.}\ }\textbf {\bibinfo {volume} {111}},\ \bibinfo
  {pages} {022501} (\bibinfo {year} {2013})},\ \Eprint
  {http://arxiv.org/abs/1305.2234} {arXiv:1305.2234 [hep-ex]} \BibitemShut
  {NoStop}%
\bibitem [{\citenamefont {Fiorentini}\ \emph {et~al.}(2013)\citenamefont
  {Fiorentini} \emph {et~al.}}]{Fiorentini:2013ezn}%
  \BibitemOpen
  \bibfield  {author} {\bibinfo {author} {\bibfnamefont {G.~A.}\ \bibnamefont
  {Fiorentini}} \emph {et~al.} (\bibinfo {collaboration} {MINERvA}),\ }\href
  {\doibase 10.1103/PhysRevLett.111.022502} {\bibfield  {journal} {\bibinfo
  {journal} {Phys. Rev. Lett.}\ }\textbf {\bibinfo {volume} {111}},\ \bibinfo
  {pages} {022502} (\bibinfo {year} {2013})},\ \Eprint
  {http://arxiv.org/abs/1305.2243} {arXiv:1305.2243 [hep-ex]} \BibitemShut
  {NoStop}%
\bibitem [{\citenamefont {Walton}\ \emph {et~al.}(2015)\citenamefont {Walton}
  \emph {et~al.}}]{Walton:2014esl}%
  \BibitemOpen
  \bibfield  {author} {\bibinfo {author} {\bibfnamefont {T.}~\bibnamefont
  {Walton}} \emph {et~al.} (\bibinfo {collaboration} {MINERvA}),\ }\href
  {\doibase 10.1103/PhysRevD.91.071301} {\bibfield  {journal} {\bibinfo
  {journal} {Phys. Rev.}\ }\textbf {\bibinfo {volume} {D91}},\ \bibinfo {pages}
  {071301} (\bibinfo {year} {2015})},\ \Eprint {http://arxiv.org/abs/1409.4497}
  {arXiv:1409.4497 [hep-ex]} \BibitemShut {NoStop}%
\bibitem [{\citenamefont {Wolcott}\ \emph
  {et~al.}(2016{\natexlab{a}})\citenamefont {Wolcott} \emph
  {et~al.}}]{Wolcott:2015hda}%
  \BibitemOpen
  \bibfield  {author} {\bibinfo {author} {\bibfnamefont {J.}~\bibnamefont
  {Wolcott}} \emph {et~al.} (\bibinfo {collaboration} {MINERvA}),\ }\href
  {\doibase 10.1103/PhysRevLett.116.081802} {\bibfield  {journal} {\bibinfo
  {journal} {Phys. Rev. Lett.}\ }\textbf {\bibinfo {volume} {116}},\ \bibinfo
  {pages} {081802} (\bibinfo {year} {2016}{\natexlab{a}})},\ \Eprint
  {http://arxiv.org/abs/1509.05729} {arXiv:1509.05729 [hep-ex]} \BibitemShut
  {NoStop}%
\bibitem [{\citenamefont {Betancourt}\ \emph {et~al.}(2017)\citenamefont
  {Betancourt} \emph {et~al.}}]{Betancourt:2017uso}%
  \BibitemOpen
  \bibfield  {author} {\bibinfo {author} {\bibfnamefont {M.}~\bibnamefont
  {Betancourt}} \emph {et~al.} (\bibinfo {collaboration} {MINERvA}),\ }\href
  {\doibase 10.1103/PhysRevLett.119.082001} {\bibfield  {journal} {\bibinfo
  {journal} {Phys. Rev. Lett.}\ }\textbf {\bibinfo {volume} {119}},\ \bibinfo
  {pages} {082001} (\bibinfo {year} {2017})},\ \Eprint
  {http://arxiv.org/abs/1705.03791} {arXiv:1705.03791 [hep-ex]} \BibitemShut
  {NoStop}%
\bibitem [{\citenamefont {Day}\ and\ \citenamefont
  {McFarland}(2012)}]{Day:2012gb}%
  \BibitemOpen
  \bibfield  {author} {\bibinfo {author} {\bibfnamefont {M.}~\bibnamefont
  {Day}}\ and\ \bibinfo {author} {\bibfnamefont {K.~S.}\ \bibnamefont
  {McFarland}},\ }\href {\doibase 10.1103/PhysRevD.86.053003} {\bibfield
  {journal} {\bibinfo  {journal} {Phys. Rev.}\ }\textbf {\bibinfo {volume}
  {D86}},\ \bibinfo {pages} {053003} (\bibinfo {year} {2012})},\ \Eprint
  {http://arxiv.org/abs/1206.6745} {arXiv:1206.6745 [hep-ph]} \BibitemShut
  {NoStop}%
\bibitem [{\citenamefont {Abe}\ \emph {et~al.}(2016{\natexlab{a}})\citenamefont
  {Abe} \emph {et~al.}}]{Abe:2016tmq}%
  \BibitemOpen
  \bibfield  {author} {\bibinfo {author} {\bibfnamefont {K.}~\bibnamefont
  {Abe}} \emph {et~al.} (\bibinfo {collaboration} {T2K}),\ }\href@noop {}
  {\bibfield  {journal} {\bibinfo  {journal} {Phys. Rev.}\ }\textbf {\bibinfo
  {volume} {D93}},\ \bibinfo {pages} {112012} (\bibinfo {year}
  {2016}{\natexlab{a}})},\ \Eprint {http://arxiv.org/abs/1602.03652}
  {arXiv:1602.03652 [hep-ex]} \BibitemShut {NoStop}%
\bibitem [{\citenamefont {Abe}\ \emph {et~al.}(2018)\citenamefont {Abe} \emph
  {et~al.}}]{Abe:2017rfw}%
  \BibitemOpen
  \bibfield  {author} {\bibinfo {author} {\bibfnamefont {K.}~\bibnamefont
  {Abe}} \emph {et~al.} (\bibinfo {collaboration} {T2K}),\ }\href {\doibase
  10.1103/PhysRevD.97.012001} {\bibfield  {journal} {\bibinfo  {journal} {Phys.
  Rev.}\ }\textbf {\bibinfo {volume} {D97}},\ \bibinfo {pages} {012001}
  (\bibinfo {year} {2018})},\ \Eprint {http://arxiv.org/abs/1708.06771}
  {arXiv:1708.06771 [hep-ex]} \BibitemShut {NoStop}%
\bibitem [{\citenamefont {Patrick}\ \emph {et~al.}(2018)\citenamefont {Patrick}
  \emph {et~al.}}]{minerva_0pi_2d}%
  \BibitemOpen
  \bibfield  {author} {\bibinfo {author} {\bibfnamefont {C.~E.}\ \bibnamefont
  {Patrick}} \emph {et~al.} (\bibinfo {collaboration} {MINERvA}),\ }\href@noop
  {} {\  (\bibinfo {year} {2018})},\ \Eprint {http://arxiv.org/abs/1801.01197}
  {arXiv:1801.01197 [hep-ex]} \BibitemShut {NoStop}%
\bibitem [{\citenamefont {Juszczak}(2009)}]{Juszczak:2009qa}%
  \BibitemOpen
  \bibfield  {author} {\bibinfo {author} {\bibfnamefont {C.}~\bibnamefont
  {Juszczak}},\ }\bibfield  {booktitle} {\emph {\bibinfo {booktitle} {{Neutrino
  interactions: From theory to Monte Carlo simulations. Proceedings, 45th
  Karpacz Winter School in Theoretical Physics, Ladek-Zdroj, Poland, February
  2-11, 2009}}},\ }\href@noop {} {\bibfield  {journal} {\bibinfo  {journal}
  {Acta Phys. Polon.}\ }\textbf {\bibinfo {volume} {B40}},\ \bibinfo {pages}
  {2507} (\bibinfo {year} {2009})},\ \Eprint {http://arxiv.org/abs/0909.1492}
  {arXiv:0909.1492 [hep-ex]} \BibitemShut {NoStop}%
\bibitem [{\citenamefont {Rodrigues}\ \emph
  {et~al.}(2016{\natexlab{a}})\citenamefont {Rodrigues} \emph
  {et~al.}}]{Rodrigues:2015hik}%
  \BibitemOpen
  \bibfield  {author} {\bibinfo {author} {\bibfnamefont {P.~A.}\ \bibnamefont
  {Rodrigues}} \emph {et~al.} (\bibinfo {collaboration} {MINERvA}),\ }\href
  {\doibase 10.1103/PhysRevLett.116.071802} {\bibfield  {journal} {\bibinfo
  {journal} {Phys. Rev. Lett.}\ }\textbf {\bibinfo {volume} {116}},\ \bibinfo
  {pages} {071802} (\bibinfo {year} {2016}{\natexlab{a}})},\ \Eprint
  {http://arxiv.org/abs/1511.05944} {arXiv:1511.05944 [hep-ex]} \BibitemShut
  {NoStop}%
\bibitem [{\citenamefont {Aguilar-Arevalo}\ \emph {et~al.}(2018)\citenamefont
  {Aguilar-Arevalo} \emph {et~al.}}]{Aguilar-Arevalo:2018ylq}%
  \BibitemOpen
  \bibfield  {author} {\bibinfo {author} {\bibfnamefont {A.~A.}\ \bibnamefont
  {Aguilar-Arevalo}} \emph {et~al.} (\bibinfo {collaboration} {MiniBooNE}),\
  }\href@noop {} {\  (\bibinfo {year} {2018})},\ \Eprint
  {http://arxiv.org/abs/1801.03848} {arXiv:1801.03848 [hep-ex]} \BibitemShut
  {NoStop}%
\bibitem [{\citenamefont {Ahrens}\ \emph {et~al.}(1987)\citenamefont {Ahrens},
  \citenamefont {Aronson}, \citenamefont {Connolly}, \citenamefont {Gibbard},
  \citenamefont {Murtagh}, \citenamefont {Murtagh}, \citenamefont {Terada},
  \citenamefont {White}, \citenamefont {Callas}, \citenamefont {Cutts},
  \citenamefont {Hoftun}, \citenamefont {Diwan}, \citenamefont {Lanou},
  \citenamefont {Shinkawa}, \citenamefont {Kurihara}, \citenamefont {Amako},
  \citenamefont {Kabe}, \citenamefont {Nagashima}, \citenamefont {Suzuki},
  \citenamefont {Tatsumi}, \citenamefont {Yamaguchi}, \citenamefont {Abe},
  \citenamefont {Beier}, \citenamefont {Doughty}, \citenamefont {Durkin},
  \citenamefont {Heagy}, \citenamefont {Hurley}, \citenamefont {Mann},
  \citenamefont {Newcomer}, \citenamefont {Williams}, \citenamefont {York},
  \citenamefont {Hedin}, \citenamefont {Marx},\ and\ \citenamefont
  {Stern}}]{BNL_NCE}%
  \BibitemOpen
  \bibfield  {author} {\bibinfo {author} {\bibfnamefont {L.~A.}\ \bibnamefont
  {Ahrens}}, \bibinfo {author} {\bibfnamefont {S.~H.}\ \bibnamefont {Aronson}},
  \bibinfo {author} {\bibfnamefont {P.~L.}\ \bibnamefont {Connolly}}, \bibinfo
  {author} {\bibfnamefont {B.~G.}\ \bibnamefont {Gibbard}},  \emph {et~al.},\
  }\href {\doibase 10.1103/PhysRevD.35.785} {\bibfield  {journal} {\bibinfo
  {journal} {Phys. Rev. D}\ }\textbf {\bibinfo {volume} {35}},\ \bibinfo
  {pages} {785} (\bibinfo {year} {1987})}\BibitemShut {NoStop}%
\bibitem [{\citenamefont {Aguilar-Arevalo}\ \emph
  {et~al.}(2010{\natexlab{c}})\citenamefont {Aguilar-Arevalo}, \citenamefont
  {Anderson}, \citenamefont {Bazarko}, \citenamefont {Brice}, \citenamefont
  {Brown}, \citenamefont {Bugel}, \citenamefont {Cao}, \citenamefont {Coney},
  \citenamefont {Conrad}, \citenamefont {Cox}, \citenamefont {Curioni},
  \citenamefont {Dharmapalan}, \citenamefont {Djurcic}, \citenamefont {Finley},
  \citenamefont {Fleming}, \citenamefont {Ford}, \citenamefont {Garcia},
  \citenamefont {Garvey}, \citenamefont {Grange}, \citenamefont {Green},
  \citenamefont {Green}, \citenamefont {Hart}, \citenamefont {Hawker},
  \citenamefont {Imlay}, \citenamefont {Johnson}, \citenamefont {Karagiorgi},
  \citenamefont {Kasper}, \citenamefont {Katori}, \citenamefont {Kobilarcik},
  \citenamefont {Kourbanis}, \citenamefont {Koutsoliotas}, \citenamefont
  {Laird}, \citenamefont {Linden}, \citenamefont {Link}, \citenamefont {Liu},
  \citenamefont {Liu}, \citenamefont {Louis}, \citenamefont {Mahn},
  \citenamefont {Marsh}, \citenamefont {Mauger}, \citenamefont {McGary},
  \citenamefont {McGregor}, \citenamefont {Metcalf}, \citenamefont {Meyers},
  \citenamefont {Mills}, \citenamefont {Mills}, \citenamefont {Monroe},
  \citenamefont {Moore}, \citenamefont {Mousseau}, \citenamefont {Nelson},
  \citenamefont {Nienaber}, \citenamefont {Nowak}, \citenamefont {Osmanov},
  \citenamefont {Ouedraogo}, \citenamefont {Patterson}, \citenamefont
  {Pavlovic}, \citenamefont {Perevalov}, \citenamefont {Polly}, \citenamefont
  {Prebys}, \citenamefont {Raaf}, \citenamefont {Ray}, \citenamefont {Roe},
  \citenamefont {Russell}, \citenamefont {Sandberg}, \citenamefont {Schirato},
  \citenamefont {Schmitz}, \citenamefont {Shaevitz}, \citenamefont {Shoemaker},
  \citenamefont {Smith}, \citenamefont {Soderberg}, \citenamefont {Sorel},
  \citenamefont {Spentzouris}, \citenamefont {Spitz}, \citenamefont {Stancu},
  \citenamefont {Stefanski}, \citenamefont {Sung}, \citenamefont {Tanaka},
  \citenamefont {Tayloe}, \citenamefont {Tzanov}, \citenamefont {Van~de Water},
  \citenamefont {Wascko}, \citenamefont {White}, \citenamefont {Wilking},
  \citenamefont {Yang}, \citenamefont {Zeller},\ and\ \citenamefont
  {Zimmerman}}]{miniboone_NCE}%
  \BibitemOpen
  \bibfield  {author} {\bibinfo {author} {\bibfnamefont {A.~A.}\ \bibnamefont
  {Aguilar-Arevalo}}, \bibinfo {author} {\bibfnamefont {C.~E.}\ \bibnamefont
  {Anderson}}, \bibinfo {author} {\bibfnamefont {A.~O.}\ \bibnamefont
  {Bazarko}}, \bibinfo {author} {\bibfnamefont {S.~J.}\ \bibnamefont {Brice}},
  \emph {et~al.} (\bibinfo {collaboration} {MiniBooNE Collaboration}),\ }\href
  {\doibase 10.1103/PhysRevD.82.092005} {\bibfield  {journal} {\bibinfo
  {journal} {Phys. Rev. D}\ }\textbf {\bibinfo {volume} {82}},\ \bibinfo
  {pages} {092005} (\bibinfo {year} {2010}{\natexlab{c}})}\BibitemShut
  {NoStop}%
\bibitem [{\citenamefont {Aguilar-Arevalo}\ \emph {et~al.}(2015)\citenamefont
  {Aguilar-Arevalo} \emph {et~al.}}]{mbAntiNCEL}%
  \BibitemOpen
  \bibfield  {author} {\bibinfo {author} {\bibfnamefont {A.}~\bibnamefont
  {Aguilar-Arevalo}} \emph {et~al.} (\bibinfo {collaboration} {MiniBooNE
  Collaboration}),\ }\href {\doibase 10.1103/PhysRevD.91.012004} {\bibfield
  {journal} {\bibinfo  {journal} {Phys. Rev.}\ }\textbf {\bibinfo {volume}
  {D91}},\ \bibinfo {pages} {012004} (\bibinfo {year} {2015})},\ \Eprint
  {http://arxiv.org/abs/1309.7257} {arXiv:1309.7257 [hep-ex]} \BibitemShut
  {NoStop}%
\bibitem [{\citenamefont {Barish}\ \emph {et~al.}(1979)\citenamefont {Barish},
  \citenamefont {Derrick}, \citenamefont {Dombeck}, \citenamefont {Hyman},
  \citenamefont {Jaeger} \emph {et~al.}}]{ANL_Barish_1979}%
  \BibitemOpen
  \bibfield  {author} {\bibinfo {author} {\bibfnamefont {S.}~\bibnamefont
  {Barish}}, \bibinfo {author} {\bibfnamefont {M.}~\bibnamefont {Derrick}},
  \bibinfo {author} {\bibfnamefont {T.}~\bibnamefont {Dombeck}}, \bibinfo
  {author} {\bibfnamefont {L.}~\bibnamefont {Hyman}}, \bibinfo {author}
  {\bibfnamefont {K.}~\bibnamefont {Jaeger}},  \emph {et~al.},\ }\href
  {\doibase 10.1103/PhysRevD.19.2521} {\bibfield  {journal} {\bibinfo
  {journal} {Phys. Rev.}\ }\textbf {\bibinfo {volume} {D19}},\ \bibinfo {pages}
  {2521} (\bibinfo {year} {1979})}\BibitemShut {NoStop}%
\bibitem [{\citenamefont {Radecky}\ \emph {et~al.}(1982)\citenamefont
  {Radecky}, \citenamefont {Barnes}, \citenamefont {Carmony}, \citenamefont
  {Garfinkel}, \citenamefont {Derrick} \emph {et~al.}}]{ANL_Radecky_1982}%
  \BibitemOpen
  \bibfield  {author} {\bibinfo {author} {\bibfnamefont {G.}~\bibnamefont
  {Radecky}}, \bibinfo {author} {\bibfnamefont {V.}~\bibnamefont {Barnes}},
  \bibinfo {author} {\bibfnamefont {D.}~\bibnamefont {Carmony}}, \bibinfo
  {author} {\bibfnamefont {A.}~\bibnamefont {Garfinkel}}, \bibinfo {author}
  {\bibfnamefont {M.}~\bibnamefont {Derrick}},  \emph {et~al.},\ }\href
  {\doibase 10.1103/PhysRevD.25.1161, 10.1103/PhysRevD.26.3297} {\bibfield
  {journal} {\bibinfo  {journal} {Phys. Rev.}\ }\textbf {\bibinfo {volume}
  {D25}},\ \bibinfo {pages} {1161} (\bibinfo {year} {1982})}\BibitemShut
  {NoStop}%
\bibitem [{\citenamefont {Kitagaki}\ \emph {et~al.}(1986)\citenamefont
  {Kitagaki}, \citenamefont {Yuta}, \citenamefont {Tanaka}, \citenamefont
  {Yamaguchi}, \citenamefont {Abe} \emph {et~al.}}]{BNL_Kitagaki_1986}%
  \BibitemOpen
  \bibfield  {author} {\bibinfo {author} {\bibfnamefont {T.}~\bibnamefont
  {Kitagaki}}, \bibinfo {author} {\bibfnamefont {H.}~\bibnamefont {Yuta}},
  \bibinfo {author} {\bibfnamefont {S.}~\bibnamefont {Tanaka}}, \bibinfo
  {author} {\bibfnamefont {A.}~\bibnamefont {Yamaguchi}}, \bibinfo {author}
  {\bibfnamefont {K.}~\bibnamefont {Abe}},  \emph {et~al.},\ }\href {\doibase
  10.1103/PhysRevD.34.2554} {\bibfield  {journal} {\bibinfo  {journal} {Phys.
  Rev.}\ }\textbf {\bibinfo {volume} {D34}},\ \bibinfo {pages} {2554} (\bibinfo
  {year} {1986})}\BibitemShut {NoStop}%
\bibitem [{\citenamefont {Kitagaki}\ \emph {et~al.}(1990)\citenamefont
  {Kitagaki}, \citenamefont {Yuta}, \citenamefont {Tanaka}, \citenamefont
  {Yamaguchi}, \citenamefont {Abe} \emph {et~al.}}]{BNL_Kitagaki_1990}%
  \BibitemOpen
  \bibfield  {author} {\bibinfo {author} {\bibfnamefont {T.}~\bibnamefont
  {Kitagaki}}, \bibinfo {author} {\bibfnamefont {H.}~\bibnamefont {Yuta}},
  \bibinfo {author} {\bibfnamefont {S.}~\bibnamefont {Tanaka}}, \bibinfo
  {author} {\bibfnamefont {A.}~\bibnamefont {Yamaguchi}}, \bibinfo {author}
  {\bibfnamefont {K.}~\bibnamefont {Abe}},  \emph {et~al.},\ }\href {\doibase
  10.1103/PhysRevD.42.1331} {\bibfield  {journal} {\bibinfo  {journal} {Phys.
  Rev.}\ }\textbf {\bibinfo {volume} {D42}},\ \bibinfo {pages} {1331} (\bibinfo
  {year} {1990})}\BibitemShut {NoStop}%
\bibitem [{\citenamefont {Wilkinson}\ \emph {et~al.}(2014)\citenamefont
  {Wilkinson}, \citenamefont {Rodrigues}, \citenamefont {Cartwright},
  \citenamefont {Thompson},\ and\ \citenamefont
  {McFarland}}]{anl_bnl_reanalysis}%
  \BibitemOpen
  \bibfield  {author} {\bibinfo {author} {\bibfnamefont {C.}~\bibnamefont
  {Wilkinson}}, \bibinfo {author} {\bibfnamefont {P.}~\bibnamefont
  {Rodrigues}}, \bibinfo {author} {\bibfnamefont {S.}~\bibnamefont
  {Cartwright}}, \bibinfo {author} {\bibfnamefont {L.}~\bibnamefont
  {Thompson}}, \ and\ \bibinfo {author} {\bibfnamefont {K.}~\bibnamefont
  {McFarland}},\ }\href {\doibase 10.1103/PhysRevD.90.112017} {\bibfield
  {journal} {\bibinfo  {journal} {Phys. Rev.}\ }\textbf {\bibinfo {volume}
  {D90}},\ \bibinfo {pages} {112017} (\bibinfo {year} {2014})},\ \Eprint
  {http://arxiv.org/abs/1411.4482} {arXiv:1411.4482 [hep-ex]} \BibitemShut
  {NoStop}%
\bibitem [{\citenamefont {Rodrigues}\ \emph
  {et~al.}(2016{\natexlab{b}})\citenamefont {Rodrigues}, \citenamefont
  {Wilkinson},\ and\ \citenamefont {McFarland}}]{Rodrigues:2016xjj}%
  \BibitemOpen
  \bibfield  {author} {\bibinfo {author} {\bibfnamefont {P.}~\bibnamefont
  {Rodrigues}}, \bibinfo {author} {\bibfnamefont {C.}~\bibnamefont
  {Wilkinson}}, \ and\ \bibinfo {author} {\bibfnamefont {K.}~\bibnamefont
  {McFarland}},\ }\href {\doibase 10.1140/epjc/s10052-016-4314-3} {\bibfield
  {journal} {\bibinfo  {journal} {Eur. Phys. J.}\ }\textbf {\bibinfo {volume}
  {C76}},\ \bibinfo {pages} {474} (\bibinfo {year} {2016}{\natexlab{b}})},\
  \Eprint {http://arxiv.org/abs/1601.01888} {arXiv:1601.01888 [hep-ex]}
  \BibitemShut {NoStop}%
\bibitem [{\citenamefont {Wu}\ \emph {et~al.}(2015)\citenamefont {Wu},
  \citenamefont {Sato},\ and\ \citenamefont {Lee}}]{sato_2014}%
  \BibitemOpen
  \bibfield  {author} {\bibinfo {author} {\bibfnamefont {J.-J.}\ \bibnamefont
  {Wu}}, \bibinfo {author} {\bibfnamefont {T.}~\bibnamefont {Sato}}, \ and\
  \bibinfo {author} {\bibfnamefont {T.~S.~H.}\ \bibnamefont {Lee}},\ }\href
  {\doibase 10.1103/PhysRevC.91.035203} {\bibfield  {journal} {\bibinfo
  {journal} {Phys. Rev.}\ }\textbf {\bibinfo {volume} {C91}},\ \bibinfo {pages}
  {035203} (\bibinfo {year} {2015})},\ \Eprint {http://arxiv.org/abs/1412.2415}
  {arXiv:1412.2415 [nucl-th]} \BibitemShut {NoStop}%
\bibitem [{\citenamefont {Salcedo}\ \emph {et~al.}(1988)\citenamefont {Salcedo}
  \emph {et~al.}}]{Salcedo-Oset}%
  \BibitemOpen
  \bibfield  {author} {\bibinfo {author} {\bibfnamefont {L.}~\bibnamefont
  {Salcedo}} \emph {et~al.},\ }\href@noop {} {\bibfield  {journal} {\bibinfo
  {journal} {Nucl. Phys.}\ }\textbf {\bibinfo {volume} {A484}},\ \bibinfo
  {pages} {557} (\bibinfo {year} {1988})}\BibitemShut {NoStop}%
\bibitem [{\citenamefont {Ieki}\ \emph {et~al.}(2015)\citenamefont {Ieki} \emph
  {et~al.}}]{Ieki:2015okz}%
  \BibitemOpen
  \bibfield  {author} {\bibinfo {author} {\bibfnamefont {K.}~\bibnamefont
  {Ieki}} \emph {et~al.} (\bibinfo {collaboration} {DUET}),\ }\href {\doibase
  10.1103/PhysRevC.92.035205} {\bibfield  {journal} {\bibinfo  {journal} {Phys.
  Rev.}\ }\textbf {\bibinfo {volume} {C92}},\ \bibinfo {pages} {035205}
  (\bibinfo {year} {2015})},\ \Eprint {http://arxiv.org/abs/1506.07783}
  {arXiv:1506.07783 [hep-ex]} \BibitemShut {NoStop}%
\bibitem [{\citenamefont {Pinzon~Guerra}\ \emph {et~al.}(2017)\citenamefont
  {Pinzon~Guerra} \emph {et~al.}}]{PinzonGuerra:2016uae}%
  \BibitemOpen
  \bibfield  {author} {\bibinfo {author} {\bibfnamefont {E.~S.}\ \bibnamefont
  {Pinzon~Guerra}} \emph {et~al.} (\bibinfo {collaboration} {DUET}),\ }\href
  {\doibase 10.1103/PhysRevC.95.045203} {\bibfield  {journal} {\bibinfo
  {journal} {Phys. Rev.}\ }\textbf {\bibinfo {volume} {C95}},\ \bibinfo {pages}
  {045203} (\bibinfo {year} {2017})},\ \Eprint
  {http://arxiv.org/abs/1611.05612} {arXiv:1611.05612 [hep-ex]} \BibitemShut
  {NoStop}%
\bibitem [{\citenamefont {Adams}\ \emph {et~al.}(2013)\citenamefont {Adams}
  \emph {et~al.}}]{lbne}%
  \BibitemOpen
  \bibfield  {author} {\bibinfo {author} {\bibfnamefont {C.}~\bibnamefont
  {Adams}} \emph {et~al.} (\bibinfo {collaboration} {LBNE Collaboration}),\
  }\href@noop {} {\  (\bibinfo {year} {2013})},\ \Eprint
  {http://arxiv.org/abs/1307.7335} {arXiv:1307.7335 [hep-ex]} \BibitemShut
  {NoStop}%
\bibitem [{\citenamefont {Abe}\ \emph {et~al.}(2015{\natexlab{b}})\citenamefont
  {Abe} \emph {et~al.}}]{t2k_sensitivity_2014}%
  \BibitemOpen
  \bibfield  {author} {\bibinfo {author} {\bibfnamefont {K.}~\bibnamefont
  {Abe}} \emph {et~al.} (\bibinfo {collaboration} {T2K Collaboration}),\ }\href
  {\doibase 10.1093/ptep/ptv031} {\bibfield  {journal} {\bibinfo  {journal}
  {PTEP}\ }\textbf {\bibinfo {volume} {2015}},\ \bibinfo {pages} {043C01}
  (\bibinfo {year} {2015}{\natexlab{b}})},\ \Eprint
  {http://arxiv.org/abs/1409.7469} {arXiv:1409.7469 [hep-ex]} \BibitemShut
  {NoStop}%
\bibitem [{\citenamefont {Aguilar-Arevalo}\ \emph
  {et~al.}(2010{\natexlab{d}})\citenamefont {Aguilar-Arevalo} \emph
  {et~al.}}]{AguilarArevalo:2009ww}%
  \BibitemOpen
  \bibfield  {author} {\bibinfo {author} {\bibfnamefont {A.~A.}\ \bibnamefont
  {Aguilar-Arevalo}} \emph {et~al.} (\bibinfo {collaboration} {MiniBooNE}),\
  }\href@noop {} {\bibfield  {journal} {\bibinfo  {journal} {Phys. Rev.}\
  }\textbf {\bibinfo {volume} {D81}},\ \bibinfo {pages} {013005} (\bibinfo
  {year} {2010}{\natexlab{d}})},\ \Eprint {http://arxiv.org/abs/0911.2063}
  {arXiv:0911.2063 [hep-ex]} \BibitemShut {NoStop}%
\bibitem [{\citenamefont {A.A.Aguilar-Arevalo}\ \emph
  {et~al.}(2011)\citenamefont {A.A.Aguilar-Arevalo} \emph
  {et~al.}}]{AguilarArevalo:2010xt}%
  \BibitemOpen
  \bibfield  {author} {\bibinfo {author} {\bibnamefont {A.A.Aguilar-Arevalo}}
  \emph {et~al.} (\bibinfo {collaboration} {MiniBooNE}),\ }\href@noop {}
  {\bibfield  {journal} {\bibinfo  {journal} {Phys. Rev.}\ }\textbf {\bibinfo
  {volume} {D83}},\ \bibinfo {pages} {052009} (\bibinfo {year} {2011})},\
  \Eprint {http://arxiv.org/abs/1010.3264} {arXiv:1010.3264 [hep-ex]}
  \BibitemShut {NoStop}%
\bibitem [{\citenamefont {Aguilar-Arevalo}\ \emph {et~al.}(2011)\citenamefont
  {Aguilar-Arevalo} \emph {et~al.}}]{AguilarArevalo:2010bm}%
  \BibitemOpen
  \bibfield  {author} {\bibinfo {author} {\bibfnamefont {A.~A.}\ \bibnamefont
  {Aguilar-Arevalo}} \emph {et~al.} (\bibinfo {collaboration} {MiniBooNE}),\
  }\href@noop {} {\bibfield  {journal} {\bibinfo  {journal} {Phys. Rev.}\
  }\textbf {\bibinfo {volume} {D83}},\ \bibinfo {pages} {052007} (\bibinfo
  {year} {2011})},\ \Eprint {http://arxiv.org/abs/1011.3572} {arXiv:1011.3572
  [hep-ex]} \BibitemShut {NoStop}%
\bibitem [{\citenamefont {Eberly}\ \emph {et~al.}(2015)\citenamefont {Eberly}
  \emph {et~al.}}]{Eberly:2014mra}%
  \BibitemOpen
  \bibfield  {author} {\bibinfo {author} {\bibfnamefont {B.}~\bibnamefont
  {Eberly}} \emph {et~al.} (\bibinfo {collaboration} {MINERvA}),\ }\href
  {\doibase 10.1103/PhysRevD.92.092008} {\bibfield  {journal} {\bibinfo
  {journal} {Phys. Rev.}\ }\textbf {\bibinfo {volume} {D92}},\ \bibinfo {pages}
  {092008} (\bibinfo {year} {2015})},\ \Eprint {http://arxiv.org/abs/1406.6415}
  {arXiv:1406.6415 [hep-ex]} \BibitemShut {NoStop}%
\bibitem [{\citenamefont {Aliaga}\ \emph {et~al.}(2015)\citenamefont {Aliaga}
  \emph {et~al.}}]{Aliaga:2015wva}%
  \BibitemOpen
  \bibfield  {author} {\bibinfo {author} {\bibfnamefont {L.}~\bibnamefont
  {Aliaga}} \emph {et~al.},\ }\href@noop {} {\bibfield  {journal} {\bibinfo
  {journal} {Phys. Lett.}\ }\textbf {\bibinfo {volume} {B749}},\ \bibinfo
  {pages} {130} (\bibinfo {year} {2015})},\ \Eprint
  {http://arxiv.org/abs/1503.02107} {arXiv:1503.02107 [hep-ex]} \BibitemShut
  {NoStop}%
\bibitem [{\citenamefont {McGivern}\ \emph {et~al.}(2016)\citenamefont
  {McGivern} \emph {et~al.}}]{McGivern:2016bwh}%
  \BibitemOpen
  \bibfield  {author} {\bibinfo {author} {\bibfnamefont {C.~L.}\ \bibnamefont
  {McGivern}} \emph {et~al.} (\bibinfo {collaboration} {MINERvA}),\ }\href
  {\doibase 10.1103/PhysRevD.94.052005} {\bibfield  {journal} {\bibinfo
  {journal} {Phys. Rev.}\ }\textbf {\bibinfo {volume} {D94}},\ \bibinfo {pages}
  {052005} (\bibinfo {year} {2016})},\ \Eprint
  {http://arxiv.org/abs/1606.07127} {arXiv:1606.07127 [hep-ex]} \BibitemShut
  {NoStop}%
\bibitem [{\citenamefont {Abe}\ \emph {et~al.}(2017{\natexlab{c}})\citenamefont
  {Abe} \emph {et~al.}}]{Abe:2016aoo}%
  \BibitemOpen
  \bibfield  {author} {\bibinfo {author} {\bibfnamefont {K.}~\bibnamefont
  {Abe}} \emph {et~al.} (\bibinfo {collaboration} {T2K}),\ }\href@noop {}
  {\bibfield  {journal} {\bibinfo  {journal} {Phys. Rev.}\ }\textbf {\bibinfo
  {volume} {D95}},\ \bibinfo {pages} {012010} (\bibinfo {year}
  {2017}{\natexlab{c}})},\ \Eprint {http://arxiv.org/abs/1605.07964}
  {arXiv:1605.07964 [hep-ex]} \BibitemShut {NoStop}%
\bibitem [{\citenamefont {Acciarri}\ \emph {et~al.}(2017)\citenamefont
  {Acciarri} \emph {et~al.}}]{Acciarri:2015ncl}%
  \BibitemOpen
  \bibfield  {author} {\bibinfo {author} {\bibfnamefont {R.}~\bibnamefont
  {Acciarri}} \emph {et~al.} (\bibinfo {collaboration} {ArgoNeuT}),\
  }\href@noop {} {\bibfield  {journal} {\bibinfo  {journal} {Phys.Rev.}\
  }\textbf {\bibinfo {volume} {D96}},\ \bibinfo {pages} {012006} (\bibinfo
  {year} {2017})},\ \Eprint {http://arxiv.org/abs/1511.00941} {arXiv:1511.00941
  [hep-ex]} \BibitemShut {NoStop}%
\bibitem [{\citenamefont {Nakayama}(2005)}]{Nakayama:2004dp}%
  \BibitemOpen
  \bibfield  {author} {\bibinfo {author} {\bibfnamefont {S.}~\bibnamefont
  {Nakayama}} (\bibinfo {collaboration} {K2K}),\ }\href@noop {} {\bibfield
  {journal} {\bibinfo  {journal} {Phys. Lett.}\ }\textbf {\bibinfo {volume}
  {B619}},\ \bibinfo {pages} {255} (\bibinfo {year} {2005})},\ \Eprint
  {http://arxiv.org/abs/hep-ex/0408134} {arXiv:hep-ex/0408134} \BibitemShut
  {NoStop}%
\bibitem [{\citenamefont {Vilain}\ \emph {et~al.}(1993)\citenamefont {Vilain}
  \emph {et~al.}}]{Vilain:1993sf}%
  \BibitemOpen
  \bibfield  {author} {\bibinfo {author} {\bibfnamefont {P.}~\bibnamefont
  {Vilain}} \emph {et~al.} (\bibinfo {collaboration} {CHARM-II}),\ }\href
  {\doibase 10.1016/0370-2693(93)91223-A} {\bibfield  {journal} {\bibinfo
  {journal} {Phys. Lett.}\ }\textbf {\bibinfo {volume} {B313}},\ \bibinfo
  {pages} {267} (\bibinfo {year} {1993})}\BibitemShut {NoStop}%
\bibitem [{\citenamefont {Olive}\ \emph {et~al.}(2014)\citenamefont {Olive}
  \emph {et~al.}}]{pdg_2014}%
  \BibitemOpen
  \bibfield  {author} {\bibinfo {author} {\bibfnamefont {K.~A.}\ \bibnamefont
  {Olive}} \emph {et~al.} (\bibinfo {collaboration} {Particle Data Group}),\
  }\href {\doibase 10.1088/1674-1137/38/9/090001} {\bibfield  {journal}
  {\bibinfo  {journal} {Chin. Phys.}\ }\textbf {\bibinfo {volume} {C38}},\
  \bibinfo {pages} {090001} (\bibinfo {year} {2014})}\BibitemShut {NoStop}%
\bibitem [{\citenamefont {Schmitz}\ \emph {et~al.}(2008)\citenamefont {Schmitz}
  \emph {et~al.}}]{AguilarArevalo:2008xs}%
  \BibitemOpen
  \bibfield  {author} {\bibinfo {author} {\bibfnamefont {D.}~\bibnamefont
  {Schmitz}} \emph {et~al.} (\bibinfo {collaboration} {MiniBooNE}),\
  }\href@noop {} {\bibfield  {journal} {\bibinfo  {journal} {Phys. Lett.}\
  }\textbf {\bibinfo {volume} {B664}},\ \bibinfo {pages} {41} (\bibinfo {year}
  {2008})},\ \Eprint {http://arxiv.org/abs/0803.3423} {arXiv:0803.3423
  [hep-ex]} \BibitemShut {NoStop}%
\bibitem [{\citenamefont {Kurimoto}\ \emph
  {et~al.}(2010{\natexlab{a}})\citenamefont {Kurimoto} \emph
  {et~al.}}]{Kurimoto:2009wq}%
  \BibitemOpen
  \bibfield  {author} {\bibinfo {author} {\bibfnamefont {Y.}~\bibnamefont
  {Kurimoto}} \emph {et~al.} (\bibinfo {collaboration} {SciBooNE}),\
  }\href@noop {} {\bibfield  {journal} {\bibinfo  {journal} {Phys. Rev.}\
  }\textbf {\bibinfo {volume} {D81}},\ \bibinfo {pages} {033004} (\bibinfo
  {year} {2010}{\natexlab{a}})},\ \Eprint {http://arxiv.org/abs/0910.5768}
  {arXiv:0910.5768 [hep-ex]} \BibitemShut {NoStop}%
\bibitem [{\citenamefont {Hasegawa}\ \emph {et~al.}(2005)\citenamefont
  {Hasegawa} \emph {et~al.}}]{Hasegawa:2005td}%
  \BibitemOpen
  \bibfield  {author} {\bibinfo {author} {\bibfnamefont {M.}~\bibnamefont
  {Hasegawa}} \emph {et~al.} (\bibinfo {collaboration} {K2K}),\ }\href
  {\doibase 10.1103/PhysRevLett.95.252301} {\bibfield  {journal} {\bibinfo
  {journal} {Phys. Rev. Lett.}\ }\textbf {\bibinfo {volume} {95}},\ \bibinfo
  {pages} {252301} (\bibinfo {year} {2005})},\ \Eprint
  {http://arxiv.org/abs/hep-ex/0506008} {arXiv:hep-ex/0506008 [hep-ex]}
  \BibitemShut {NoStop}%
\bibitem [{\citenamefont {Hiraide}\ \emph {et~al.}(2008)\citenamefont {Hiraide}
  \emph {et~al.}}]{Hiraide:2008eu}%
  \BibitemOpen
  \bibfield  {author} {\bibinfo {author} {\bibfnamefont {K.}~\bibnamefont
  {Hiraide}} \emph {et~al.} (\bibinfo {collaboration} {SciBooNE}),\ }\href
  {\doibase 10.1103/PhysRevD.78.112004} {\bibfield  {journal} {\bibinfo
  {journal} {Phys. Rev.}\ }\textbf {\bibinfo {volume} {D78}},\ \bibinfo {pages}
  {112004} (\bibinfo {year} {2008})},\ \Eprint {http://arxiv.org/abs/0811.0369}
  {arXiv:0811.0369 [hep-ex]} \BibitemShut {NoStop}%
\bibitem [{\citenamefont {Tanaka}(2009)}]{Tanaka:2009ag}%
  \BibitemOpen
  \bibfield  {author} {\bibinfo {author} {\bibfnamefont {H.-K.}\ \bibnamefont
  {Tanaka}},\ }\bibfield  {booktitle} {\emph {\bibinfo {booktitle}
  {{Proceedings, 6th International Workshop on Neutrino-nucleus interactions in
  the few GeV region (NUINT 09): Sitges, Spain, May 18-22, 2009}}},\ }\href
  {\doibase 10.1063/1.3274167} {\bibfield  {journal} {\bibinfo  {journal} {AIP
  Conf. Proc.}\ }\textbf {\bibinfo {volume} {1189}},\ \bibinfo {pages} {255}
  (\bibinfo {year} {2009})},\ \Eprint {http://arxiv.org/abs/0910.4754}
  {arXiv:0910.4754 [hep-ex]} \BibitemShut {NoStop}%
\bibitem [{\citenamefont {Kurimoto}\ \emph
  {et~al.}(2010{\natexlab{b}})\citenamefont {Kurimoto} \emph
  {et~al.}}]{Kurimoto:2010rc}%
  \BibitemOpen
  \bibfield  {author} {\bibinfo {author} {\bibfnamefont {Y.}~\bibnamefont
  {Kurimoto}} \emph {et~al.} (\bibinfo {collaboration} {SciBooNE}),\
  }\href@noop {} {\bibfield  {journal} {\bibinfo  {journal} {Phys. Rev.}\
  }\textbf {\bibinfo {volume} {D81}},\ \bibinfo {pages} {111102} (\bibinfo
  {year} {2010}{\natexlab{b}})},\ \Eprint {http://arxiv.org/abs/1005.0059}
  {arXiv:1005.0059 [hep-ex]} \BibitemShut {NoStop}%
\bibitem [{\citenamefont {Higuera}\ \emph {et~al.}(2014)\citenamefont {Higuera}
  \emph {et~al.}}]{minerva_coh_2014}%
  \BibitemOpen
  \bibfield  {author} {\bibinfo {author} {\bibfnamefont {A.}~\bibnamefont
  {Higuera}} \emph {et~al.} (\bibinfo {collaboration} {{MINER$\nu$A}
  Collaboration}),\ }\href {\doibase 10.1103/PhysRevLett.113.261802} {\bibfield
   {journal} {\bibinfo  {journal} {Phys. Rev. Lett.}\ }\textbf {\bibinfo
  {volume} {113}},\ \bibinfo {pages} {261802} (\bibinfo {year} {2014})},\
  \Eprint {http://arxiv.org/abs/1409.3835} {arXiv:1409.3835 [hep-ex]}
  \BibitemShut {NoStop}%
\bibitem [{\citenamefont {Mislivec}\ \emph {et~al.}(2017)\citenamefont
  {Mislivec} \emph {et~al.}}]{Mislivec:2017qfz}%
  \BibitemOpen
  \bibfield  {author} {\bibinfo {author} {\bibfnamefont {A.}~\bibnamefont
  {Mislivec}} \emph {et~al.} (\bibinfo {collaboration} {Minerva}),\ }\href@noop
  {} {\  (\bibinfo {year} {2017})},\ \Eprint {http://arxiv.org/abs/1711.01178}
  {arXiv:1711.01178 [hep-ex]} \BibitemShut {NoStop}%
\bibitem [{\citenamefont {Abe}\ \emph {et~al.}(2016{\natexlab{b}})\citenamefont
  {Abe} \emph {et~al.}}]{Abe:2016fic}%
  \BibitemOpen
  \bibfield  {author} {\bibinfo {author} {\bibfnamefont {K.}~\bibnamefont
  {Abe}} \emph {et~al.} (\bibinfo {collaboration} {T2K}),\ }\href@noop {}
  {\bibfield  {journal} {\bibinfo  {journal} {Phys. Rev. Lett.}\ }\textbf
  {\bibinfo {volume} {117}},\ \bibinfo {pages} {192501} (\bibinfo {year}
  {2016}{\natexlab{b}})},\ \Eprint {http://arxiv.org/abs/1604.04406}
  {arXiv:1604.04406 [hep-ex]} \BibitemShut {NoStop}%
\bibitem [{\citenamefont {Adamson}\ \emph {et~al.}(2016)\citenamefont {Adamson}
  \emph {et~al.}}]{Adamson:2016hyz}%
  \BibitemOpen
  \bibfield  {author} {\bibinfo {author} {\bibfnamefont {P.}~\bibnamefont
  {Adamson}} \emph {et~al.} (\bibinfo {collaboration} {MINOS}),\ }\href
  {\doibase 10.1103/PhysRevD.94.072006} {\bibfield  {journal} {\bibinfo
  {journal} {Phys. Rev.}\ }\textbf {\bibinfo {volume} {D94}},\ \bibinfo {pages}
  {072006} (\bibinfo {year} {2016})},\ \Eprint
  {http://arxiv.org/abs/1608.05702} {arXiv:1608.05702 [hep-ex]} \BibitemShut
  {NoStop}%
\bibitem [{\citenamefont {Acciarri}\ \emph
  {et~al.}(2014{\natexlab{a}})\citenamefont {Acciarri} \emph
  {et~al.}}]{Acciarri:2014eit}%
  \BibitemOpen
  \bibfield  {author} {\bibinfo {author} {\bibfnamefont {R.}~\bibnamefont
  {Acciarri}} \emph {et~al.} (\bibinfo {collaboration} {ArgoNeuT}),\ }\href
  {\doibase 10.1103/PhysRevLett.113.261801, 10.1103/PhysRevLett.114.039901}
  {\bibfield  {journal} {\bibinfo  {journal} {Phys. Rev. Lett.}\ }\textbf
  {\bibinfo {volume} {113}},\ \bibinfo {pages} {261801} (\bibinfo {year}
  {2014}{\natexlab{a}})},\ \bibinfo {note} {[erratum: Phys. Rev.
  Lett.114,no.3,039901(2015)]},\ \Eprint {http://arxiv.org/abs/1408.0598}
  {arXiv:1408.0598 [hep-ex]} \BibitemShut {NoStop}%
\bibitem [{\citenamefont {Wolcott}\ \emph
  {et~al.}(2016{\natexlab{b}})\citenamefont {Wolcott} \emph
  {et~al.}}]{Wolcott:2016hws}%
  \BibitemOpen
  \bibfield  {author} {\bibinfo {author} {\bibfnamefont {J.}~\bibnamefont
  {Wolcott}} \emph {et~al.} (\bibinfo {collaboration} {MINERvA}),\ }\href
  {\doibase 10.1103/PhysRevLett.117.111801} {\bibfield  {journal} {\bibinfo
  {journal} {Phys. Rev. Lett.}\ }\textbf {\bibinfo {volume} {117}},\ \bibinfo
  {pages} {111801} (\bibinfo {year} {2016}{\natexlab{b}})},\ \Eprint
  {http://arxiv.org/abs/1604.01728} {arXiv:1604.01728 [hep-ex]} \BibitemShut
  {NoStop}%
\bibitem [{\citenamefont {Rein}(1986)}]{Rein:1986cd}%
  \BibitemOpen
  \bibfield  {author} {\bibinfo {author} {\bibfnamefont {D.}~\bibnamefont
  {Rein}},\ }\href {\doibase 10.1016/0550-3213(86)90106-9} {\bibfield
  {journal} {\bibinfo  {journal} {Nucl. Phys.}\ }\textbf {\bibinfo {volume}
  {B278}},\ \bibinfo {pages} {61} (\bibinfo {year} {1986})}\BibitemShut
  {NoStop}%
\bibitem [{\citenamefont {Marshall}\ \emph {et~al.}(2016)\citenamefont
  {Marshall}, \citenamefont {Aliaga}, \citenamefont {Altinok}, \citenamefont
  {Bellantoni}, \citenamefont {Bercellie}, \citenamefont {Betancourt},
  \citenamefont {Bodek}, \citenamefont {Bravar}, \citenamefont {Budd},
  \citenamefont {Cai}, \citenamefont {Carneiro}, \citenamefont {Chvojka},
  \citenamefont {da~Motta}, \citenamefont {Devan}, \citenamefont {Dytman},
  \citenamefont {D\'{\i}az}, \citenamefont {Eberly}, \citenamefont {Endress},
  \citenamefont {Felix}, \citenamefont {Fields}, \citenamefont {Filkins},
  \citenamefont {Fine}, \citenamefont {Gago}, \citenamefont {Galindo},
  \citenamefont {Gallagher}, \citenamefont {Ghosh}, \citenamefont {Golan},
  \citenamefont {Gran}, \citenamefont {Griswold}, \citenamefont {Harris},
  \citenamefont {Higuera}, \citenamefont {Hurtado}, \citenamefont {Kiveni},
  \citenamefont {Kleykamp}, \citenamefont {Kordosky}, \citenamefont {Le},
  \citenamefont {Maher}, \citenamefont {Majoros}, \citenamefont {Manly},
  \citenamefont {Mann}, \citenamefont {Martinez~Caicedo}, \citenamefont
  {McFarland}, \citenamefont {McGivern}, \citenamefont {McGowan}, \citenamefont
  {Messerly}, \citenamefont {Miller}, \citenamefont {Mislivec}, \citenamefont
  {Morf\'{\i}n}, \citenamefont {Mousseau}, \citenamefont {Naples},
  \citenamefont {Nelson}, \citenamefont {Norrick}, \citenamefont {Nuruzzaman},
  \citenamefont {Osta}, \citenamefont {Paolone}, \citenamefont {Park},
  \citenamefont {Patrick}, \citenamefont {Perdue}, \citenamefont
  {Rakotondravohitra}, \citenamefont {Ramirez}, \citenamefont {Ransome},
  \citenamefont {Ray}, \citenamefont {Ren}, \citenamefont {Rimal},
  \citenamefont {Rodrigues}, \citenamefont {Rosenberg}, \citenamefont
  {Ruterbories}, \citenamefont {Schellman}, \citenamefont {Schmitz},
  \citenamefont {Shadler}, \citenamefont {Simon}, \citenamefont
  {Solano~Salinas}, \citenamefont {S\'anchez}, \citenamefont {Tice},
  \citenamefont {Valencia}, \citenamefont {Walton}, \citenamefont {Wang},
  \citenamefont {Watkins}, \citenamefont {Wiley}, \citenamefont {Wolcott},
  \citenamefont {Wospakrik},\ and\ \citenamefont {Zhang}}]{minerva_cckplus}%
  \BibitemOpen
  \bibfield  {author} {\bibinfo {author} {\bibfnamefont {C.~M.}\ \bibnamefont
  {Marshall}}, \bibinfo {author} {\bibfnamefont {L.}~\bibnamefont {Aliaga}},
  \bibinfo {author} {\bibfnamefont {O.}~\bibnamefont {Altinok}}, \bibinfo
  {author} {\bibfnamefont {L.}~\bibnamefont {Bellantoni}},  \emph {et~al.}
  (\bibinfo {collaboration} {MINERvA Collaboration}),\ }\href {\doibase
  10.1103/PhysRevD.94.012002} {\bibfield  {journal} {\bibinfo  {journal} {Phys.
  Rev. D}\ }\textbf {\bibinfo {volume} {94}},\ \bibinfo {pages} {012002}
  (\bibinfo {year} {2016})}\BibitemShut {NoStop}%
\bibitem [{\citenamefont {Marshall}\ \emph {et~al.}(2017)\citenamefont
  {Marshall}, \citenamefont {Aliaga}, \citenamefont {Altinok}, \citenamefont
  {Bellantoni}, \citenamefont {Bercellie}, \citenamefont {Betancourt},
  \citenamefont {Bodek}, \citenamefont {Bravar}, \citenamefont {Cai},
  \citenamefont {Carneiro}, \citenamefont {da~Motta}, \citenamefont {Dytman},
  \citenamefont {D\'{\i}az}, \citenamefont {Dunkman}, \citenamefont {Eberly},
  \citenamefont {Endress}, \citenamefont {Felix}, \citenamefont {Fields},
  \citenamefont {Fine}, \citenamefont {Gago}, \citenamefont {Galindo},
  \citenamefont {Gallagher}, \citenamefont {Ghosh}, \citenamefont {Golan},
  \citenamefont {Gran}, \citenamefont {Harris}, \citenamefont {Higuera},
  \citenamefont {Hurtado}, \citenamefont {Kleykamp}, \citenamefont {Kordosky},
  \citenamefont {Le}, \citenamefont {Maher}, \citenamefont {Manly},
  \citenamefont {Mann}, \citenamefont {Caicedo}, \citenamefont {McFarland},
  \citenamefont {McGivern}, \citenamefont {McGowan}, \citenamefont {Messerly},
  \citenamefont {Miller}, \citenamefont {Mislivec}, \citenamefont
  {Morf\'{\i}n}, \citenamefont {Mousseau}, \citenamefont {Naples},
  \citenamefont {Nelson}, \citenamefont {Norrick}, \citenamefont {Nuruzzaman},
  \citenamefont {Paolone}, \citenamefont {Patrick}, \citenamefont {Perdue},
  \citenamefont {Ram\'{\i}rez}, \citenamefont {Ransome}, \citenamefont {Ray},
  \citenamefont {Ren}, \citenamefont {Rimal}, \citenamefont {Rodrigues},
  \citenamefont {Ruterbories}, \citenamefont {Schmitz}, \citenamefont
  {Solano~Salinas}, \citenamefont {Sultana}, \citenamefont {S\'anchez~Falero},
  \citenamefont {Valencia}, \citenamefont {Walton}, \citenamefont {Wolcott},
  \citenamefont {Wospakrik}, \citenamefont {Yaeggy},\ and\ \citenamefont
  {Zhang}}]{minerva_nckplus}%
  \BibitemOpen
  \bibfield  {author} {\bibinfo {author} {\bibfnamefont {C.~M.}\ \bibnamefont
  {Marshall}}, \bibinfo {author} {\bibfnamefont {L.}~\bibnamefont {Aliaga}},
  \bibinfo {author} {\bibfnamefont {O.}~\bibnamefont {Altinok}}, \bibinfo
  {author} {\bibfnamefont {L.}~\bibnamefont {Bellantoni}},  \emph {et~al.}
  (\bibinfo {collaboration} {MINERvA Collaboration}),\ }\href {\doibase
  10.1103/PhysRevLett.119.011802} {\bibfield  {journal} {\bibinfo  {journal}
  {Phys. Rev. Lett.}\ }\textbf {\bibinfo {volume} {119}},\ \bibinfo {pages}
  {011802} (\bibinfo {year} {2017})}\BibitemShut {NoStop}%
\bibitem [{\citenamefont {Wang}\ \emph {et~al.}(2016)\citenamefont {Wang},
  \citenamefont {Marshall}, \citenamefont {Aliaga}, \citenamefont {Altinok},
  \citenamefont {Bellantoni}, \citenamefont {Bercellie}, \citenamefont
  {Betancourt}, \citenamefont {Bodek}, \citenamefont {Bravar}, \citenamefont
  {Budd}, \citenamefont {Cai}, \citenamefont {Carneiro}, \citenamefont
  {da~Motta}, \citenamefont {Dytman}, \citenamefont {D\'{\i}az}, \citenamefont
  {Eberly}, \citenamefont {Endress}, \citenamefont {Felix}, \citenamefont
  {Fields}, \citenamefont {Fine}, \citenamefont {Galindo}, \citenamefont
  {Gallagher}, \citenamefont {Ghosh}, \citenamefont {Golan}, \citenamefont
  {Gran}, \citenamefont {Harris}, \citenamefont {Higuera}, \citenamefont
  {Hurtado}, \citenamefont {Kiveni}, \citenamefont {Kleykamp}, \citenamefont
  {Kordosky}, \citenamefont {Le}, \citenamefont {Maher}, \citenamefont {Manly},
  \citenamefont {Mann}, \citenamefont {Martinez~Caicedo}, \citenamefont
  {McFarland}, \citenamefont {McGivern}, \citenamefont {McGowan}, \citenamefont
  {Messerly}, \citenamefont {Miller}, \citenamefont {Mislivec}, \citenamefont
  {Morf\'{\i}n}, \citenamefont {Mousseau}, \citenamefont {Naples},
  \citenamefont {Nelson}, \citenamefont {Norrick}, \citenamefont {Nuruzzaman},
  \citenamefont {Paolone}, \citenamefont {Park}, \citenamefont {Patrick},
  \citenamefont {Perdue}, \citenamefont {Rakotondravohitra}, \citenamefont
  {Ramirez}, \citenamefont {Ransome}, \citenamefont {Ray}, \citenamefont {Ren},
  \citenamefont {Rimal}, \citenamefont {Rodrigues}, \citenamefont
  {Ruterbories}, \citenamefont {Schellman}, \citenamefont {Schmitz},
  \citenamefont {Simon}, \citenamefont {Solano~Salinas}, \citenamefont {Tice},
  \citenamefont {Valencia}, \citenamefont {Walton}, \citenamefont {Wolcott},
  \citenamefont {Wospakrik}, \citenamefont {Zavala},\ and\ \citenamefont
  {Zhang}}]{minerva_cohkplus}%
  \BibitemOpen
  \bibfield  {author} {\bibinfo {author} {\bibfnamefont {Z.}~\bibnamefont
  {Wang}}, \bibinfo {author} {\bibfnamefont {C.~M.}\ \bibnamefont {Marshall}},
  \bibinfo {author} {\bibfnamefont {L.}~\bibnamefont {Aliaga}}, \bibinfo
  {author} {\bibfnamefont {O.}~\bibnamefont {Altinok}},  \emph {et~al.}
  (\bibinfo {collaboration} {MINERvA Collaboration}),\ }\href {\doibase
  10.1103/PhysRevLett.117.061802} {\bibfield  {journal} {\bibinfo  {journal}
  {Phys. Rev. Lett.}\ }\textbf {\bibinfo {volume} {117}},\ \bibinfo {pages}
  {061802} (\bibinfo {year} {2016})}\BibitemShut {NoStop}%
\bibitem [{\citenamefont {Tzanov}\ \emph
  {et~al.}(2006{\natexlab{b}})\citenamefont {Tzanov} \emph
  {et~al.}}]{Tzanov:2005kr}%
  \BibitemOpen
  \bibfield  {author} {\bibinfo {author} {\bibfnamefont {M.}~\bibnamefont
  {Tzanov}} \emph {et~al.} (\bibinfo {collaboration} {NuTeV}),\ }\href
  {\doibase 10.1103/PhysRevD.74.012008} {\bibfield  {journal} {\bibinfo
  {journal} {Phys. Rev.}\ }\textbf {\bibinfo {volume} {D74}},\ \bibinfo {pages}
  {012008} (\bibinfo {year} {2006}{\natexlab{b}})},\ \Eprint
  {http://arxiv.org/abs/hep-ex/0509010} {arXiv:hep-ex/0509010 [hep-ex]}
  \BibitemShut {NoStop}%
\bibitem [{\citenamefont {Wu}\ \emph {et~al.}(2008)\citenamefont {Wu} \emph
  {et~al.}}]{Wu:2007ab}%
  \BibitemOpen
  \bibfield  {author} {\bibinfo {author} {\bibfnamefont {Q.}~\bibnamefont {Wu}}
  \emph {et~al.} (\bibinfo {collaboration} {NOMAD}),\ }\href {\doibase
  10.1016/j.physletb.2007.12.027} {\bibfield  {journal} {\bibinfo  {journal}
  {Phys. Lett.}\ }\textbf {\bibinfo {volume} {B660}},\ \bibinfo {pages} {19}
  (\bibinfo {year} {2008})},\ \Eprint {http://arxiv.org/abs/0711.1183}
  {arXiv:0711.1183 [hep-ex]} \BibitemShut {NoStop}%
\bibitem [{\citenamefont {Adamson}\ \emph
  {et~al.}(2010{\natexlab{b}})\citenamefont {Adamson} \emph
  {et~al.}}]{Adamson:2009ju}%
  \BibitemOpen
  \bibfield  {author} {\bibinfo {author} {\bibfnamefont {P.}~\bibnamefont
  {Adamson}} \emph {et~al.} (\bibinfo {collaboration} {MINOS}),\ }\href
  {\doibase 10.1103/PhysRevD.81.072002} {\bibfield  {journal} {\bibinfo
  {journal} {Phys. Rev.}\ }\textbf {\bibinfo {volume} {D81}},\ \bibinfo {pages}
  {072002} (\bibinfo {year} {2010}{\natexlab{b}})},\ \Eprint
  {http://arxiv.org/abs/0910.2201} {arXiv:0910.2201 [hep-ex]} \BibitemShut
  {NoStop}%
\bibitem [{\citenamefont {Anderson}\ \emph {et~al.}(2012)\citenamefont
  {Anderson} \emph {et~al.}}]{Anderson:2011ce}%
  \BibitemOpen
  \bibfield  {author} {\bibinfo {author} {\bibfnamefont {C.}~\bibnamefont
  {Anderson}} \emph {et~al.} (\bibinfo {collaboration} {ArgoNeuT}),\ }\href
  {\doibase 10.1103/PhysRevLett.108.161802} {\bibfield  {journal} {\bibinfo
  {journal} {Phys. Rev. Lett.}\ }\textbf {\bibinfo {volume} {108}},\ \bibinfo
  {pages} {161802} (\bibinfo {year} {2012})},\ \Eprint
  {http://arxiv.org/abs/1111.0103} {arXiv:1111.0103 [hep-ex]} \BibitemShut
  {NoStop}%
\bibitem [{\citenamefont {Abe}\ \emph {et~al.}(2013{\natexlab{b}})\citenamefont
  {Abe} \emph {et~al.}}]{Abe:2013jth}%
  \BibitemOpen
  \bibfield  {author} {\bibinfo {author} {\bibfnamefont {K.}~\bibnamefont
  {Abe}} \emph {et~al.} (\bibinfo {collaboration} {T2K}),\ }\href {\doibase
  10.1103/PhysRevD.87.092003} {\bibfield  {journal} {\bibinfo  {journal} {Phys.
  Rev.}\ }\textbf {\bibinfo {volume} {D87}},\ \bibinfo {pages} {092003}
  (\bibinfo {year} {2013}{\natexlab{b}})},\ \Eprint
  {http://arxiv.org/abs/1302.4908} {arXiv:1302.4908 [hep-ex]} \BibitemShut
  {NoStop}%
\bibitem [{\citenamefont {Acciarri}\ \emph
  {et~al.}(2014{\natexlab{b}})\citenamefont {Acciarri} \emph
  {et~al.}}]{Acciarri:2014isz}%
  \BibitemOpen
  \bibfield  {author} {\bibinfo {author} {\bibfnamefont {R.}~\bibnamefont
  {Acciarri}} \emph {et~al.} (\bibinfo {collaboration} {ArgoNeuT}),\ }\href
  {\doibase 10.1103/PhysRevD.89.112003} {\bibfield  {journal} {\bibinfo
  {journal} {Phys. Rev.}\ }\textbf {\bibinfo {volume} {D89}},\ \bibinfo {pages}
  {112003} (\bibinfo {year} {2014}{\natexlab{b}})},\ \Eprint
  {http://arxiv.org/abs/1404.4809} {arXiv:1404.4809 [hep-ex]} \BibitemShut
  {NoStop}%
\bibitem [{\citenamefont {Nakajima}\ \emph {et~al.}(2011)\citenamefont
  {Nakajima} \emph {et~al.}}]{Nakajima:2010fp}%
  \BibitemOpen
  \bibfield  {author} {\bibinfo {author} {\bibfnamefont {Y.}~\bibnamefont
  {Nakajima}} \emph {et~al.} (\bibinfo {collaboration} {SciBooNE}),\ }\href
  {\doibase 10.1103/PhysRevD.83.012005} {\bibfield  {journal} {\bibinfo
  {journal} {Phys. Rev.}\ }\textbf {\bibinfo {volume} {D83}},\ \bibinfo {pages}
  {012005} (\bibinfo {year} {2011})},\ \Eprint {http://arxiv.org/abs/1011.2131}
  {arXiv:1011.2131 [hep-ex]} \BibitemShut {NoStop}%
\bibitem [{\citenamefont {Abe}\ \emph {et~al.}(2016{\natexlab{c}})\citenamefont
  {Abe} \emph {et~al.}}]{Abe:2015biq}%
  \BibitemOpen
  \bibfield  {author} {\bibinfo {author} {\bibfnamefont {K.}~\bibnamefont
  {Abe}} \emph {et~al.} (\bibinfo {collaboration} {T2K}),\ }\href {\doibase
  10.1103/PhysRevD.93.072002} {\bibfield  {journal} {\bibinfo  {journal} {Phys.
  Rev.}\ }\textbf {\bibinfo {volume} {D93}},\ \bibinfo {pages} {072002}
  (\bibinfo {year} {2016}{\natexlab{c}})},\ \Eprint
  {http://arxiv.org/abs/1509.06940} {arXiv:1509.06940 [hep-ex]} \BibitemShut
  {NoStop}%
\bibitem [{\citenamefont {Abe}\ \emph {et~al.}(2017{\natexlab{d}})\citenamefont
  {Abe} \emph {et~al.}}]{Abe:2017ufe}%
  \BibitemOpen
  \bibfield  {author} {\bibinfo {author} {\bibfnamefont {K.}~\bibnamefont
  {Abe}} \emph {et~al.} (\bibinfo {collaboration} {T2K}),\ }\href {\doibase
  10.1103/PhysRevD.96.052001} {\bibfield  {journal} {\bibinfo  {journal} {Phys.
  Rev.}\ }\textbf {\bibinfo {volume} {D96}},\ \bibinfo {pages} {052001}
  (\bibinfo {year} {2017}{\natexlab{d}})},\ \Eprint
  {http://arxiv.org/abs/1706.04257} {arXiv:1706.04257 [hep-ex]} \BibitemShut
  {NoStop}%
\bibitem [{\citenamefont {Devan}\ \emph
  {et~al.}(2016{\natexlab{b}})\citenamefont {Devan} \emph
  {et~al.}}]{DeVan:2016rkm}%
  \BibitemOpen
  \bibfield  {author} {\bibinfo {author} {\bibfnamefont {J.}~\bibnamefont
  {Devan}} \emph {et~al.} (\bibinfo {collaboration} {MINERvA}),\ }\href
  {\doibase 10.1103/PhysRevD.94.112007} {\bibfield  {journal} {\bibinfo
  {journal} {Phys. Rev.}\ }\textbf {\bibinfo {volume} {D94}},\ \bibinfo {pages}
  {112007} (\bibinfo {year} {2016}{\natexlab{b}})},\ \Eprint
  {http://arxiv.org/abs/1610.04746} {arXiv:1610.04746 [hep-ex]} \BibitemShut
  {NoStop}%
\bibitem [{\citenamefont {Ren}\ \emph {et~al.}(2017)\citenamefont {Ren} \emph
  {et~al.}}]{Ren:2017xov}%
  \BibitemOpen
  \bibfield  {author} {\bibinfo {author} {\bibfnamefont {L.}~\bibnamefont
  {Ren}} \emph {et~al.} (\bibinfo {collaboration} {MINERvA}),\ }\href {\doibase
  10.1103/PhysRevD.95.072009} {\bibfield  {journal} {\bibinfo  {journal} {Phys.
  Rev.}\ }\textbf {\bibinfo {volume} {D95}},\ \bibinfo {pages} {072009}
  (\bibinfo {year} {2017})},\ \Eprint {http://arxiv.org/abs/1701.04857}
  {arXiv:1701.04857 [hep-ex]} \BibitemShut {NoStop}%
\bibitem [{\citenamefont {Chen}\ \emph {et~al.}(2007)\citenamefont {Chen} \emph
  {et~al.}}]{Chen:2007ae}%
  \BibitemOpen
  \bibfield  {author} {\bibinfo {author} {\bibfnamefont {H.}~\bibnamefont
  {Chen}} \emph {et~al.} (\bibinfo {collaboration} {MicroBooNE}),\ }\href@noop
  {} {\  (\bibinfo {year} {2007})}\BibitemShut {NoStop}%
\bibitem [{\citenamefont {Kin}\ \emph {et~al.}(2017)\citenamefont {Kin},
  \citenamefont {Harada}, \citenamefont {Seiya},\ and\ \citenamefont
  {Yamamoto}}]{Kin:2017way}%
  \BibitemOpen
  \bibfield  {author} {\bibinfo {author} {\bibfnamefont {K.}~\bibnamefont
  {Kin}}, \bibinfo {author} {\bibfnamefont {J.}~\bibnamefont {Harada}},
  \bibinfo {author} {\bibfnamefont {Y.}~\bibnamefont {Seiya}}, \ and\ \bibinfo
  {author} {\bibfnamefont {K.}~\bibnamefont {Yamamoto}} (\bibinfo
  {collaboration} {WAGASCI}),\ }\bibfield  {booktitle} {\emph {\bibinfo
  {booktitle} {{Proceedings, 27th International Conference on Neutrino Physics
  and Astrophysics (Neutrino 2016): London, United Kingdom, July 4-9, 2016}}},\
  }\href {\doibase 10.1088/1742-6596/888/1/012125} {\bibfield  {journal}
  {\bibinfo  {journal} {J. Phys. Conf. Ser.}\ }\textbf {\bibinfo {volume}
  {888}},\ \bibinfo {pages} {012125} (\bibinfo {year} {2017})}\BibitemShut
  {NoStop}%
\bibitem [{\citenamefont {Fukuda}(2017)}]{Fukuda:2017ezx}%
  \BibitemOpen
  \bibfield  {author} {\bibinfo {author} {\bibfnamefont {T.}~\bibnamefont
  {Fukuda}} (\bibinfo {collaboration} {NINJA}),\ }\bibfield  {booktitle} {\emph
  {\bibinfo {booktitle} {{Proceedings, 3rd International Symposium on Quest for
  the Origin of Particles and the Universe (KMI2017): Nagoya, Japan, January
  5-7, 2017}}},\ }\href@noop {} {\bibfield  {journal} {\bibinfo  {journal}
  {PoS}\ }\textbf {\bibinfo {volume} {KMI2017}},\ \bibinfo {pages} {012}
  (\bibinfo {year} {2017})}\BibitemShut {NoStop}%
\bibitem [{\citenamefont {Lu}\ \emph {et~al.}(2016)\citenamefont {Lu},
  \citenamefont {Pickering}, \citenamefont {Dolan}, \citenamefont {Barr},
  \citenamefont {Coplowe}, \citenamefont {Uchida}, \citenamefont {Wark},
  \citenamefont {Wascko}, \citenamefont {Weber},\ and\ \citenamefont
  {Yuan}}]{Lu:2015tcr}%
  \BibitemOpen
  \bibfield  {author} {\bibinfo {author} {\bibfnamefont {X.~G.}\ \bibnamefont
  {Lu}}, \bibinfo {author} {\bibfnamefont {L.}~\bibnamefont {Pickering}},
  \bibinfo {author} {\bibfnamefont {S.}~\bibnamefont {Dolan}}, \bibinfo
  {author} {\bibfnamefont {G.}~\bibnamefont {Barr}},  \emph {et~al.},\ }\href
  {\doibase 10.1103/PhysRevC.94.015503} {\bibfield  {journal} {\bibinfo
  {journal} {Phys. Rev.}\ }\textbf {\bibinfo {volume} {C94}},\ \bibinfo {pages}
  {015503} (\bibinfo {year} {2016})},\ \Eprint
  {http://arxiv.org/abs/1512.05748} {arXiv:1512.05748 [nucl-th]} \BibitemShut
  {NoStop}%
\bibitem [{\citenamefont {Cavanna}\ \emph {et~al.}(2014)\citenamefont
  {Cavanna}, \citenamefont {Kordosky}, \citenamefont {Raaf},\ and\
  \citenamefont {Rebel}}]{Cavanna:2014iqa}%
  \BibitemOpen
  \bibfield  {author} {\bibinfo {author} {\bibfnamefont {F.}~\bibnamefont
  {Cavanna}}, \bibinfo {author} {\bibfnamefont {M.}~\bibnamefont {Kordosky}},
  \bibinfo {author} {\bibfnamefont {J.}~\bibnamefont {Raaf}}, \ and\ \bibinfo
  {author} {\bibfnamefont {B.}~\bibnamefont {Rebel}} (\bibinfo {collaboration}
  {LArIAT}),\ }\href@noop {} {\  (\bibinfo {year} {2014})},\ \Eprint
  {http://arxiv.org/abs/1406.5560} {arXiv:1406.5560 [physics.ins-det]}
  \BibitemShut {NoStop}%
\bibitem [{\citenamefont {Abi}\ \emph {et~al.}(2017)\citenamefont {Abi} \emph
  {et~al.}}]{Abi:2017aow}%
  \BibitemOpen
  \bibfield  {author} {\bibinfo {author} {\bibfnamefont {B.}~\bibnamefont
  {Abi}} \emph {et~al.} (\bibinfo {collaboration} {DUNE}),\ }\href@noop {} {\
  (\bibinfo {year} {2017})},\ \Eprint {http://arxiv.org/abs/1706.07081}
  {arXiv:1706.07081 [physics.ins-det]} \BibitemShut {NoStop}%
\bibitem [{\citenamefont {Antonello}\ \emph {et~al.}(2015)\citenamefont
  {Antonello} \emph {et~al.}}]{Antonello:2015lea}%
  \BibitemOpen
  \bibfield  {author} {\bibinfo {author} {\bibfnamefont {M.}~\bibnamefont
  {Antonello}} \emph {et~al.} (\bibinfo {collaboration} {LAr1-ND, ICARUS-WA104,
  MicroBooNE}),\ }\href@noop {} {\  (\bibinfo {year} {2015})},\ \Eprint
  {http://arxiv.org/abs/1503.01520} {arXiv:1503.01520 [physics.ins-det]}
  \BibitemShut {NoStop}%
\bibitem [{\citenamefont {Berns}\ \emph {et~al.}(2013)\citenamefont {Berns}
  \emph {et~al.}}]{Berns:2013usa}%
  \BibitemOpen
  \bibfield  {author} {\bibinfo {author} {\bibfnamefont {H.}~\bibnamefont
  {Berns}} \emph {et~al.} (\bibinfo {collaboration} {CAPTAIN}),\ }in\ \href
  {https://inspirehep.net/record/1253116/files/arXiv:1309.1740.pdf} {\emph
  {\bibinfo {booktitle} {{Proceedings, 2013 Community Summer Study on the
  Future of U.S. Particle Physics: Snowmass on the Mississippi (CSS2013):
  Minneapolis, MN, USA, July 29-August 6, 2013}}}}\ (\bibinfo {year} {2013})\
  \Eprint {http://arxiv.org/abs/1309.1740} {arXiv:1309.1740 [physics.ins-det]}
  \BibitemShut {NoStop}%
\bibitem [{\citenamefont {Anghel}\ \emph {et~al.}(2015)\citenamefont {Anghel}
  \emph {et~al.}}]{Anghel:2015xxt}%
  \BibitemOpen
  \bibfield  {author} {\bibinfo {author} {\bibfnamefont {I.}~\bibnamefont
  {Anghel}} \emph {et~al.} (\bibinfo {collaboration} {ANNIE}),\ }\href@noop {}
  {\  (\bibinfo {year} {2015})},\ \Eprint {http://arxiv.org/abs/1504.01480}
  {arXiv:1504.01480 [physics.ins-det]} \BibitemShut {NoStop}%
\bibitem [{\citenamefont {Bhadra}\ \emph {et~al.}(2014)\citenamefont {Bhadra}
  \emph {et~al.}}]{Bhadra:2014oma}%
  \BibitemOpen
  \bibfield  {author} {\bibinfo {author} {\bibfnamefont {S.}~\bibnamefont
  {Bhadra}} \emph {et~al.} (\bibinfo {collaboration} {nuPRISM}),\ }\href@noop
  {} {\  (\bibinfo {year} {2014})},\ \Eprint {http://arxiv.org/abs/1412.3086}
  {arXiv:1412.3086 [physics.ins-det]} \BibitemShut {NoStop}%
\bibitem [{\citenamefont {Nieves}\ \emph {et~al.}(2004)\citenamefont {Nieves},
  \citenamefont {Amaro},\ and\ \citenamefont {Valverde}}]{Nieves:2004wx}%
  \BibitemOpen
  \bibfield  {author} {\bibinfo {author} {\bibfnamefont {J.}~\bibnamefont
  {Nieves}}, \bibinfo {author} {\bibfnamefont {J.~E.}\ \bibnamefont {Amaro}}, \
  and\ \bibinfo {author} {\bibfnamefont {M.}~\bibnamefont {Valverde}},\ }\href
  {\doibase 10.1103/PhysRevC.70.055503, 10.1103/PhysRevC.72.019902} {\bibfield
  {journal} {\bibinfo  {journal} {Phys. Rev.}\ }\textbf {\bibinfo {volume}
  {C70}},\ \bibinfo {pages} {055503} (\bibinfo {year} {2004})},\ \bibinfo
  {note} {[Erratum: Phys. Rev.C72,019902(2005)]},\ \Eprint
  {http://arxiv.org/abs/nucl-th/0408005} {arXiv:nucl-th/0408005 [nucl-th]}
  \BibitemShut {NoStop}%
\bibitem [{\citenamefont {Bodek}\ and\ \citenamefont
  {Ritchie}(1981)}]{BodekRitchie}%
  \BibitemOpen
  \bibfield  {author} {\bibinfo {author} {\bibfnamefont {A.}~\bibnamefont
  {Bodek}}\ and\ \bibinfo {author} {\bibfnamefont {J.~L.}\ \bibnamefont
  {Ritchie}},\ }\href {\doibase 10.1103/PhysRevD.23.1070} {\bibfield  {journal}
  {\bibinfo  {journal} {Phys. Rev. D}\ }\textbf {\bibinfo {volume} {23}},\
  \bibinfo {pages} {1070} (\bibinfo {year} {1981})}\BibitemShut {NoStop}%
\bibitem [{\citenamefont {Nieves}\ \emph {et~al.}(2011)\citenamefont {Nieves},
  \citenamefont {Simo},\ and\ \citenamefont {Vacas}}]{nieves_2011}%
  \BibitemOpen
  \bibfield  {author} {\bibinfo {author} {\bibfnamefont {J.}~\bibnamefont
  {Nieves}}, \bibinfo {author} {\bibfnamefont {I.~R.}\ \bibnamefont {Simo}}, \
  and\ \bibinfo {author} {\bibfnamefont {M.~J.~V.}\ \bibnamefont {Vacas}},\
  }\href {\doibase 10.1103/PhysRevC.83.045501} {\bibfield  {journal} {\bibinfo
  {journal} {Phys. Rev. C}\ }\textbf {\bibinfo {volume} {83}},\ \bibinfo
  {pages} {045501} (\bibinfo {year} {2011})}\BibitemShut {NoStop}%
\bibitem [{\citenamefont {Gran}\ \emph {et~al.}(2013)\citenamefont {Gran},
  \citenamefont {Nieves}, \citenamefont {Sanchez},\ and\ \citenamefont
  {Vicente~Vacas}}]{Gran:2013kda}%
  \BibitemOpen
  \bibfield  {author} {\bibinfo {author} {\bibfnamefont {R.}~\bibnamefont
  {Gran}}, \bibinfo {author} {\bibfnamefont {J.}~\bibnamefont {Nieves}},
  \bibinfo {author} {\bibfnamefont {F.}~\bibnamefont {Sanchez}}, \ and\
  \bibinfo {author} {\bibfnamefont {M.~J.}\ \bibnamefont {Vicente~Vacas}},\
  }\href {\doibase 10.1103/PhysRevD.88.113007} {\bibfield  {journal} {\bibinfo
  {journal} {Phys. Rev.}\ }\textbf {\bibinfo {volume} {D88}},\ \bibinfo {pages}
  {113007} (\bibinfo {year} {2013})},\ \Eprint {http://arxiv.org/abs/1307.8105}
  {arXiv:1307.8105 [hep-ph]} \BibitemShut {NoStop}%
\bibitem [{\citenamefont {Schwehr}\ \emph {et~al.}(2016)\citenamefont
  {Schwehr}, \citenamefont {Cherdack},\ and\ \citenamefont
  {Gran}}]{Schwehr:2016pvn}%
  \BibitemOpen
  \bibfield  {author} {\bibinfo {author} {\bibfnamefont {J.}~\bibnamefont
  {Schwehr}}, \bibinfo {author} {\bibfnamefont {D.}~\bibnamefont {Cherdack}}, \
  and\ \bibinfo {author} {\bibfnamefont {R.}~\bibnamefont {Gran}},\ }\href@noop
  {} {\  (\bibinfo {year} {2016})},\ \Eprint {http://arxiv.org/abs/1601.02038}
  {arXiv:1601.02038 [hep-ph]} \BibitemShut {NoStop}%
\bibitem [{\citenamefont {Rein}\ and\ \citenamefont
  {Sehgal}(1981)}]{Rein-Sehgal}%
  \BibitemOpen
  \bibfield  {author} {\bibinfo {author} {\bibfnamefont {D.}~\bibnamefont
  {Rein}}\ and\ \bibinfo {author} {\bibfnamefont {L.~M.}\ \bibnamefont
  {Sehgal}},\ }\href {\doibase 10.1016/0003-4916(81)90242-6} {\bibfield
  {journal} {\bibinfo  {journal} {Annals Phys.}\ }\textbf {\bibinfo {volume}
  {133}},\ \bibinfo {pages} {79} (\bibinfo {year} {1981})}\BibitemShut
  {NoStop}%
\bibitem [{\citenamefont {Rein}\ and\ \citenamefont
  {Sehgal}(2007)}]{Rein:2006di}%
  \BibitemOpen
  \bibfield  {author} {\bibinfo {author} {\bibfnamefont {D.}~\bibnamefont
  {Rein}}\ and\ \bibinfo {author} {\bibfnamefont {L.~M.}\ \bibnamefont
  {Sehgal}},\ }\href {\doibase 10.1016/j.physletb.2007.10.025} {\bibfield
  {journal} {\bibinfo  {journal} {Phys. Lett.}\ }\textbf {\bibinfo {volume}
  {B657}},\ \bibinfo {pages} {207} (\bibinfo {year} {2007})},\ \Eprint
  {http://arxiv.org/abs/hep-ph/0606185} {arXiv:hep-ph/0606185 [hep-ph]}
  \BibitemShut {NoStop}%
\bibitem [{\citenamefont {Berger}\ and\ \citenamefont
  {Sehgal}(2009)}]{Berger:2008xs}%
  \BibitemOpen
  \bibfield  {author} {\bibinfo {author} {\bibfnamefont {C.}~\bibnamefont
  {Berger}}\ and\ \bibinfo {author} {\bibfnamefont {L.~M.}\ \bibnamefont
  {Sehgal}},\ }\href {\doibase 10.1103/PhysRevD.79.053003} {\bibfield
  {journal} {\bibinfo  {journal} {Phys. Rev.}\ }\textbf {\bibinfo {volume}
  {D79}},\ \bibinfo {pages} {053003} (\bibinfo {year} {2009})},\ \Eprint
  {http://arxiv.org/abs/0812.2653} {arXiv:0812.2653 [hep-ph]} \BibitemShut
  {NoStop}%
\bibitem [{\citenamefont {A.Bodek}\ and\ \citenamefont
  {Yang}(2003)}]{bodek-yang}%
  \BibitemOpen
  \bibfield  {author} {\bibinfo {author} {\bibnamefont {A.Bodek}}\ and\
  \bibinfo {author} {\bibfnamefont {U.}~\bibnamefont {Yang}},\ }\href@noop {}
  {\bibfield  {journal} {\bibinfo  {journal} {AIP Conf. Proc.}\ }\textbf
  {\bibinfo {volume} {670}},\ \bibinfo {pages} {110} (\bibinfo {year}
  {2003})}\BibitemShut {NoStop}%
\bibitem [{\citenamefont {Yang}\ \emph {et~al.}(2009)\citenamefont {Yang},
  \citenamefont {Andreopoulos}, \citenamefont {Gallagher}, \citenamefont
  {Hoffmann},\ and\ \citenamefont {Kehayias}}]{Yang:2009zx}%
  \BibitemOpen
  \bibfield  {author} {\bibinfo {author} {\bibfnamefont {T.}~\bibnamefont
  {Yang}}, \bibinfo {author} {\bibfnamefont {C.}~\bibnamefont {Andreopoulos}},
  \bibinfo {author} {\bibfnamefont {H.}~\bibnamefont {Gallagher}}, \bibinfo
  {author} {\bibfnamefont {K.}~\bibnamefont {Hoffmann}}, \ and\ \bibinfo
  {author} {\bibfnamefont {P.}~\bibnamefont {Kehayias}},\ }\href {\doibase
  10.1140/epjc/s10052-009-1094-z} {\bibfield  {journal} {\bibinfo  {journal}
  {Eur. Phys. J.}\ }\textbf {\bibinfo {volume} {C63}},\ \bibinfo {pages} {1}
  (\bibinfo {year} {2009})},\ \Eprint {http://arxiv.org/abs/0904.4043}
  {arXiv:0904.4043 [hep-ph]} \BibitemShut {NoStop}%
\end{thebibliography}%
